\documentclass{article}
\pdfoutput=1
\usepackage{graphicx}
\usepackage{epstopdf, epsfig}
\usepackage{verbatim}
\usepackage{siunitx}
\usepackage{natbib}

\usepackage{amsmath,amssymb}
\usepackage{amsfonts}
\usepackage{float,epsfig,graphicx}
\graphicspath{./fig/}
\usepackage{enumitem}
\usepackage[hidelinks,colorlinks=true,citecolor=blue,linkcolor=red]{hyperref}

\usepackage{comment}

\usepackage[top=1in, bottom=1.25in, left=1.25in, right=1.25in]{geometry}
\usepackage{authblk}

\graphicspath{{./fig/}}

\newcommand{\br}{\mathbf r}
\newcommand{\bx}{\mathbf x}
\newcommand{\by}{\mathbf y}
\newcommand{\bz}{\mathbf z}
\newcommand{\bt}{\mathbf t}
\newcommand{\bn}{\mathbf n}
\newcommand{\bb}{\mathbf b}
\newcommand{\bF}{\mathbf F}
\newcommand{\bM}{\mathbf M}

\newcommand{\bomega}{\boldsymbol{\omega}}
\newcommand{\bu}{\mathbf u}

\newcommand\Rey{\mbox{\textit{Re}}}  
\newcommand\Str{\mbox{\textit{St}}}   

\title{Effect of leading-edge curvature actuation on flapping fin performance} 

\author[1]{David Fern\'andez-Guti\'errez}
\author[1]{Wim M.\ van Rees\thanks{wvanrees@mit.edu}}
\affil[1]{Department of Mechanical Engineering, Massachusetts Institute of Technology,
Cambridge, MA 02139, USA}

\date{}

\begin{document}

\maketitle

\begin{abstract}
Ray-finned fish are able to adapt the curvature of their fins through musculature at the base of the fin. In this work we numerically investigate the effects of such leading-edge curvature actuation on the hydrodynamic performance of a heaving and pitching fin. We present a geometric and numerical framework for constructing the shape of ray-membrane type fins with imposed leading-edge curvatures, under the constraint of membrane inextensibility. This algorithm is coupled with a 3D Navier-Stokes solver, enabling us to assess the hydrodynamic performance of such fins. To determine the space of possible shapes, we present a simple model for leading-edge curvature actuation through two coefficients that determine chordwise and spanwise curvature, respectively. We systematically vary these two parameters through regimes that mimic both passive elastic deformations and active curvatures against the hydrodynamic loading, and compute thrust and power coefficients, as well as hydrodynamic efficiency. Our results demonstrate that both thrust and efficiency are predominantly affected by chordwise curvature, with some small additional benefits of spanwise curvature on efficiency. The main improvements in performance are explained by the altered trailing-edge kinematics arising from leading-edge curvature actuation, which can largely be reproduced by a rigid fin whose trailing-edge kinematics follow that of the curving fin. Changes in fin camber, for fixed trailing-edge kinematics, mostly benefit efficiency. Based on our results, we discuss the use of leading-edge curvature actuation as a robust and versatile way to improve flapping fin performance. 
\end{abstract}


\section{Introduction}

The potential of biologically-inspired flapping fin propulsion for practical applications lies in its predicted ability to provide high efficiency at a range of speeds, high maneuverability, and a concealed profile. 
This has spurred a tremendous scientific effort over the last few decades \citep{triantafyllou2000, smits2019} to understand and design bio-inspired propulsion examples that deliver on this potential. A significant development within this design landscape is driven by recent developments in additive manufacturing and smart structures, so that robotic swimmers increasingly incorporate soft, flexible materials \citep{Chu:2012, Katzschmann:2018, Christianson:2018}. This leads to increasingly complex systems, whose behavior is characterized both by passive elastic deformation of the structure as well as actuation degrees-of-freedom that can induce actively-controlled shape changes. Consequently, there is a need to understand to what extent such dynamic shape changes affect hydrodynamic performance of flapping fin propulsion.

Passively deforming elastic surfaces have been studied extensively, as their input parameters and performance can be easily tested and controlled in experimental and numerical settings \citep{katz1978, prempraneerach2003}. Using 2D flat plates, \citet{dewey2013} and \citet{quinn2014,quinn2015} show how the largest thrust is attained when a combination of heave and pitch of the leading edge is imposed so resonance with structural natural frequencies occurs. This conclusion is shared by \citet{tytell2016} and \citet{floryan2018a}, who also show how maximum values of hydrodynamic efficiency, defined as the ratio of thrust power extracted and power required to actuate the fin, are obtained in a localized zone not correlated with the natural modes. 
In 3D, however, \citet{liu1997} observed how passive spanwise deformation alone can be detrimental for the fin efficiency. Combined elastic chordwise and spanwise deformations were studied through potential flow simulations in \citet{zhu2007, zhu2008}. They conclude that curvature can improve hydrodynamic efficiency and reduce the sensitivity on the flow parameters. 
\citet{lin2018} simulated rectangular plates in heaving motion imposing a chordwise flexural motion, reporting how efficiency increases with the degree of flexibility. 

All the above works rely on a structural model of the fin to model purely passive, elastic deformations due to the hydrodynamic loading. Natural rayed fish fins, though, are composed of collagen-membranes supported by bony rays that can be actively curved through a set of muscles at the base of each ray~\citep{lauder2004, alben2007}. As a result, dynamic curvature changes of real fish fins can consist of passive bending due to hydrodynamic loading, as well as active actuation of the individual rays against the flow~\citep{fish2006}. Biological observations show that this musculature is active even during steady swimming \citep{flammang2008}, and how these combined effects lead to complex 3D fin shapes \citep{bainbridge1963, lauder2000, lauder2007b} consisting of both chordwise (along rays) and spanwise (across rays) curvature components. This was quantified in \citet{lauder2005} and \citet{bozkurttas2008}, who used a proper-orthogonal decomposition to break down the fin motion into various modes, observed also experimentally in real fish by \citet{flammang2008, flammang2009}. Their results show how a discrete number of modes capture properly the fin motion gaits. Using a robotic rayed caudal fin model, \citet{lauder2007a, tangorra2009, esposito2012} analyzed the contribution to the thrust production for each of the individual deformation modes. They identified active cupping as the mode that produces the largest amount of thrust, where rays follow a parabolic shape in phase with the pitch, and with the top and bottom rays leading the motion. 

The above body of literature provides a picture that passive elastic deformations of flapping fins can improve their efficiency and thrust production, though finding the best structural design for a given hydrodynamic condition can be challenging. For spanwise elastic deformations the hydrodynamic trends and structural design criteria are not yet systematically investigated. Further, there is an indication that actively curving fins against the hydrodynamic loading can improve hydrodynamic performance, but this needs further investigation. 

Our current work is motivated by the wish to further understand the role of both passive and active curvature changes on the hydrodynamic performance of flapping fin propulsion. However, as opposed to the studies above, we do not explicitly consider any specific elastic model of the fin. Instead, we parametrize the dynamic chordwise and spanwise curvature variations of the fin geometry, and directly explore the effect of imposed curvature variations on hydrodynamic performance. This enables us to side-step the fluid-structure interaction problem, and avoid making any assumptions about materials, elastic properties, and actuation techniques. Instead, our approach aims to identify hydrodynamically beneficial curvature variations of the fin, and understand the underlying flow mechanisms. In a future step, this information can then be used as a target state for a fluid-structure interaction design study, aided by the capability of modern actuation mechanisms for shape-changing structures \citep{Boley:2019}. 

In the rest of this article we detail the proposed mathematical representation of the fin geometry in section~\ref{ss:fin_shape}, showing its capability to reproduce typical swimming modes observed in nature. 
The 3D Navier-Stokes solver used and its integration with the fin-shape generation algorithm is described next in section~\ref{ss:3D_solver}. 
The particular problem definition of a deforming fin subject to heave and pitch solid-body velocities, and the numerical setup adopted to simulate it, are then explained in section~\ref{s:res_problem_definition}. 
Simulation results from the parametric analysis of chordwise and spanwise curvature effects are presented in section~\ref{s:results}, discussing in detail the impact of each curvature type in sections~\ref{ss:results_chordwise} and~\ref{ss:results_spanwise}. Finally, we present some concluding remarks in section~\ref{s:conclusions}.

\section{Methodology} \label{s:methodology}
\subsection{Description of fin shape} \label{ss:fin_shape}

Our description and parametrization of the fin shape builds on our earlier work \citep{fernandezgutierrez2020}, though for clarity we will concisely describe here the complete shape definition and its derivation. 

\subsubsection{Geometric model} \label{sss:geometry_model}
\begin{figure}
\centering
\includegraphics[width=0.8\linewidth]{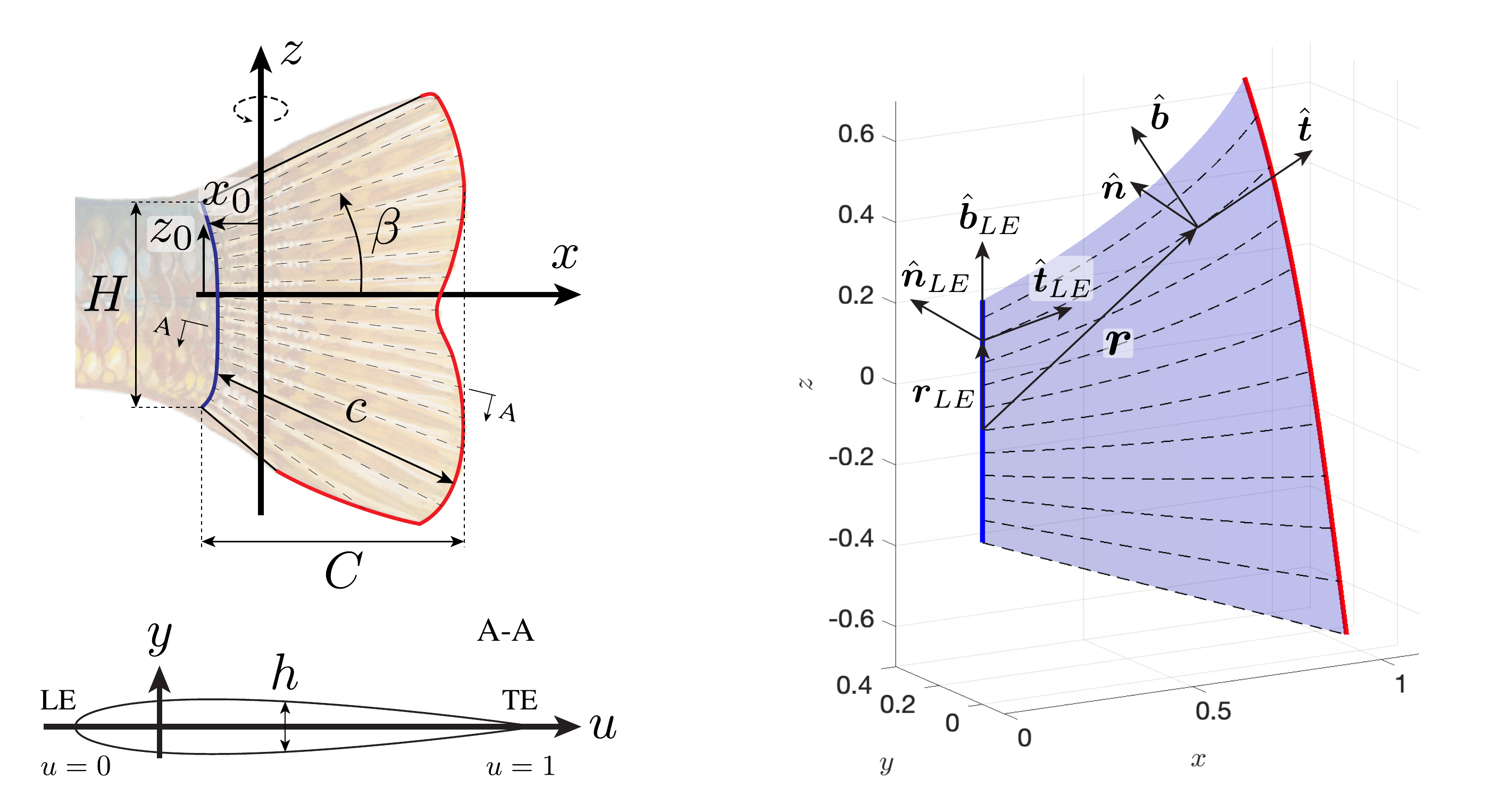}
\caption{Notation and conventions for the geometric representation of the fin \textit{(left)}, and the local coordinate system \textit{(right)}.}
\label{fig:fin_shape_variables}
\end{figure}

We represent any fin geometry by a parametric three-dimensional mid-surface definition combined with a thickness distribution over it. 
Starting with the mid-surface, we introduce parameters $(u,v)$ where $u \in [0,1]$ and $v \in [-1, 1]$. The undeformed mid-surface is defined as
\begin{align}
\label{eq:r_undeformed} \br_0(u, v) = \br_{LE}(v) + u \: c(v) \left[\cos(\beta(v)) \hat{\bx} +  \sin(\beta(v)) \hat{\bz} \right]\,,
\end{align}
where 
\[
\br_{LE}(v) = x_{LE}(v)\hat{\bx} + v H/2 \hat{\bz} \,,
\]
is the leading edge position vector and $x_{LE}(v)$ is the profile of the leading edge, as shown in figure~\ref{fig:fin_shape_variables}. Further, $\beta(v)$ is the angle of the rays along the chordwise direction, $c(v)$ is the length of the chord as measured along a ray, and $H$ the height of the fin at the leading edge (figure~\ref{fig:fin_shape_variables}). The mapping and leading-edge position vector are defined such that, in $\mathbb{R}^3$, the $\hat{\bz}$-axis corresponds to the axis of rotation of the fin. With the mid-surface defined, the description of the rest configuration of the fin can be completed by the thickness function, $h(u,v)$, providing the orthogonal distance between the outer fin surfaces at each side of the fin's mid-surface. Throughout this work, we use the fin overall chord $C$ as length scale, defined as
\[
C = \max_{u,v}(\br_0 \cdot \hat{\bx}) - \min_{u,v}(\br_0 \cdot \hat{\bx}) \,.
\]

To describe the deformed configuration of the mid-surface, we establish a Darboux frame at any location along the rays as shown on the right side of figure~\ref{fig:fin_shape_variables}. The frame is characterized by the tangent unit vector along the rays, $\hat{\bt}$, the normal unit vector to the mid-surface, $\hat{\bn}$, and the bi-normal unit vector $\hat{\bb} = \hat{\bt} \times \hat{\bn}$. Note that our vectors $\hat{\bb}$ and $\hat{\bn}$ are rotated compared to the normal and binormal vectors arising using a Frenet framing of a space curve, due to the fact that here $\hat{\bn}$ corresponds to the mid-surface normal vector. 
Using the Darboux framing, we can then define three non-dimensional curvatures corresponding to the directions of the local coordinate system, defined as
\begin{align}
\label{eq:curvatures} \left. \begin{array}{l}
\dfrac{\mathrm{d}\hat{\bt}}{\mathrm{d}u} =  + \kappa^{n} \hat{\bn} + \kappa^{g} \hat{\bb} \vspace{5pt} \\
\dfrac{\mathrm{d} \hat{\bn}}{\mathrm{d}u} =  -\kappa^{n} \hat{\bt} +  \kappa^{t} \hat{\bb} \vspace{5pt} \\
\dfrac{\mathrm{d} \hat{\bb}}{\mathrm{d}u} =  -\kappa^{g} \hat{\bt} - \kappa^{t} \hat{\bn} 
\end{array}  \: \right\} 
\hspace{10pt} \leftrightarrow \ \hspace{10pt} 
\left. \begin{array}{l}
\kappa^{t} =  \dfrac{\mathrm{d}\hat{\bn}}{\mathrm{d}u} \cdot  \hat{\bb} =  - \dfrac{\mathrm{d} \hat{\bb}}{\mathrm{d}u} \cdot \hat{\bn} \vspace{5pt} \\
\kappa^{g} =  \dfrac{\mathrm{d}\hat{\bt}}{\mathrm{d}u} \cdot  \hat{\bb} = - \dfrac{\mathrm{d} \hat{\bb}}{\mathrm{d}u} \cdot \hat{\bt} \vspace{5pt} \\
\kappa^{n} =  \dfrac{\mathrm{d}\hat{\bt}}{\mathrm{d}u} \cdot  \hat{\bn} = - \dfrac{\mathrm{d} \hat{\bn}}{\mathrm{d}u} \cdot \hat{\bt} 
\end{array} \: \right\} \,,
\end{align}
where the curvatures are non-dimensional; the dimensionalized forms can be found when multiplying with the local chord $c(v)$. More precisely, the values of $\kappa^t$, $\kappa^g$, and $\kappa^n$, respectively, correspond to the geodesic torsion, geodesic curvature, and normal curvature of the constant-$v$ curve on the mid-surface. 

For the \textit{deformed} configuration, we can then write the position of the mid-surface as
\begin{equation}
\label{eq:shape_exactsol} \br(u, v) = \br_{LE}(v) + c(v) \int_{0}^{u} \hat{\bt}(u^*,v) \:\mathrm{d}u^*,
\end{equation}
with $\br_{LE}(v)$ defined as above, and $u^*$ an integration variable. We can in turn express $\hat{\bt}$ in terms of the curvatures from equation~\eqref{eq:curvatures} as
\begin{equation}
\label{eq:shape_exactsol2}
\hat{\bt}(u,v) = \hat{\bt}_{LE}(v) + \int_0^u \left[ \kappa^{n} \hat{\bn} + \kappa^{g} \hat{\bb} \right](u^*,v) \:\mathrm{d}u^*,
\end{equation}
where $\hat{\bt}_{LE}(v) = \cos(\beta(v)) \hat{\bx} + \sin(\beta(v)) \hat{\bz}$ is the tangent unit vector at the undeformed LE. The problem of finding the deformed mid-surface is then reduced to finding the functional form of the three curvatures, or, equivalently, the basis $(\hat{\bt}, \hat{\bn}, \hat{\bb})$ along each ray. Note that when $\kappa^{n} = \kappa^{g} = 0$, we recover the flat configuration described in equation~\eqref{eq:r_undeformed}.  

Mechanically, fish can actuate the rays at the LE to balance the hydrodynamic loading, acting as control mechanism of $\kappa^{n}$ for each ray \citep{alben2007}. Thus, $\kappa^{n}$ becomes a controllable degree of freedom, allowing us to consider it as a known, user-defined input whose specific form will be discussed further in section~\ref{ss:curvature_parametrization}. 

To find corresponding expressions for $\kappa^{g}$ and $\kappa^{t}$, we use two assumptions. First, we treat the membrane connecting the rays as inextensible based on its material properties \citep{alben2007,nguyen2017}, so  $\|\mathrm{d}\br(u,v)\|=\|\mathrm{d}\br_0(u,v)\|$ where $\mathrm{d}\br$ and $\mathrm{d}\br_0$ are the differential of the deformed and undeformed mid-surface position, respectively. 
Second, we assume that the membrane remains smooth, which discretely implies that the mid-surface normals as obtained from integrating the Darboux frame along each ray are consistent with the mid-surface normals as obtained from differentiating the position vector across rays, as further explained in the next section.

Lastly, to obtain the volumetric shape of the deformed fin, we neglect the effect of transverse normal and shear strains, similar to the Kirchhoff-Love assumptions in plate and shell theory, so that the thickness function remains unchanged in the deformed configuration.

\subsubsection{Discrete representation and solution algorithm for the fin geometry} \label{sss:discrete_geom}
The exact solution to the mid-surface shape formulation described in section~\ref{sss:geometry_model} is difficult to find, so we propose here an iterative solution technique that maintains the discrete error in satisfying the aforementioned constraints below a user-specified threshold. 

We start by discretizing the mid-surface into a structured mesh with $N_v$ rays in the spanwise direction, each of which is represented through a set of $N_u$ equidistant nodes. Throughout we assume a known functional form of $\kappa^n$, and impose zero curvature at the tips ($\kappa^{g}_{i,N_u} = \kappa^{t}_{i,N_u} = 0$) and symmetric $\kappa^{t}$ across the $i_c$-th central element ($i_c = \left\lceil N_v/2 \right\rceil$), 
\begin{align*}
\kappa^{t}_{i_c,j} / c_{i_c} &= \left\{ \begin{array}{ll}
0 & \text{$N_v$ odd} \\
-\kappa^{t}_{i_c+1,j}/c_{i_c+1} & \text{$N_v$ even} 
\end{array} \right. \,.
\end{align*}

We then assume initial values for the remaining values of $\kappa_{i,j}^g$ and $\kappa_{i,j}^t$, and determine the location of the ray nodes by discretely integrating the Darboux frame along each ray, according to equations~\eqref{eq:curvatures}-\eqref{eq:shape_exactsol2}. Using a finite-difference approximation of the derivatives, and noting that the resulting vector after applying the transformation needs to be re-normalized, this leads to a marching algorithm for the $i$-th~ray:
\begin{align}
\label{eq:K_matrix} K_{i,j} &= \left[ \begin{array}{rrr}
0 & -\kappa_{i,j}^{n} & -\kappa_{i,j}^{g} \\
\kappa_{i,j}^{n} & 0 & -\kappa_{i,j}^{t} \\
\kappa_{i,j}^{g} & \kappa_{i,j}^{t} & 0
\end{array} \right]  \,,\\
\left[ \bt^{*} \:\: \bn^{*} \:\: \bb^{*} \right]_{i,j+1} &= \left[ \hat{\bt} \:\: \hat{\bn} \:\: \hat{\bb} \right]_{i,j} \left(  \mathbb{I}_3 + \frac{K_{i,j+1} + K_{i,j}}{2}\, \Delta u  \right) \,, \\
\left[ \hat{\bt} \:\: \hat{\bn} \:\: \hat{\bb} \right]_{i,j+1} &= \left[ \frac{\bt^{*}}{\|\bt^{*}\|} \:\: \frac{\bn^{*}}{\|\bn^{*}\|} \:\: \frac{\bb^{*}}{\|\bb^{*}\|} \right]_{i,j+1} \,, \\
\label{eq:rij_pos} \br_{i,j+1} &= \br_{i,j} + \frac{\hat{\bt}_{i,j} + \hat{\bt}_{i,j+1}}{\| \hat{\bt}_{i,j} + \hat{\bt}_{i,j+1}\|} \, c_{i} \, \Delta u \,,
\end{align}
where $\mathbb{I}_3$ is the $3 \times 3$ identity matrix and $\Delta u=1/(N_u -1)$. For each ray, we use as initial values the known LE position $\br_{i,1}$ and direction vectors $[\hat{\bt},\hat{\bn},\hat{\bb}]_{i,1}$ from the rigid-body kinematics. 

Given the above procedure to compute the Darboux frame and position vector for each ray, we can then update our initial guesses for $\kappa^g$ and $\kappa^t$ using a Newton-Raphson algorithm. The goal of the algorithm is to minimize deviation from the inextensibility and smoothness constraints, quantified by the signed error metrics $\mathcal{E}_l^{\mathrm{dist}}$ and $\mathcal{E}_l^{\mathrm{smth}}$, respectively:
\begin{align}
\label{eq:error_metric_1} \mathcal{E}_l^{\mathrm{dist}} &= \left\{ \begin{array}{ll}
\dfrac{\| \br_{i+1,j} - \br_{i,j} \|}{d_{i,j}} -1 & i<i_c :\: l=i+(j-2) (N_v-1) \,,\\
\dfrac{\| \br_{i,j} - \br_{i-1,j} \|}{d_{i-1,j}} -1  & i>i_c :\: l=(i-1)+(j-2) (N_v-1) \,,\\
\end{array}\right. \\
\label{eq:error_metric_2} \mathcal{E}_l^{\mathrm{smth}} &= \left( \hat{\bn}_{i,j} \times \hat{\bn}^{(\br)}_{i,j} \right) \cdot \hat{\bt}_{i,j}, \quad l=i+(j-2) N_v + (N_v-1)(N_u-1) \,,\\
\label{eq:n_hat_r} \hat{\bn}^{(\br)}_{i,j} &= \frac{\br_{i+1,j} - \br_{i-1,j}}{\| \br_{i+1,j} - \br_{i-1,j} \|} \times \bt_{i,j} \,, 
\end{align}
where $l$ is a global index to identify each unknown curvature, $d_{i,j}$ the spanwise distance between adjacent nodes in the undeformed configuration computed analytically from equation~\eqref{eq:r_undeformed}, and $\hat{\bn}^{(\br)}_{i,j} = [(\partial \br / \partial v) / \| (\partial \br / \partial v) \| \times \hat{\bt} ]_{i,j}$ the normal direction from adjacent ray nodes based on the smoothness constraint that analytically should match $\hat{\bn}_{i,j}$. 

We numerically differentiate these error metrics with respect to the unknown curvature variables to determine the Jacobian of the system:
{\small
\begin{align}
\mathcal{J}_{l,m} &\approx \left[ \begin{array}{cc}
\dfrac{\Delta \mathcal{E}^{\mathrm{dist}}_l}{\Delta \kappa^{g}_m} & \dfrac{\Delta \mathcal{E}^{\mathrm{dist}}_l}{\Delta \kappa^{t}_m} \vspace{5pt}\\
\dfrac{\Delta \mathcal{E}^{\mathrm{smth}}_l}{\Delta \kappa^{g}_m} & \dfrac{\Delta \mathcal{E}^{\mathrm{smth}}_l}{\Delta \kappa^{t}_m}
\end{array} \right] \,, 
\begin{array}{l} \kappa^{g}_m \equiv \kappa^{g}_{i,j} \left\{ \begin{array}{l}
i<i_c :\: m=i+(j-1) (N_v-1)\,, \vspace{5pt}\\
i>i_c :\: m=(i-1)+(j-1) (N_v-1) \,,
\end{array} \right. \vspace{5pt}\\
\kappa^{t}_m \equiv \kappa^{t}_{i,j} \quad m=i+(j-1) N_v + (N_v-1)(N_u-1) \,.
\end{array} 
\end{align}
}
In each Newton-Raphson step we then invert the Jacobian matrix using an LU decomposition with partial pivoting to update the curvature values:
\begin{align}
\label{eq:newton_raphson_phipsi} 
\left[ \begin{array}{c}
\kappa^{g}_m \\
\kappa^{t}_m
\end{array}\right]^{(k+1)}
 = \left[ \begin{array}{c}
\kappa^{g}_m \\
\kappa^{t}_m
\end{array}\right]^{(k)} - \left[\mathcal{J}_{l,m}^{(k)}\right]^{-1} \left[ \begin{array}{c}
\mathcal{E}_l^{\mathrm{dist}} \\
\mathcal{E}_l^{\mathrm{smth}}
\end{array}\right]^{(k)} \,.
\end{align}
where $k$ denotes the Newton-Raphson iteration. Given the new curvature values $\kappa^{g,(k+1)}$ and $\kappa^{t,(k+1)}$, we can again evaluate equations~\eqref{eq:K_matrix}-\eqref{eq:rij_pos} to compute the corresponding new Darboux frame and position vectors, and evaluate the associated error metrics~\eqref{eq:error_metric_1}-\eqref{eq:error_metric_2}. If they are below a given threshold, $\left| \mathcal{E}_l^{\mathrm{dist}} \right| < \epsilon^{\mathrm{dist}}$ and $\left| \mathcal{E}_l^{\mathrm{smth}} \right| < \epsilon^{\mathrm{smth}} \:\: \forall l$, the solution has been found and we stop. Otherwise, we start a new iteration by updating the Jacobian matrix with the new ray configuration.

\subsubsection{Interpolation to reduce computational cost} \label{sss:interpolation}
We can significantly improve the algorithm's performance by solving for the values of $\kappa^{t}$ and $\kappa^{g}$ on a coarser mesh, with $N_r \ll N_v$ rays and $N_s \ll N_u$ nodes along them, and use interpolation to determine the intermediate values in the finer mesh taking advantage of the smooth nature of the mid-surface. 

The interpolation scheme uses first a quadratic approximation to determine the chordwise derivatives of $\kappa^t$ and $\kappa^g$ at each coarse grid node using 3-point stencils with values at the node and its closest neighbors. Then, the interpolated values $\kappa_{i,j}^t$, $\kappa_{i,j}^g$ are determined between each pair of nodes using a cubic interpolation using the curvatures and its derivatives at the nodes. 
Using the interpolated curvatures along each ray, we can determine the fine-grid node locations along each ray following equations~\eqref{eq:K_matrix}-\eqref{eq:rij_pos}. 
Then, we can obtain the fine-grid node locations between rays following a similar interpolation procedure, now in the spanwise direction, determining the derivative values using a quadratic fit and then interpolating the node coordinates $\mathbf{r}_{i,j}$ with a cubic spline. Note that under this approach, the spanwise position derivatives are computed explicitly for each node and therefore are continuous across nodes.

With this adjustment, we still follow the iterative algorithm described in section~\ref{sss:discrete_geom}, substituting $(i,j) \rightarrow (p,q)$ where $p \in [1,N_r]$ and $q \in [1,N_s]$. In addition, we can use the fine-grid interpolated nodes to determine the distance between nodes, as well as the spanwise derivative values at the nodes for $\mathcal{E}^{\mathrm{dist}}$, so they are defined now as
\begin{align}
\| \br_{p+1,q} - \br_{p,q} \| \hspace{10pt} &\rightarrow \hspace{10pt} 
\sum_{i=i_p}^{i_{p+1}-1} \| \br_{i+1,j_q} - \br_{i,j_q} \| \,, \\
\bn^{(\br)}_{p,q} &= \left( \frac{\mathrm{d}\br}{\mathrm{d}v} \right)_{p,q} \times \bt_{p,q} \,,
\end{align}
where $i_p$ and $j_q$ are the $i$ and $j$ indexes corresponding to the $p$-th ray and $q$-th chordwise location in the coarser grid, respectively.

\subsection{3D Navier-Stokes solver and integration of fin shape} \label{ss:3D_solver}

We use in this work the remeshed vortex method with a penalization technique \citep{gazzola2011}, which solves the 3D viscous incompressible Navier-Stokes equations in vorticity-velocity form:
\begin{align}
    \frac{\partial \bomega}{\partial t} + (\bu \cdot \nabla) \bomega = (\bomega \cdot \nabla) \bu + \nu \nabla^2 \bomega + \lambda \nabla \times \left[ \chi (\bu_s - \bu) \right] \,,
\end{align}
where $\bomega = \nabla \times \bu$ is the vorticity vector and $\bu$ is the fluid velocity vector. The last term on the right-hand side is responsible for enforcing the solid-body boundary conditions, with $\chi$ the characteristic function representing the body ($\chi=1$ inside the body, $\chi=0$ outside, and mollified at the interface), $\bu_s$ the imposed velocity inside the body, and $\lambda \gg 1$ the penalization factor that dynamically forces the flow inside the body to follow the imposed body motion. As explained in \cite{gazzola2011}, we solve the velocity from the vorticity by inverting a Poisson's equation with free-space boundary conditions, enabling the use of a compact domain. This framework has been validated extensively in the past for simulations and optimizations related to self-propelled 2D and 3D swimmers \citep{gazzola2011, gazzola2012, rees2013, rees2015}. In the context of this work, we also verified our method in appendix~A in the supplementary data 
for flapping fin propulsion specifically.

To integrate our model, we can decompose the body velocity field at any point $\br$ inside the body as $\bu_s(\br,t) = \bu_T(t) + \bu_R(\br,t) + \bu_{\text{def}}(\br,t)$, where $ \bu_T(t)$ is the translational velocity, $ \bu_R(\br,t) = \boldsymbol{\dot{\theta}}(t) \times \br$ is the rigid-body rotational velocity \footnote{Note that the fin pitches around the $z$-axis, so the origin of the position vector $\br$ is always at the center of rotation.}, and $\bu_{\text{def}}(\br,t)$ is the deformation velocity field arising from a time-varying curvature distribution. In this work, $\bu_T(t)$ and $\boldsymbol{\dot{\theta}}(t)$ are imposed through the heave and pitch kinematics of the fin, and $\chi(\br,t)$ and $\bu_{\text{def}}(\br,t)$ are determined from the geometric model characterizing the fin shape described in section~\ref{ss:fin_shape}.

As in \citet{bernier2019}, we compute the overall hydrodynamic force and moment acting on the body from the projection and penalization components, such that
\begin{align}
    \bF &= \int_{\Omega_b} \nabla \cdot \sigma \:\mathrm{d}V 
    = \overbrace{\frac{\mathrm{D}}{\mathrm{D}t} \int_{\mathcal{V}_b} \rho \bu \:\mathrm{d}V}^{\bF_{\mathrm{proj}}}  +  \overbrace{\int_{\Omega_b} \rho \lambda \chi \left( \bu - \bu_S \right) \:\mathrm{d}V}^{\bF_{\mathrm{penal}}} \,,\\ 
    \bM &= \int_{\Omega} \br \times \left(\nabla \cdot \sigma \right) \:\mathrm{d}V   = \overbrace{\frac{\mathrm{D}}{\mathrm{D}t} \int_{\mathcal{V}_b} \br \times \left( \rho \bu \right) \:\mathrm{d}V}^{\bM_{\mathrm{proj}}}  +  \overbrace{\int_{\Omega_b} \br \times \left[\rho \lambda \left( \bu - \bu_S \right)\right] \:\mathrm{d}V}^{\bM_{\mathrm{penal}}} \,.
\end{align}
where $\Omega_b$ and $\mathcal{V}_{b}$ and are the control and material volume of the solid body. 
We further identify the horizontal component opposite to the incident flow as thrust, and the transverse component in the direction of heave as lift, 
\begin{align}
    \label{eq:thrust} T &= - \bF \cdot \hat{\bx} \,,\\
    \label{eq:lift} L &= \bF \cdot \hat{\by} \,.
\end{align}

Following a similar approach, we can compute the power required to overcome the hydrodynamic loads and actuate the fin. Starting from the general definition \citep{winter1987} applied to a control volume coinciding with the body ,
\begin{align}
    \label{eq:power_general} P = -\int_{\Omega_b} \nabla \cdot \left(\sigma \bu \right) \mathrm{d}V 
    = - \int_{\Omega_b} \left[ \left(\nabla \cdot \sigma\right) \cdot \bu  + \nabla \bu :  \sigma \right] \: \mathrm{d}V \,,
\end{align}
we can use the incompressible Newtonian stress tensor $\sigma = -p \mathbb{I} + \mu \left( \nabla \bu + \nabla \bu^T \right)$, where $p$ is the fluid pressure, $\mathbb{I}$ the identity tensor and $^T$ the transpose operator, to express the power as 
\begin{align}
    \label{eq:power} P = - \int_{\Omega_b} \mu \nabla\bu : \left( \nabla\bu + \nabla\bu^T \right) \:\mathrm{d}V 
- \frac{\mathrm{D}}{\mathrm{D}t} \int_{\mathcal{V}_b} \frac{\rho}{2} \bu \cdot \bu \:\mathrm{d}V 
- \int_{\Omega_b} \lambda \chi \left( \bu - \bu_S \right) \cdot \bu \:\mathrm{d}V \,.
\end{align}

\section{Problem definition} \label{s:res_problem_definition}

In this section we will first explain our choice of flow regime and fin details, determined by Reynolds and Strouhal number, the fin geometry, and the rigid-body fin kinematics. We will then explain our parametrization choices for the fin curvature through $\kappa^{n}$. Finally we will discuss the numerical settings and performance metrics used to generate the results.

\subsection{Flow regime and fin details}

We model the fin shape as a simple trapezoidal planform pitching around the leading edge, to simplify the large variety of fin shapes observed in nature. 
As discussed more in depth in our previous work \citep{fernandezgutierrez2020}, we choose $H=0.6C$ as leading edge height and $1.35C$ as trailing edge height inspired by the caudal fin of a bluegill sunfish as a representative ray-finned fish. 
The fin moves with rigid-body kinematics consisting of the following harmonic heaving and pitching motion:
\begin{align} 
\label{eq:ref_heave} y(t) &= A_y \sin(2 \pi f t) \,,\\
\label{eq:ref_pitch} \theta(t) &= A_\theta \sin(2 \pi f t + \varphi_\theta),
\end{align}
where $f$ is the flapping frequency, $A_y$ the heaving amplitude, $A_\theta$ the pitching amplitude and $\varphi_\theta$ the phase angle between heave and pitch. The rigid-body components of the body velocity $\bu_s$ are then imposed as
\begin{align}
    \label{eq:u_t} \bu_T(t) &= \dot{y}(t) \hat{\by} \,,\\ 
    \label{eq:u_r} \bu_R(\br,t) &= \dot{\theta}(t) \hat{\bz} \times \br  \,.
\end{align}

The free parameters are chosen based on a review of existing studies in this realm. Specifically, we set $\tilde{A}_y = A_y / C = 0.4$, consistent with the suggestion of \citet{triantafyllou2000} of amplitudes of heave motion comparable to the chord lengths; we use $A_{\theta} = 30^{\circ}$, following the biological observations shown by \citet{hu2016}; and we choose $\varphi_{\theta} = -90^{\circ}$, as suggested by \citet{read2003} for optimum efficiency. 

The flow regime, characterized by the Reynolds number $\Rey = U_{\infty} C/\nu$, is limited by the computational requirements of the solver. In this work we set it to $\Rey=1500$, which is lower than most adult fish but representative of smaller and early-stage fishes. Lastly, the flapping frequency is non-dimensionalized through the Strouhal number $\Str=2 f A_y / U_{\infty}$, where $U_\infty$ is the free-stream velocity magnitude. We fix the Strouhal number $\Str = 0.3$, consistent with experimental observations of real fish and theoretical scaling laws at this Reynolds number~\citep{triantafyllou2000, gazzola2014, floryan2018b}.

\subsection{Curvature parametrization} \label{ss:curvature_parametrization}

Though the algorithm presented in section \ref{sss:discrete_geom} is general, we choose here a simple parametrization of $\kappa^n$ that enables us to investigate a representative range of curvature variations. First, we set the curvature to a constant along each ray, so that $\kappa^{n}(u, v, t) = \kappa^{n}_0(v, t)$, which mimics the type of leading-edge control demonstrated in real fish \citep{alben2007}. Second, we define the leading-edge curvature as a linear combination of uniform and parabolic curvature profiles across the span of the fin. Based on experimental observations \citep{flammang2008}, we further choose to apply the uniform curvature variations in-phase with the heave, and the parabolic curvature variations with a $\ang{90}$ phase-shift, so that the top and bottom rays lead the center ray.
Mathematically, this leads to the following non-dimensional normal curvature parametrization 
\begin{align}
\label{eq:normal_curvature} \kappa^{n}_0(v, t) = \frac{c(v)}{C} \left[ a_{0}^{\kappa} \cos(\beta(v)) \sin (2\pi f t) + a_{2}^{\kappa} \: v^2  \cos (2\pi f t) \right] \,,
\end{align}{}
reducing the curvature characterization to two coefficients modulating the chordwise ($a_0^{\kappa}$) and spanwise ($a_2^{\kappa}$) curvature variations, respectively. The inclusion of the overall chord in equation \eqref{eq:normal_curvature} makes the the imposed LE curvature distribution independent of the chord length distribution across rays. Further, the $\cos(\beta(v))$ factor in the first term accounts for the orientation of each ray, so that $a_0^\kappa$ controls purely cylindrical deformation modes of the fin (see appendix~B in the supplementary data 
for more details). 

\begin{figure}
\centering
\includegraphics[width=\linewidth]{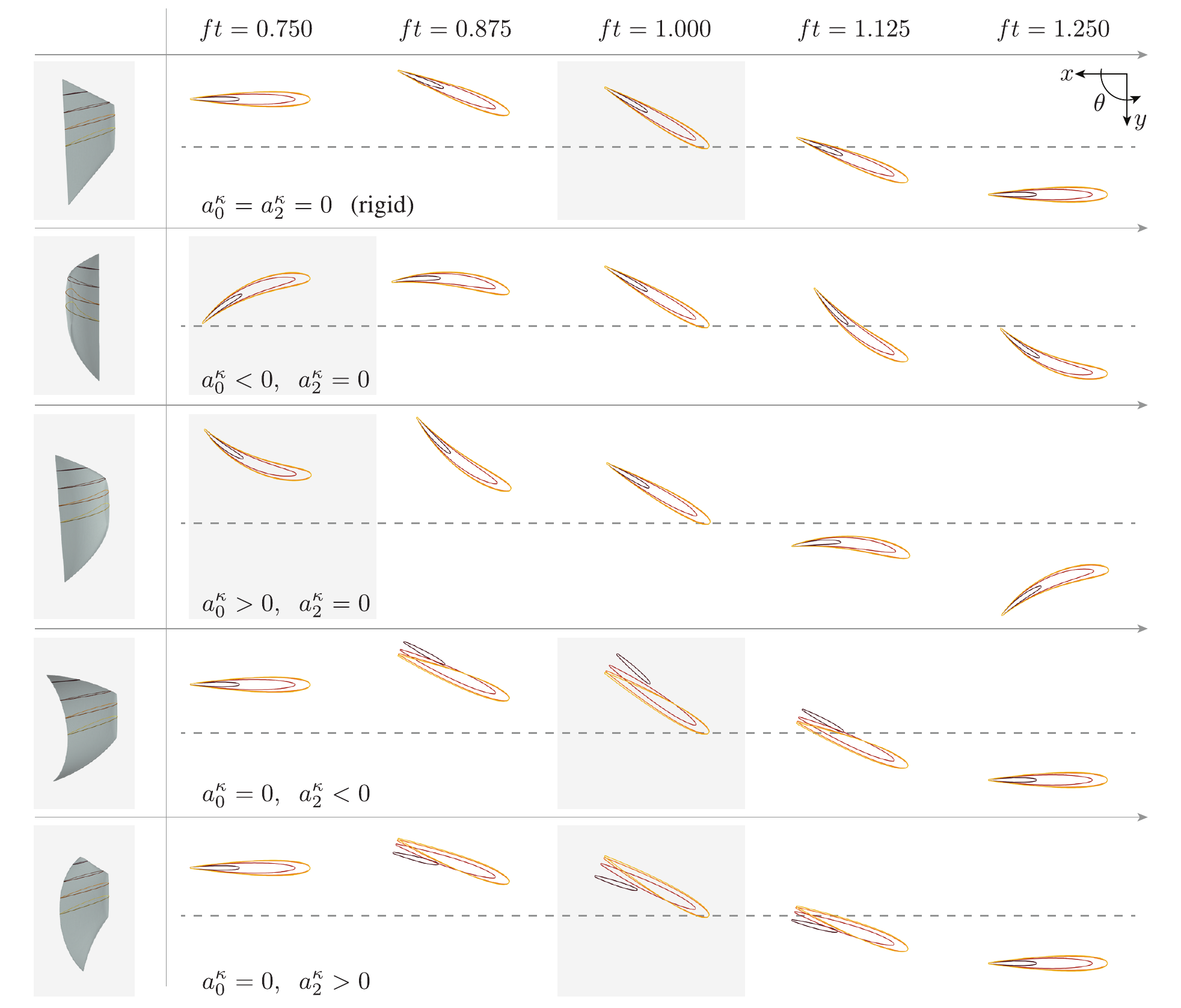}
\caption{Horizontal cross sections taken at $z/C = \{0.000,\, 0.175,\, 0.350,\, 0.525\}$ under various curvature regimes obtained within the two-dimensional parametrization ($a_0^\kappa$, $a_2^\kappa$). Shaded background indicates the time step plotted in the 3D view on the left.
}
\label{fig:active_passive_curvatures}
\end{figure}

Combined with our choice of heaving and pitching kinematics, figure~\ref{fig:active_passive_curvatures} demonstrates the effect of positive and negative values of our two parameters $a_0^\kappa$ and $a_2^\kappa$ on the fin shape variations, with $a_0^{\kappa} = a_2^\kappa = 0$ corresponding to a rigid fin.

\subsection{Numerical settings}

The spatial resolution throughout the simulations is set by a uniform grid spacing of $\Delta x = C/200$, following the grid convergence analysis presented in appendix~C in the supplementary data. 
The temporal resolution is fixed by a Lagrangian CFL time step constraint of $\text{LCFL} = 0.1$ \citep{rees2011}. The computational domain increases dynamically to capture the support of the vorticity field as the wake grows. 

The imposed rigid-body kinematics are ramped-up during the first flapping period through multiplication with a quarter period of a sine function, which lets the flow reach periodic conditions after this first cycle as shown in appendix~A in the supplementary data. 
The time-varying thrust, lift, and power coefficients are defined as
\begin{align}
    C_T(t) &= \frac{T(t)}{0.5 \rho A U_{\infty}^2} \,,\\
    C_L(t) &= \frac{L(t)}{0.5 \rho A U_{\infty}^2} \,,\\
    C_P(t) &= \frac{P(t)}{0.5 \rho A U_{\infty}^3} \,,
\end{align}
where $T$, $L$, and $P$ are the thrust, lift, and power computed from the flow field at a given time step following equations~\eqref{eq:thrust}, \eqref{eq:lift}, and~\eqref{eq:power}, and $A$ is the reference fin area taken as twice the mid-surface area. 

Since the thrust and power coefficients have periodicity of $2f$, we simulate until non-dimensional time $ft = 1.5$, and compute the cycle-averaged thrust and power coefficients ($\overline{C_T}$ and $\overline{C_{P}}$, respectively) over the last simulated half cycle ($1 \le ft \le 1.5$):
%
\begin{align}
    \overline{C_T} &= \int_{t=1/f}^{1.5/f} C_T(t) \:\mathrm{d}t \,,\\
    \overline{C_P} &= \int_{t=1}^{1.5/f} C_P(t)  \:\mathrm{d}t \,.
\end{align}
We can then define the propulsive efficiency as 
\begin{align}
    \label{eq:efficiency} \eta &= \frac{ \overline{C_T}}{\overline{C_P}} \,.
\end{align}    

In the following, we will primarily rely on $\overline{C_T}$, $\overline{C_P}$, and $\eta$, as defined above, as metrics for hydrodynamic performance.

\section{Effect of curvature variations on hydrodynamic performance} \label{s:results}

\begin{figure}
\centering
\includegraphics[width=0.49\linewidth]{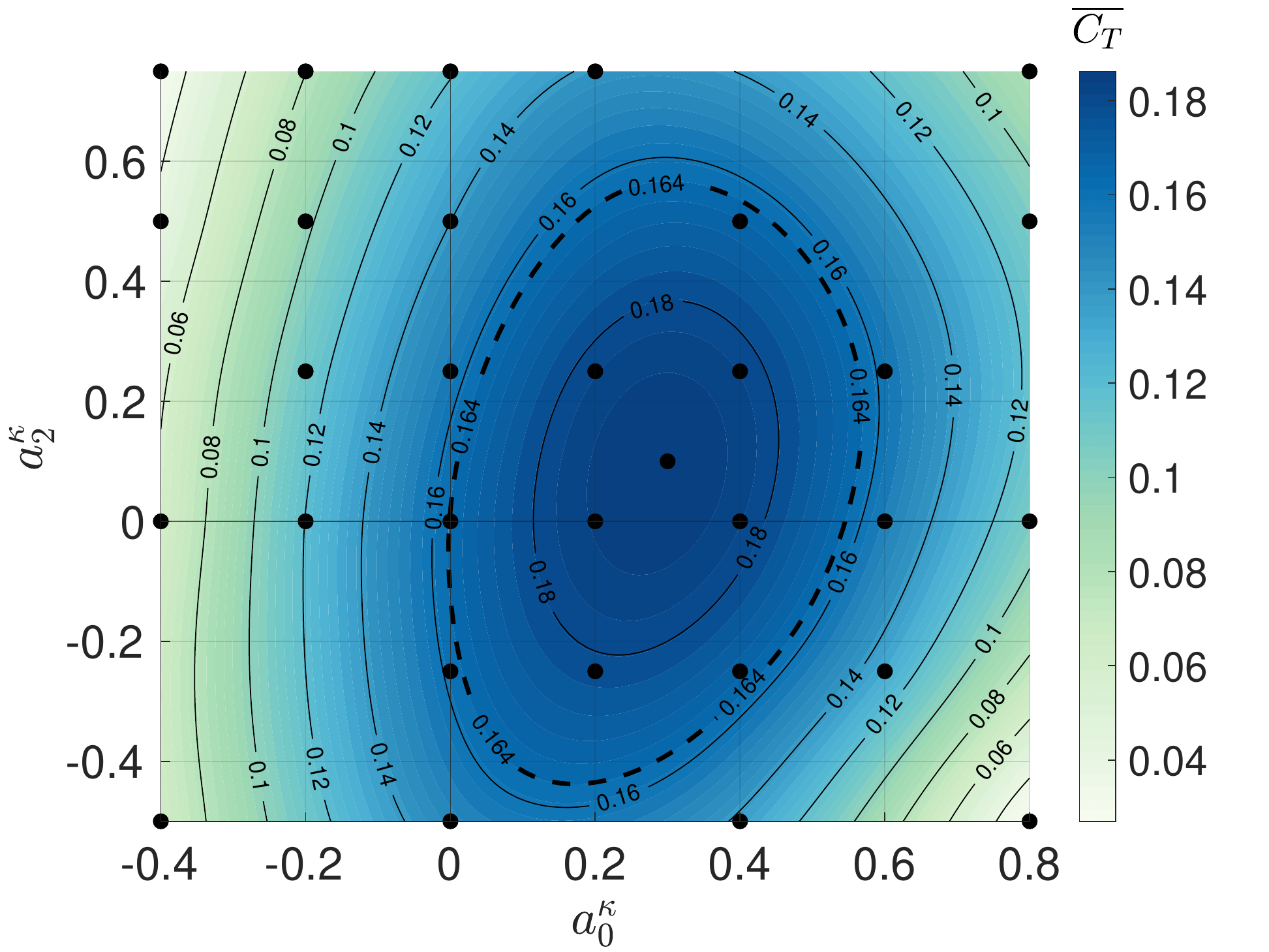}
\includegraphics[width=0.49\linewidth]{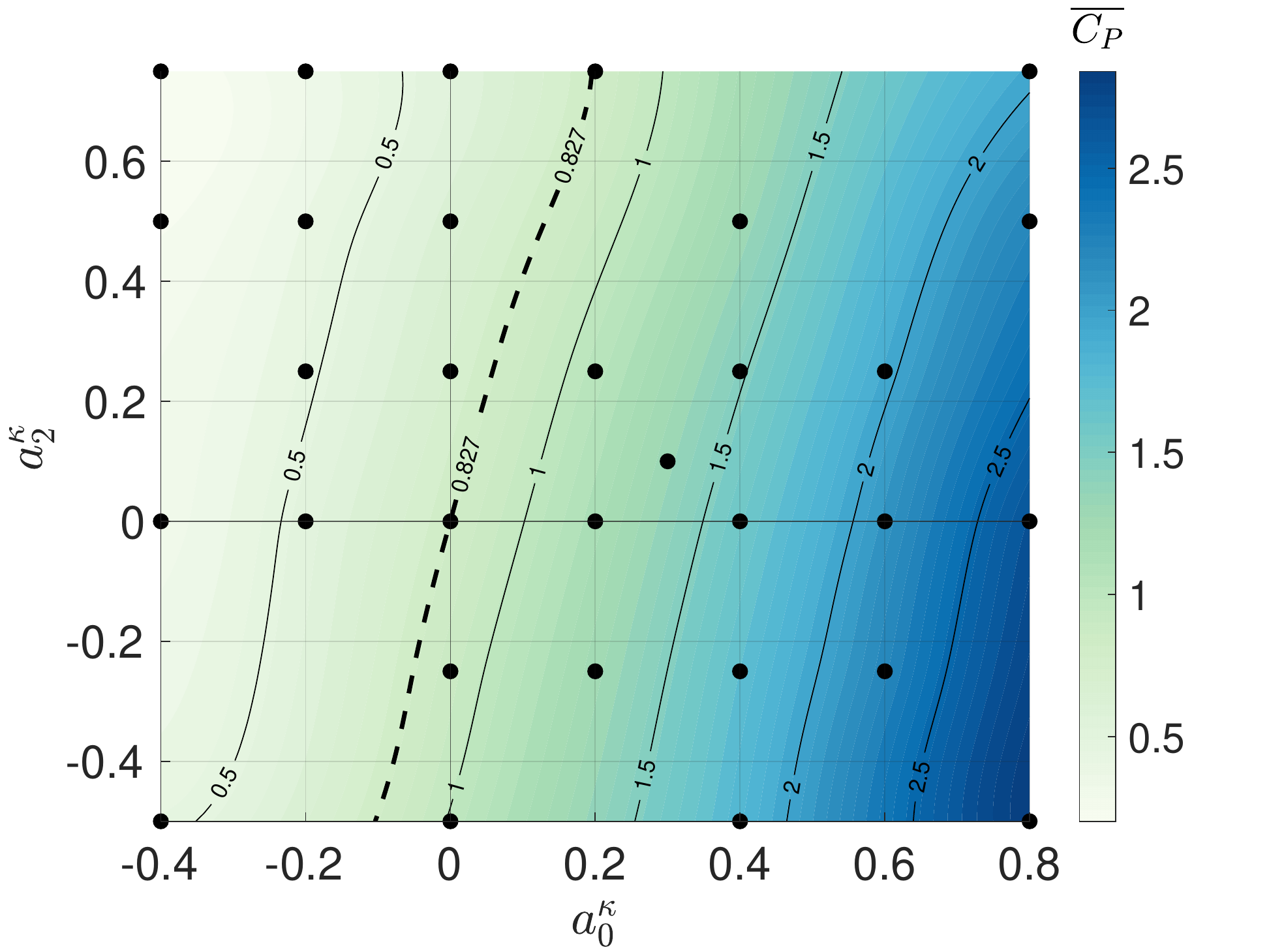}
\includegraphics[width=0.49\linewidth]{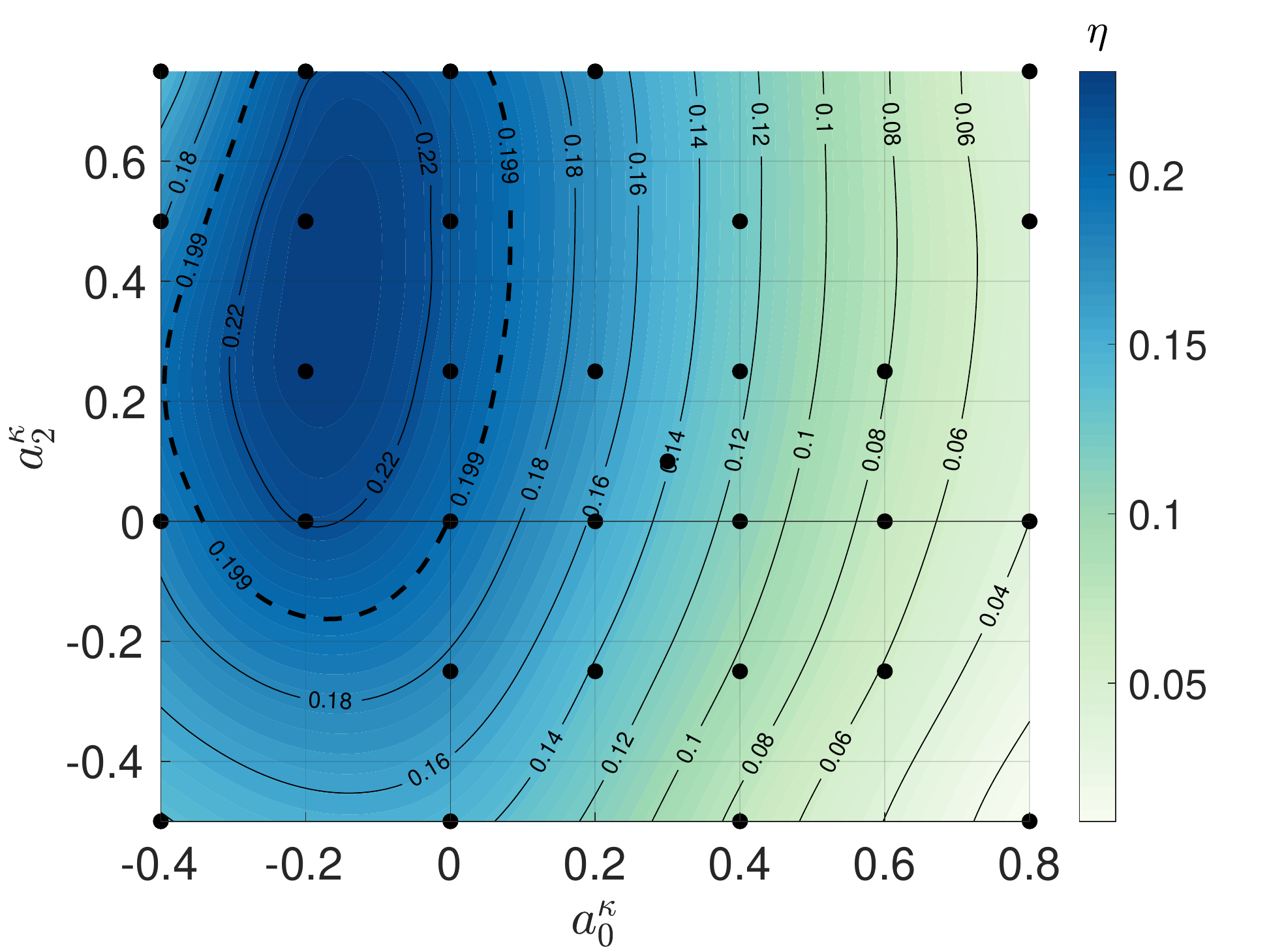}
\caption{Cycle-averaged thrust (\textit{top-left}) and power (\textit{top-right}) coefficients and efficiency (\textit{bottom}) results from Navier-Stokes simulations (\textit{black dots}), and an interpolated contour plot based on these results, as a function of the two curvature parameters $a_0^\kappa$ and $a_2^\kappa$.}
\label{fig:caudalfin_ctm_eff}
\end{figure}

Using the numerical framework and heave/pitch kinematics as described above, we simulated a set of flapping fins with curvature parameter variations $a_0^{\kappa}\in [-0.4,0.8]$ and $a_2^{\kappa} \in [-0.5,0.75]$, with $a_0^{\kappa}=a_2^{\kappa}=0$ corresponding to a rigid fin. For each simulation, we recorded the mean thrust and power coefficients, and computed the propulsive efficiency. These results are shown as contour plots in figure~\ref{fig:caudalfin_ctm_eff}, visualizing the effect of changing the curvature parameters on the hydrodynamic performance metrics. A variation of these plots is provided in appendix~D in the supplementary data, 
highlighting the curve that maximizes efficiency for a given range of thrust coefficients. 

Based on figure~\ref{fig:caudalfin_ctm_eff}, the maximum computed thrust occurs at $a_0^{\kappa} = 0.3$ and $a_2^{\kappa} = 0.1$, and is about 15\% larger than that for a rigid fin. Further, we can see that positive values of $a_0^\kappa$ generally improve the thrust coefficient up until the maximum, after which the thrust coefficient decays. The effect of spanwise curvature variations, as measured by $a_2^\kappa$, is less pronounced than the chordwise curvature effect.

For efficiency, the maximum occurs at $a_0^\kappa = -0.2$ and $a_2^\kappa = 0.25$, leading to about 18\% improvement over the rigid fin. The increase in efficiency is driven by a strong decrease of the power coefficient as $a_0^\kappa$ decreases. We also observe a small reduction of the power with increasing spanwise curvature parameter, so that the maximum efficiency is achieved at positive $a_2^\kappa$. 

\begin{figure}
    \centering
    \includegraphics[width=\linewidth]{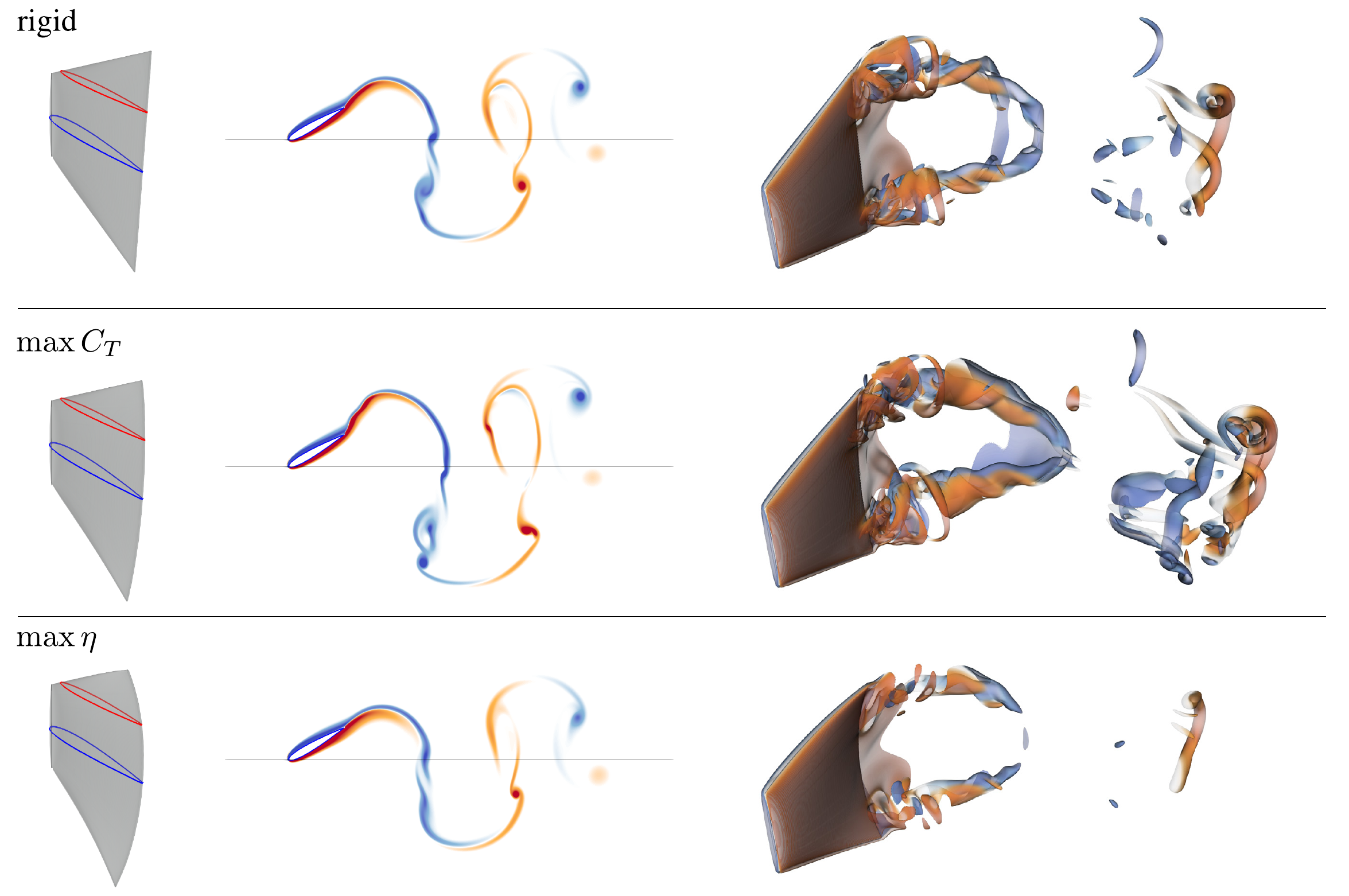}
    \caption{Fin shape (\textit{left}), vorticity field at $v=0$ (\textit{middle}), and 3D vorticity field (\textit{right}) for the rigid configuration (\textit{top}, $a_0^\kappa=0.0$, $a_2^\kappa=0.0$), the maximum thrust configuration (\textit{middle}, $a_0^\kappa=0.3$, $a_2^\kappa=0.1$), and the maximum efficiency configuration (\textit{bottom}, $a_0^\kappa=-0.2$, $a_2^\kappa=0.5$), all at $ft=1.5$. Both the 2D and 3D vorticity fields visualize, and are colored by, $\omega_z$. }
    \label{fig:caudalfin_flow}
\end{figure}

Figure~\ref{fig:caudalfin_flow} shows the vortical structures at $ft =1.5$ for the rigid fin, and the conditions corresponding to maximum computed thrust and maximum computed efficiency, respectively. We observe an increase in the intensity of the vortices shed from the fin for the maximum thrust, whereas the maximum efficiency case has a much weaker wake signature.  

All these results are relatively invariant to the planform shape, as demonstrated in appendix~E in the supplementary data, 
which shows similar numbers for a square planform with $H = C$. 

The next section provides an in-depth analysis of the effect of $a_0^\kappa$ and $a_2^\kappa$ on the thrust and efficiency of the fin, based on the described results. 

\section{Analysis of the effect of curvature variations}
\label{s:discussion}

In the following two subsections, we investigate in detail the effect of our chordwise and spanwise curvature parameters on the hydrodynamic performance of the fin, guided by the above observations.

\subsection{Effect of chordwise curvature parameter \texorpdfstring{$a_0^\kappa$}{a0k}} \label{ss:results_chordwise}

As shown in the previous section, chordwise deformation has the largest impact on both thrust and power, which is  qualitatively consistent with previous results \citep{zhu2008,esposito2012}. In this section we focus on the underlying mechanisms by considering only configurations with $a_2^{\kappa}=0$. 

\begin{figure}
\centering
\includegraphics[width=\linewidth]{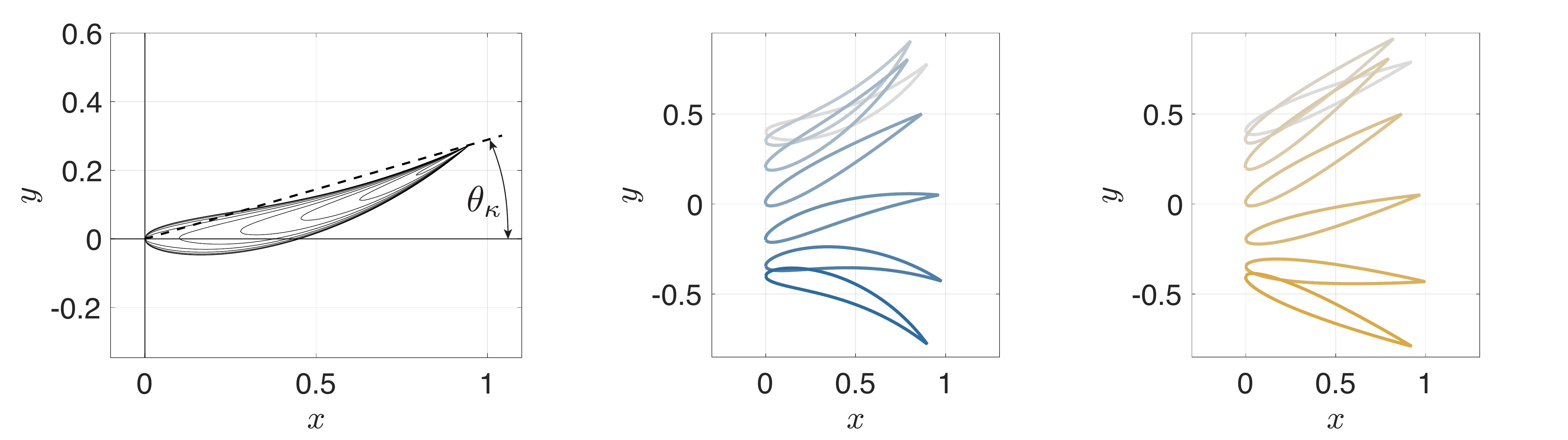}
\caption{\textit{Left}: Horizontal cross sections at $f t=1.25$ of the fin with $a_0^{\kappa}=0.8$ and $a_2^\kappa = 0$. \textit{Right}: Cross sections at $v=0$ of the curved and $\kappa$-pitch configurations during the down-stroke half-cycle, visualized at seven equidistant time instances between $ft = 0.25$ (lightest) and $ft = 0.75$ (darkest).}
\label{fig:kpitch_geometry}
\end{figure}

Geometrically, by varying $a_0^\kappa$, the mid-surface plane rolls over a vertical cylinder of radius $C/a_0^\kappa$, as shown in appendix~B in the supplementary data. 
As $a_0^\kappa$ increases, this means the curving fin is different from the reference rigid fin in two aspects. First, the line connecting leading and trailing edge of the fin also undergoes additional lateral trailing-edge excursions (see figure~\ref{fig:kpitch_geometry}). Second, on top of the modified trailing edge kinematics, the fin experiences a camber-like deformation. The former effect can be described as an additional pitching contribution, on top of the reference pitching kinematics~\eqref{eq:ref_pitch}. Based on the deformation mode considered, this additional pitching term can be derived as $\theta_{\kappa}(t) = 0.5 a_0^{\kappa} \sin(2\pi f t)$. 
With this insight, we can then decompose the effect of $a_0^\kappa$ into two characteristics: the first increases the pitch variations of the reference rigid fin with $\theta_{\kappa}(t)$, and the second adds the chordwise curvature on top of this rigid-body motion without affecting the leading- and trailing-edge locations. 

We investigate the first effect by simulating a rigid fin undergoing altered pitch kinematics given by
\begin{align}
\label{eq:kappa_pitch} \theta^{\kappa\text{-pitch}}(t) &= - A_\theta \cos(2 \pi f t) + 0.5 a_0^{\kappa} \sin(2\pi f t) \,,
\end{align}
while keeping the geometry and heave kinematics the same as the reference rigid fin. This configuration, which we denote as the $\kappa$-pitch case, is also parametrized by $a_0^{\kappa}$, though the fin does not undergo any curvature variations. 

\begin{figure}
\centering
\includegraphics[width=0.32\linewidth]{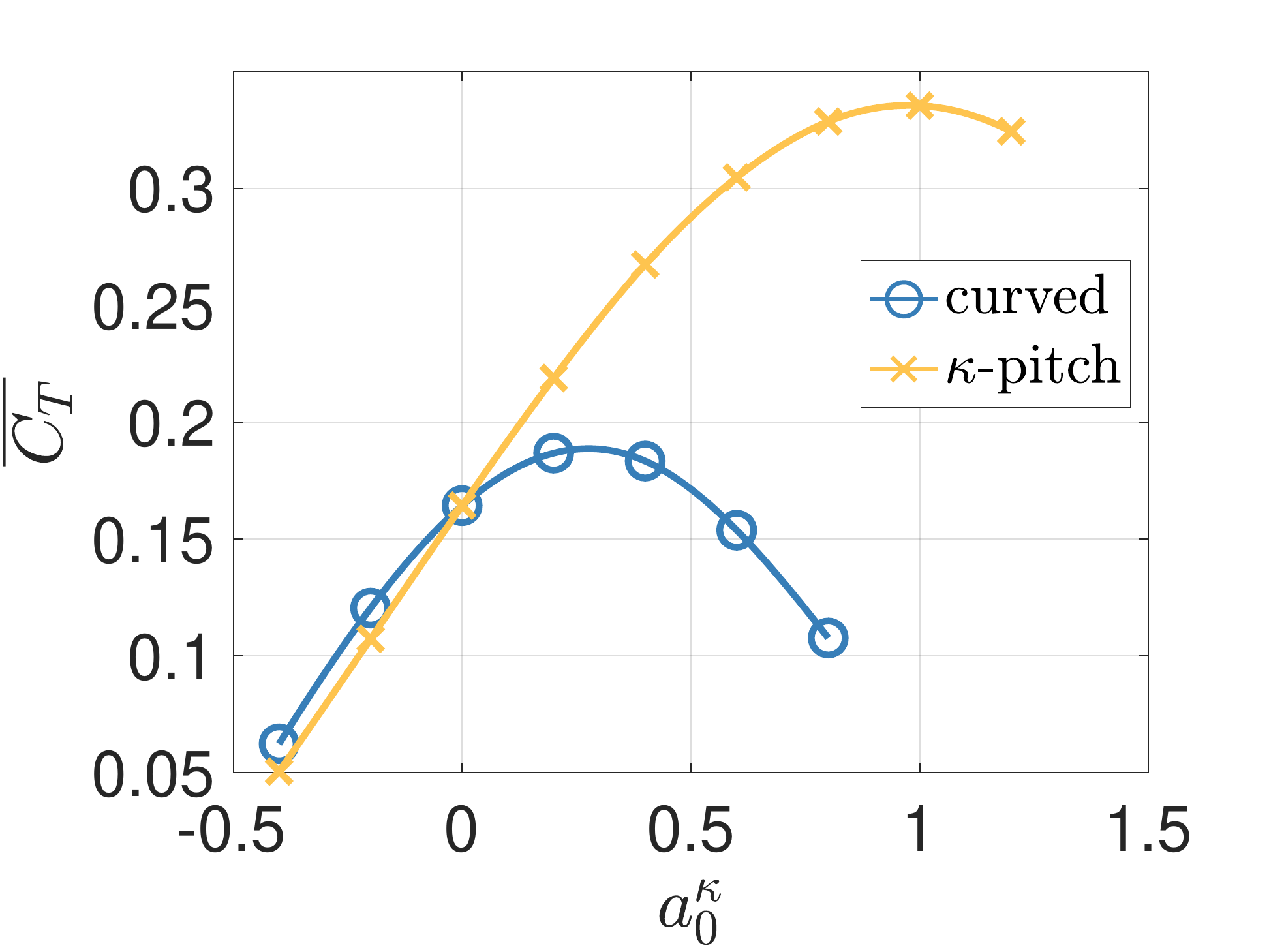}
\includegraphics[width=0.32\linewidth]{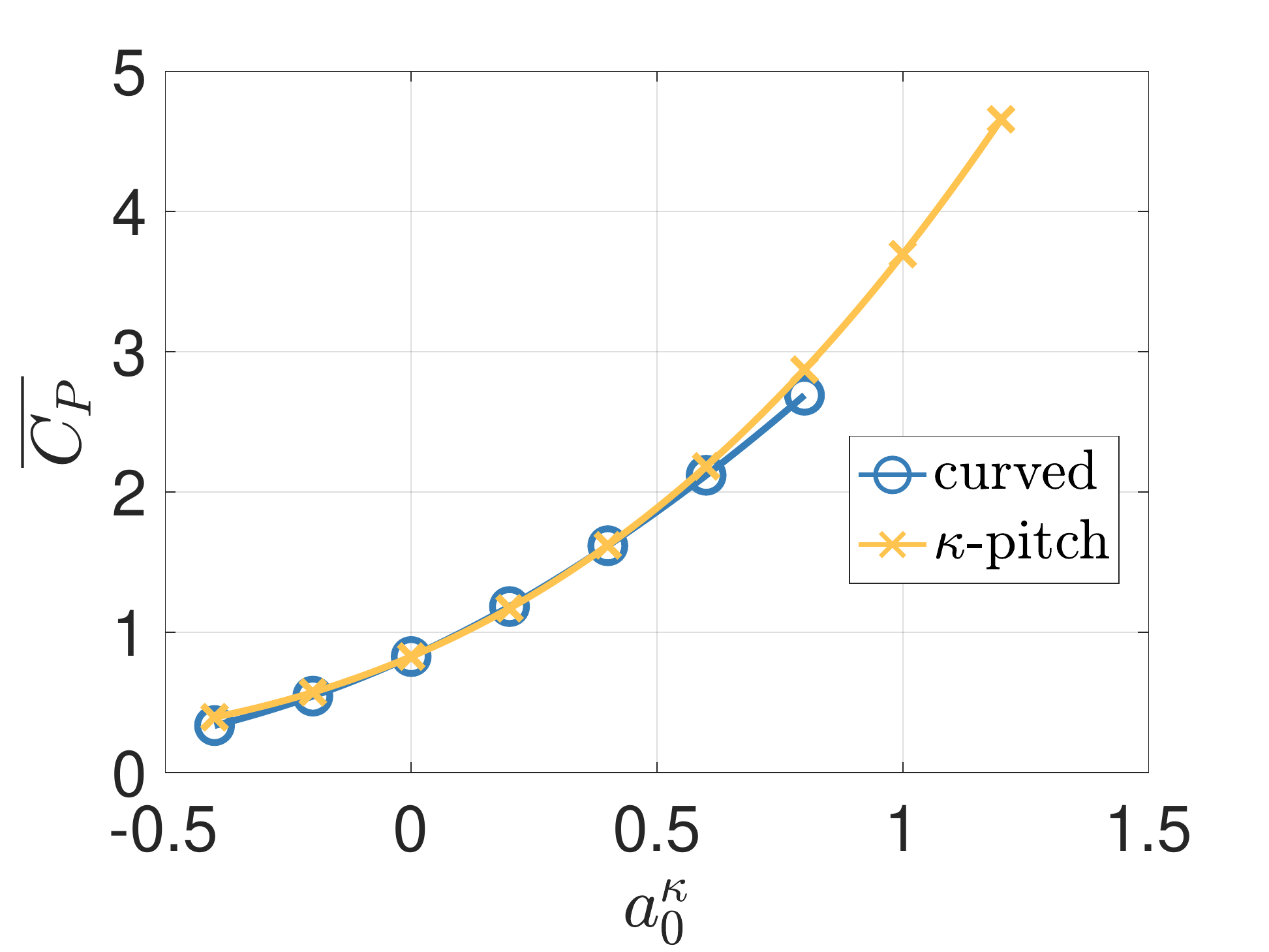}
\includegraphics[width=0.32\linewidth]{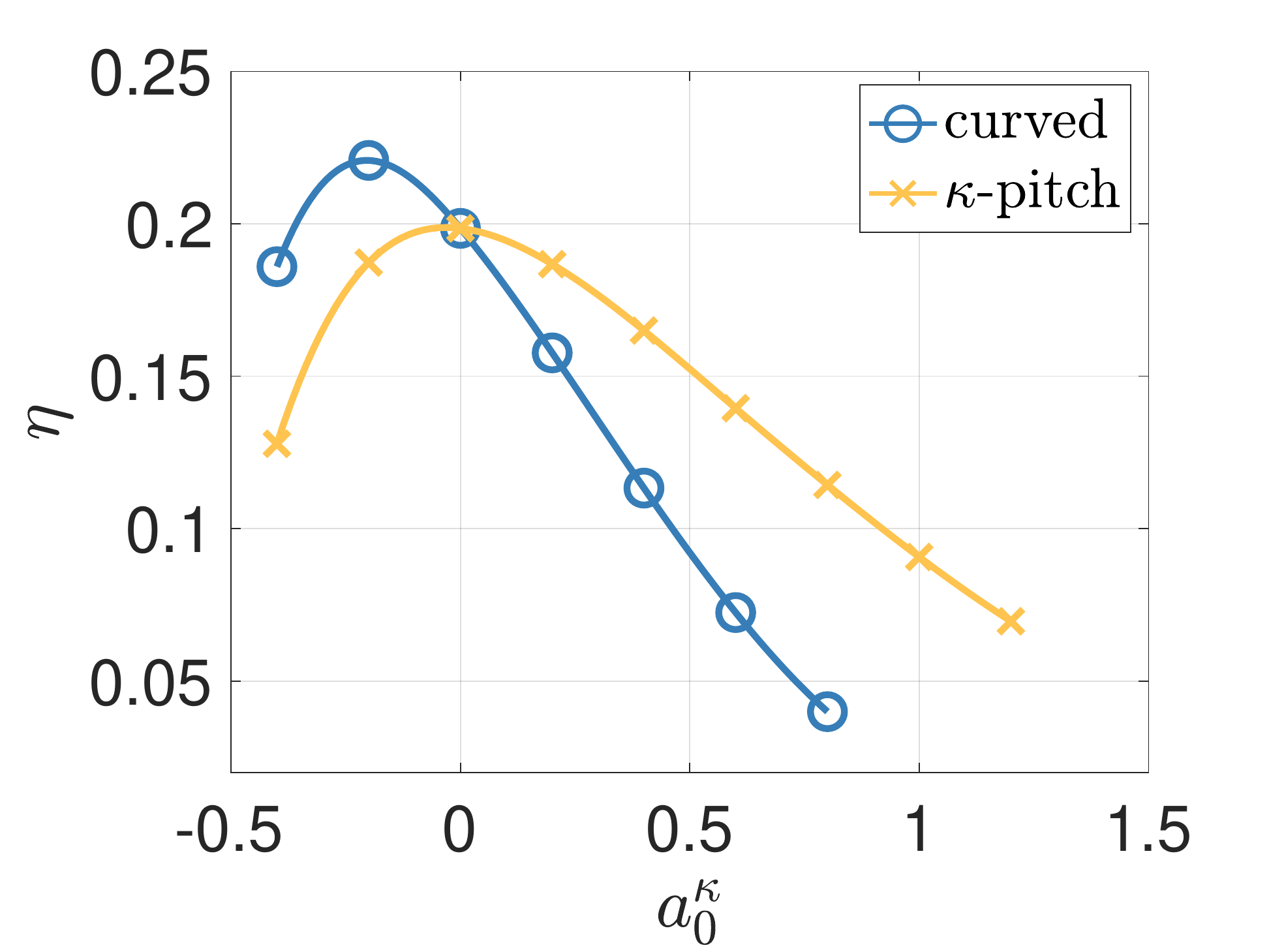}
\caption{Cycle-averaged thrust coefficient (\textit{left}), power coefficient (\textit{center}), and efficiency (\textit{right}) as a function of the chordwise curvature parameter $a_0^\kappa$, for the fin with curvature variations (\textit{in blue}) and the rigid fin with $\kappa$-pitch kinematics (\textit{in orange}). }
\label{fig:caudalfin_kpitch}
\end{figure}

Figure~\ref{fig:caudalfin_kpitch} compares the thrust, power, and efficiency of the curved and $\kappa$-pitch configurations for the range of $a_0^{\kappa}$ studied, where again $a_0^\kappa = 0$ corresponds to the rigid fin with unaltered pitching kinematics. We observe that the $\kappa$-pitch case qualitatively reproduces the effect of $a_0^\kappa$ on the mean thrust coefficient, leading to a decrease in thrust for negative values and the existence of a maximum at finite $a_0^\kappa > 0$. The effect of $a_0^\kappa$ on power and efficiency are also qualitatively comparable between the curved and $\kappa$-pitch configurations. This provides our first insight into why the chordwise curvature variations lead to increased thrust coefficient.

However, quantitatively there is a significant increase in the maximum thrust coefficient achieved by the $\kappa$-pitch case over the optimally curved case. Further, the peak thrust for the $\kappa$-pitch fin occurs at $a_0^\kappa = 0.95$, versus $a_0^\kappa = 0.28$ for the curved fin. Since power consumption is about equal between the two cases, the efficiency of the $\kappa$-pitch fin at high thrust values ($a_0^\kappa > 0$) is significantly higher than for the curved fin. The optimal efficiency, on the other hand, is achieved at much lower thrust values -- here the curved fin outperforms the $\kappa$-pitch one slightly, which we will discuss more at the end of this subsection.

\begin{figure}
\includegraphics[width=0.49\linewidth]{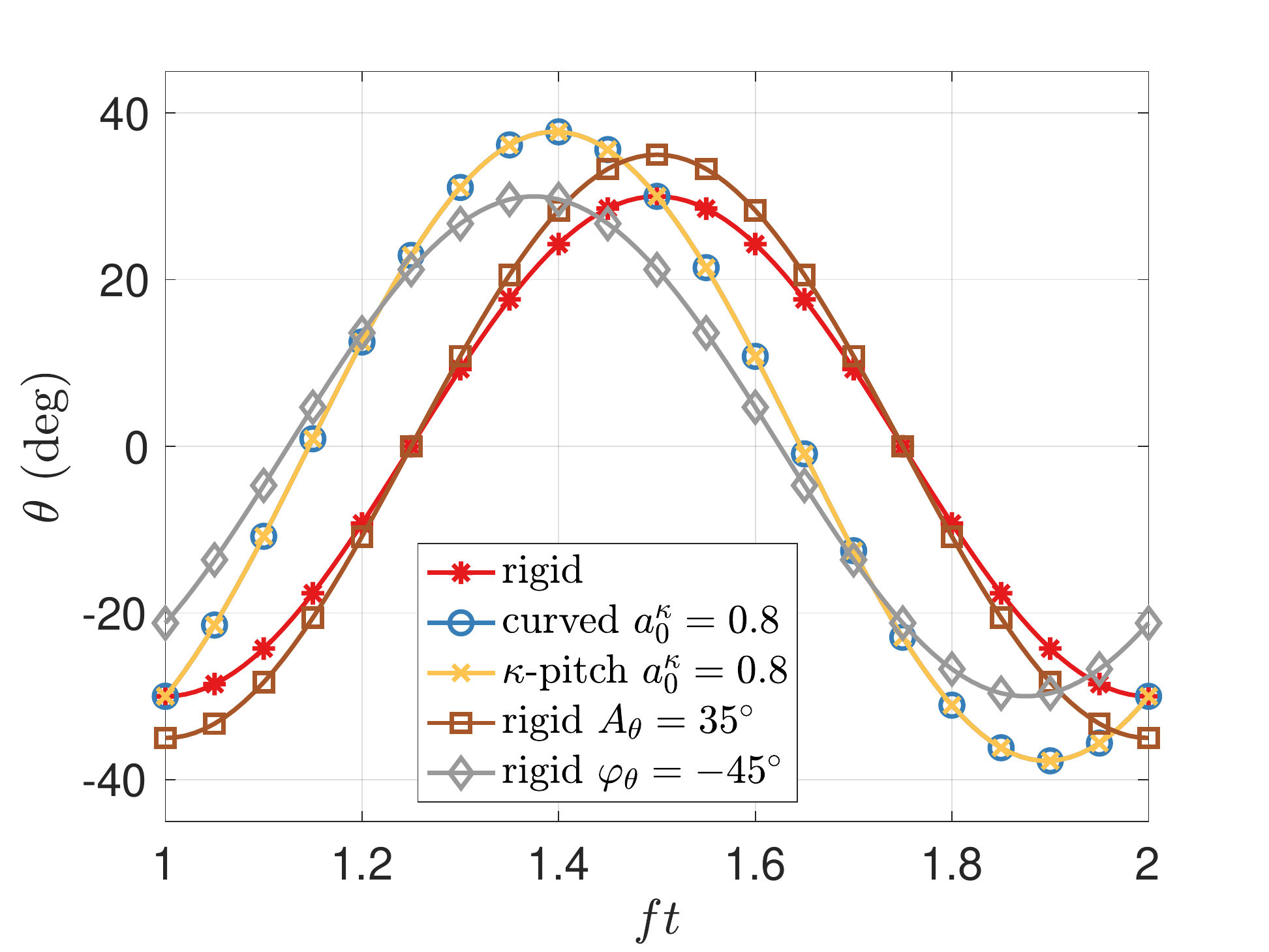}
\includegraphics[width=0.49\linewidth]{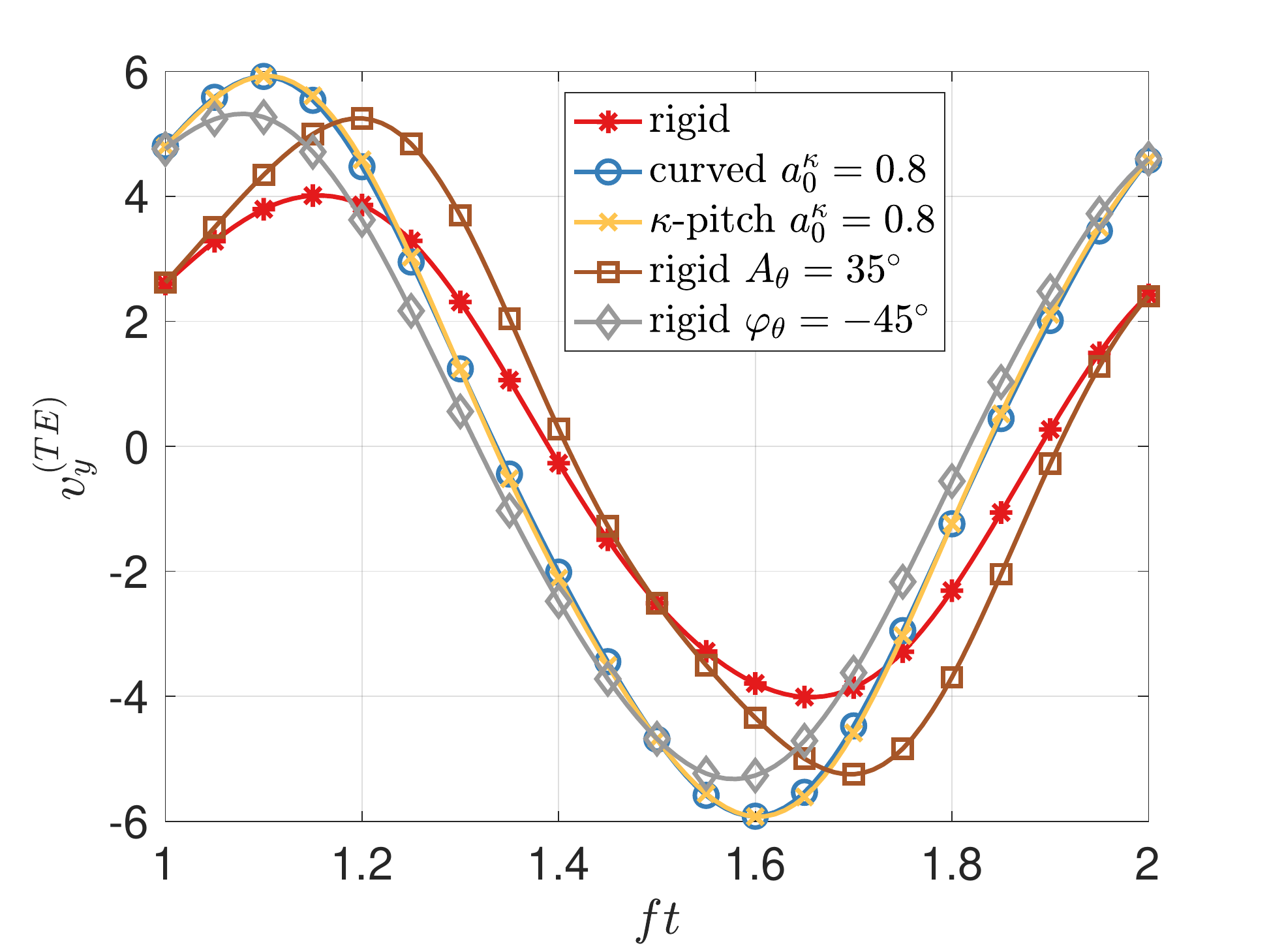}
\centering
\caption{Pitch angle (\textit{left}) and trailing edge lateral velocity (\textit{right}) during a flapping cycle, for the rigid fin with reference kinematics (\textit{red}), the fin with curvature variations $a_0^\kappa = 0.8$ (\textit{blue}), and the rigid fin with $\kappa$-pitch kinematics using $a_0^\kappa = 0.8$ (\textit{orange}). Also shown are the rigid fin results with harmonic pitch variations with amplitude $A_\theta = \ang{35}$ (\textit{brown}) and with phase shift $\varphi = -\ang{45}$ (\textit{gray}). }
\label{fig:caudalfin_kpitch_theta_vyTE}
\end{figure}

To understand why the $\kappa$-pitch kinematics are able to practically double the thrust coefficient (at $a_0^\kappa = 0.8$) of the reference rigid fin ($a_0^\kappa = 0$), the left panel of figure~\ref{fig:caudalfin_kpitch_theta_vyTE} compares the pitch angle variations as a function of time for the reference rigid fin (in red) and the $\kappa$-pitch fin (in orange). Note that, by construction, the pitch angle variations of the $\kappa$-pitch configuration (in orange) are identical to that of the curving fin (in blue) at equal values of $a_0^\kappa$. The plot shows how, compared to the reference rigid fin, the $\kappa$-pitch configuration not only achieves an increase in maximum pitch angle, but also affects the phase shift with the heave motion. In fact, we can estimate the effective pitch amplitude and phase values of the $\kappa$-pitch kinematics, using equation~\eqref{eq:kappa_pitch}, as follows:
\begin{align}
    A_{\theta}^{\kappa\text{-pitch}} &\approx \max(\theta) = \sqrt{A_{\theta}^2 + \left(0.5 a_0^{\kappa}\right)^2}\,,\\
    \varphi_{\theta}^{\kappa\text{-pitch}} &\approx 2\pi \left( t_{\max(\theta)} - t_{\max(y)} \right) =  \frac{\pi}{2} - \arctan\left(\frac{0.5 a_0^{\kappa}}{A_{\theta}}\right) \,.
\end{align}
For $a_0^{\kappa}=0.8$, where the $\kappa$-pitch kinematics achieve maximum thrust, we then find $A_{\theta}^{\kappa\text{-pitch}} = 37.8^{\circ}$ and $\varphi_{\theta}^{\kappa\text{-pitch}} \approx -52.6^{\circ}$. 

When analyzing the isolated effect of pitch amplitude and phase angle variations on our reference rigid fin, we see why the altered kinematics of the $\kappa$-pitch configuration are virtuous. Appendix~F in the supplementary data 
shows that changing the phase shift from $\ang{-90}$ to $\ang{-45}$ doubles the thrust coefficient of the reference rigid fin, and an independent increase in pitch amplitude from $\ang{30}$ to $\ang{35}$ also leads to a modest increase in thrust. The corresponding pitch angle variations are shown in figure~\ref{fig:caudalfin_kpitch_theta_vyTE} (left) as the silver and brown lines, respectively. The $\kappa$-pitch configuration then combines a pitch amplitude and pitch phase shift that are very close combinations of the individual optimal values for the reference rigid fin with sinusoidal pitch variations. As a side note, we observe also in appendix~F in the supplementary data 
that in terms of efficiency, the $\ang{-90}$ phase angle is optimum, consistent with the findings of \citet{read2003}. 

To summarize until here, we have observed that the original curvature variation, as dictated by $a_0^\kappa$, provides an altered pitching kinematics that increases the mean thrust coefficient achieved by the fin. We can reproduce this effect with a rigid fin, both using a combined effective amplitude and phase shift, as well as through independent variations of amplitude and phase shift. Both indicate that the significant driver in thrust increase is the phase shift change from $\ang{-90}$ to $\approx\ang{-50}$. In the remainder of this subsection we will focus on two open questions: the first asks why this altered pitching kinematics improves performance, and the second asks why the $\kappa$-pitch fin provide significantly larger thrust values for all $a_0^\kappa > 0$ compared with the curving fin.

We answer the first question by examining the trailing edge (TE) lateral velocity as shown in the right panel of figure~\ref{fig:caudalfin_kpitch_theta_vyTE} for all cases discussed above. We observe that the amplitude of the TE lateral velocity increases about 1.5 fold between the reference rigid fin and the $\kappa$-pitch configuration with $a_0^\kappa = 0.8$, leading to an increase in mean thrust coefficient by a factor of 2.1. This is consistent with the added mass effect for pitching fin propulsion \citep{Garrick:1936, Gazzola:2014, smits2019} which predicts that the thrust coefficient is proportional to the square of the lateral velocity. 
The TE velocity amplitude does not solely predict performance: the timing of maximum TE velocity compared to the fixed heaving kinematics also affects the thrust coefficient. This is a much more subtle interaction, though, that would require further investigation.

\begin{figure}
\centering
\includegraphics[width=0.90\linewidth]{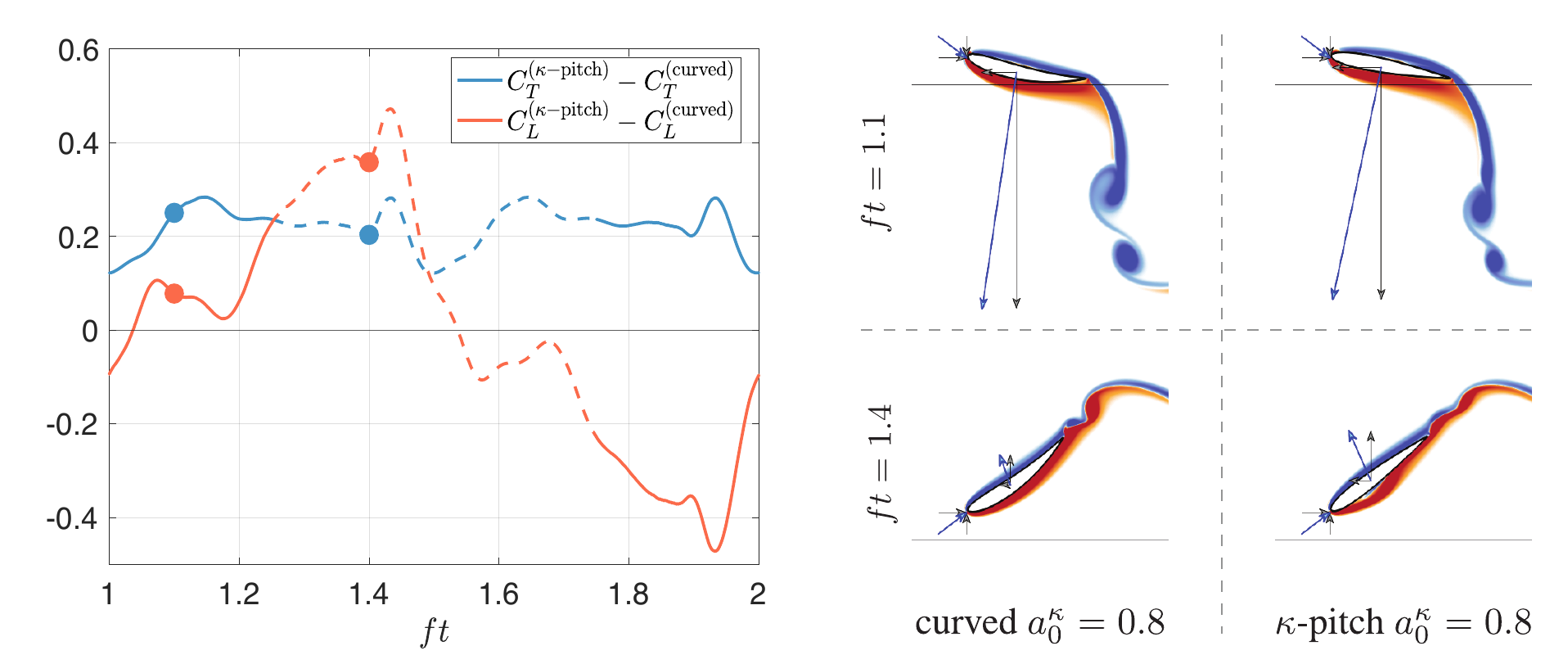}
\caption{\textit{Left}: Difference in thrust and lift coefficients between the curved and $\kappa$-pitch configurations with $a_0^{\kappa}=0.8$. Solid and dashed lines identify the upstroke and downstroke half-cycles, respectively. \textit{Right}: Vorticity contours at the center plane. Incident velocity vector and its horizontal and vertical components annotated at the LE ($\bu = [U_{\infty}, -\dot{y}]$). Fluid force vector and its horizontal and vertical components annotated at fin centroid ($\bF = [-T,L]$).}
\label{fig:caudalfin_kpitch_dct_dcl}
\end{figure}

The second open question concerns the difference between the fin with curvature variations and the $\kappa$-pitch configuration, for the same value of $a_0^\kappa$. To address this, we plot the time evolution of the \textit{difference} in thrust and lift coefficients between the curving fin and the $\kappa$-pitch configuration in figure~\ref{fig:caudalfin_kpitch_dct_dcl}. For reference, the time evolution of the individual force coefficients is included in appendix~G in the supplementary data. 
From figure~\ref{fig:caudalfin_kpitch_dct_dcl}, we can identify two reasons for the lower thrust coefficient of the chordwise curving fin compared to the $\kappa$-pitch fin. First, for times $1 \le ft \le 1.2$, corresponding to the second half of the upstroke just before reversal of the heave kinematics, the difference in $C_T$ is large whereas the difference in $C_L$ is relatively small. This implies an increased drag force on the curving fin, consistent with the curved profile in this part of the stroke where the fin becomes aligned with the inflow. The top-right panel of figure~\ref{fig:caudalfin_kpitch_dct_dcl} confirms that the total force vector is angled more vertically for the curved case compared with the $\kappa$-pitch case. Second, for times $1.25 \le ft \le 1.5$, corresponding to the first part of the downstroke after the heave motion has reversed, we observe that the $\kappa$-pitch configuration experiences both larger thrust and larger lift coefficients. This means that the overall force vector on the fin is larger for the $\kappa$-pitch fin. We attribute the decreased force of the chordwise curving fin to the camber, which essentially is `reversed' as the cross-section slopes upwards in the direction of the force resultant.  The bottom-right panel of figure~\ref{fig:caudalfin_kpitch_dct_dcl} corroborates this visually. Both of these effects are repeated every $ft = 0.5$ times due to the symmetry of the up- and down-strokes. These two reasons (additional profile drag and reverse camber) lead to the reduced performance of the chordwise curving fin compared to the $\kappa$-pitch rigid fin. 

So far, this subsection has focused on the regime $a_0^\kappa > 0$, where significant gains in the mean thrust coefficient are observed. However, our results also show that negative values of $a_0^\kappa$ monotonically decrease the power required to move the fin, and increase the efficiency $\eta$. The power reduction is apparent from figure~\ref{fig:caudalfin_cpw}, showing the power components associated with heave and pitch, defined as $P^{(L)} = -L \dot{y}$ and $P^{(M)} = - \bM \cdot \dot{\boldsymbol{\theta}}$, respectively. The plot demonstrates that the power reduction is approximately equally distributed between the heave and pitch kinematics. The deformation-related power coefficient, $C_P^\text{def} = C_P - C_P^T - C_P^M$, decreases as well, but this reduction is relatively insignificant compared to the other two components. To distinguish the effects of fin camber and trailing edge kinematics in the regime $a_0^\kappa < 0$, we can revisit figure~\ref{fig:caudalfin_kpitch}. Both the $\kappa$-pitch and the curving fins reduce their power coefficients equally, indicating that the power reduction at negative $a_0^\kappa$ is due to the reduced trailing edge velocity. However, only the curving fin demonstrates a peak in efficiency at $a_0^\kappa < 0$, since the fin camber leads to a slight increase in thrust coefficient over the $\kappa$-pitch configuration for the same values of $a_0^\kappa$. Consequently, the efficiency peak of the curving fin is higher than that of any of the rigid fins, and achieved at a negative $a_0^\kappa$ value. 

Overall, this behavior is consistent with intuition -- negative values of $a_0^\kappa$ correspond to curvature `with the flow', i.e.\ qualitatively similar to elastic deformation, as well as a hydrodynamically beneficial camber induced during the thrust-generation part of the stroke.

\subsection{Effect of spanwise curvature parameter \texorpdfstring{$a_2^\kappa$}{a2k}} \label{ss:results_spanwise}

\begin{figure}
\centering
\includegraphics[width=0.49\linewidth]{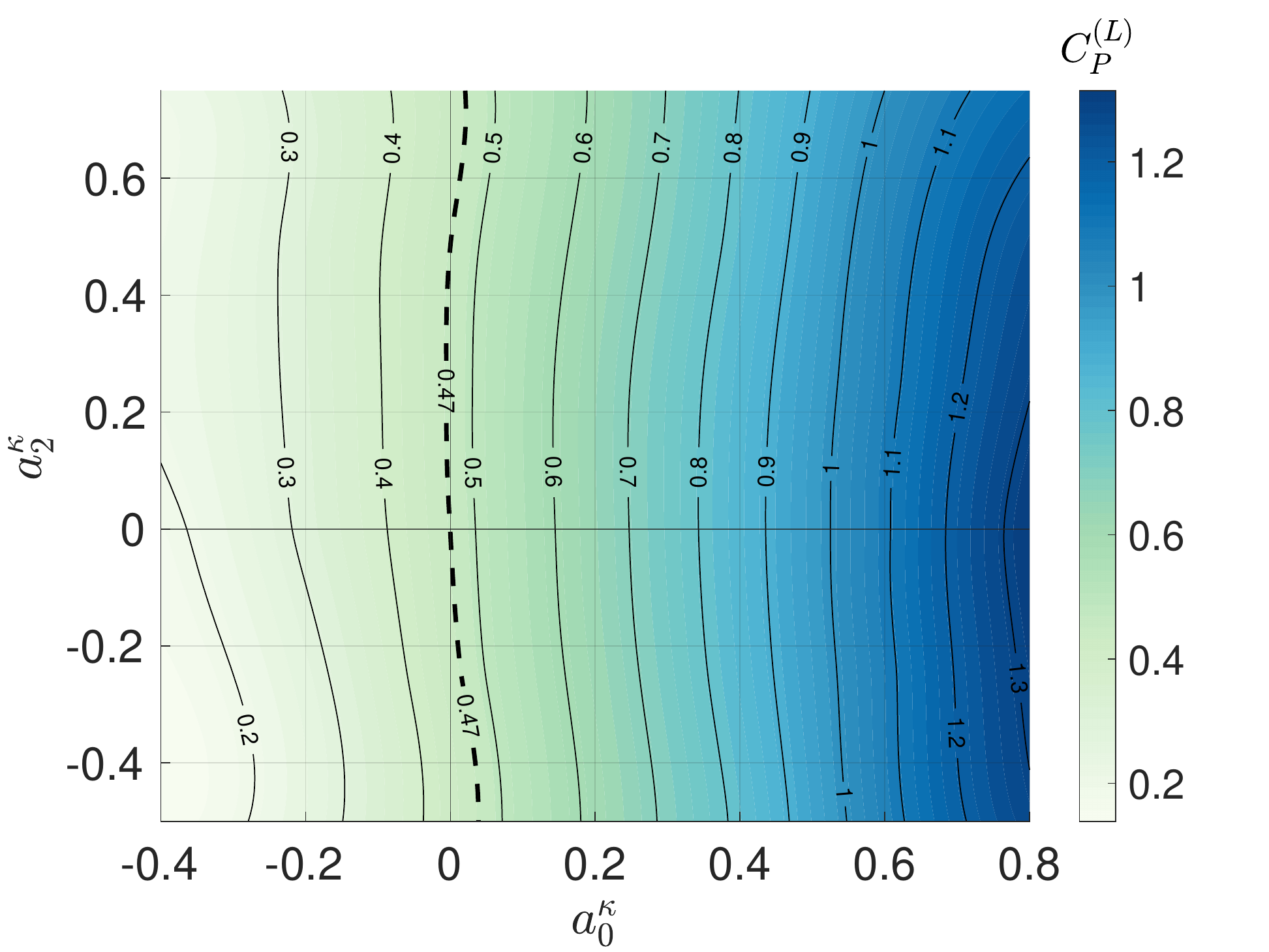}
\includegraphics[width=0.49\linewidth]{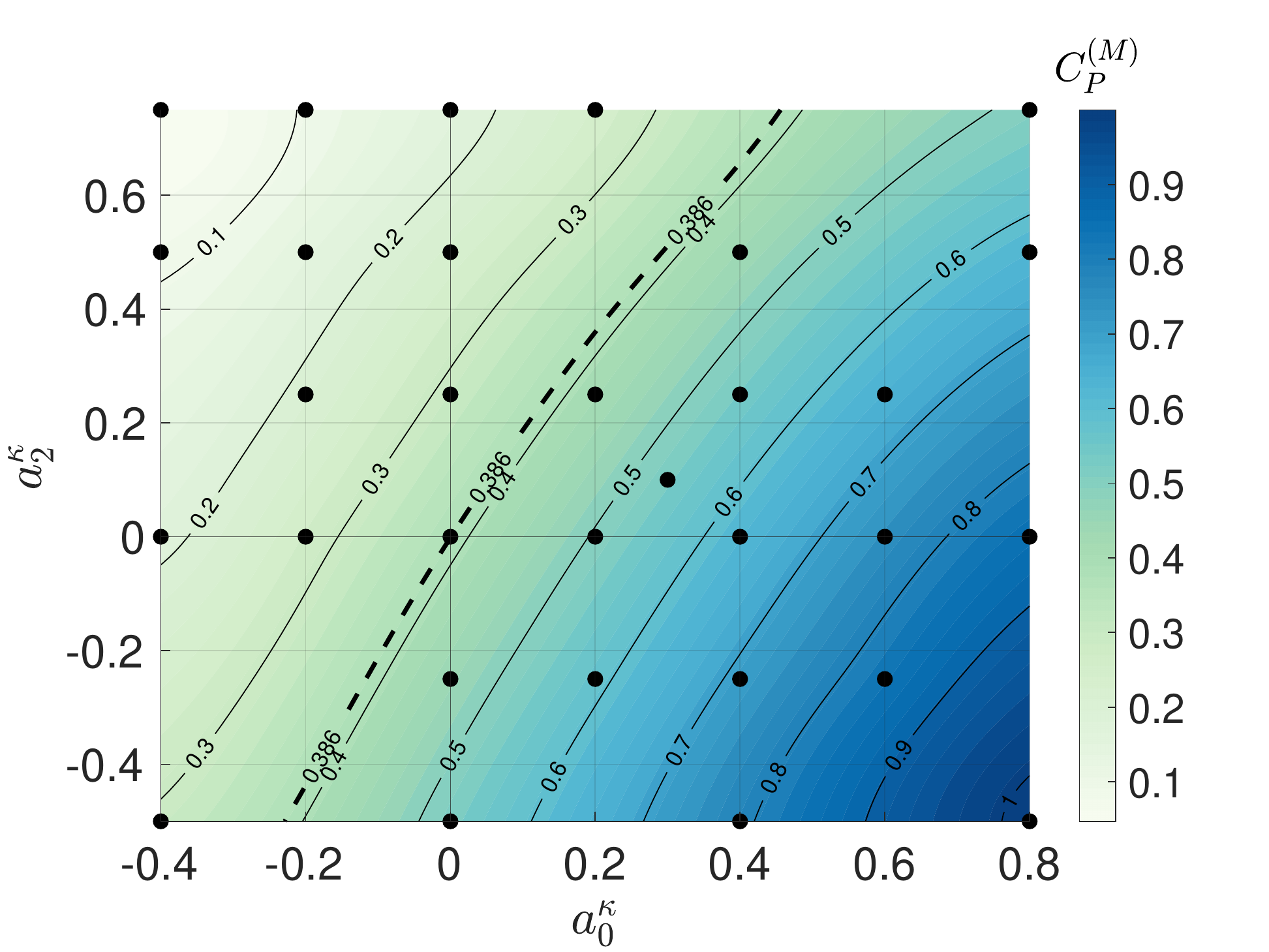}
\caption{Cycle-averaged power coefficient components linked to heave (\textit{left}) and pitch (\textit{right}) computed from Navier-Stokes simulations (\textit{black dots}), and an interpolated contour plot based on these results, as a function of the two curvature parameters $a_0^\kappa$ and $a_2^\kappa$.}
\label{fig:caudalfin_cpw}
\end{figure}

As discussed previously, spanwise curvature variations as parametrized by $a_2^\kappa$ predominantly affect the cycle-averaged power coefficient, which  monotonically decreases with increasing values of $a_2^{\kappa}$ within the range of curvatures simulated. 
Figure~\ref{fig:caudalfin_cpw} shows that this power reduction originates almost exclusively from the pitch kinematics. In this section we will investigate this effect further, considering only configurations with $a_0^{\kappa}=0$. 

Similar to the chordwise curvature in the previous section, the spanwise curvature can be decomposed into two components: the spanwise twisting of otherwise straight rays, and the actual curving of the rays without further affecting their trailing edge locations. We can isolate the former component starting from a rigid fin, and adjust the pitch variation across the height of the fin to match the LE-TE direction associated with the $a_2^\kappa$ curvature profile
\begin{align}
\label{eq:kappa_twist} \theta^{\kappa\text{-twist}}(v, t) &= \left( - A_\theta + 0.5 a_2^{\kappa} v^2 \right) \cos(2 \pi f t) \,.
\end{align}
We name this configuration $\kappa$-twist, and note that we have to relax the membrane inextensibility constraint to accomplish the resulting shape.

\begin{figure}
\centering
\includegraphics[width=0.32\linewidth]{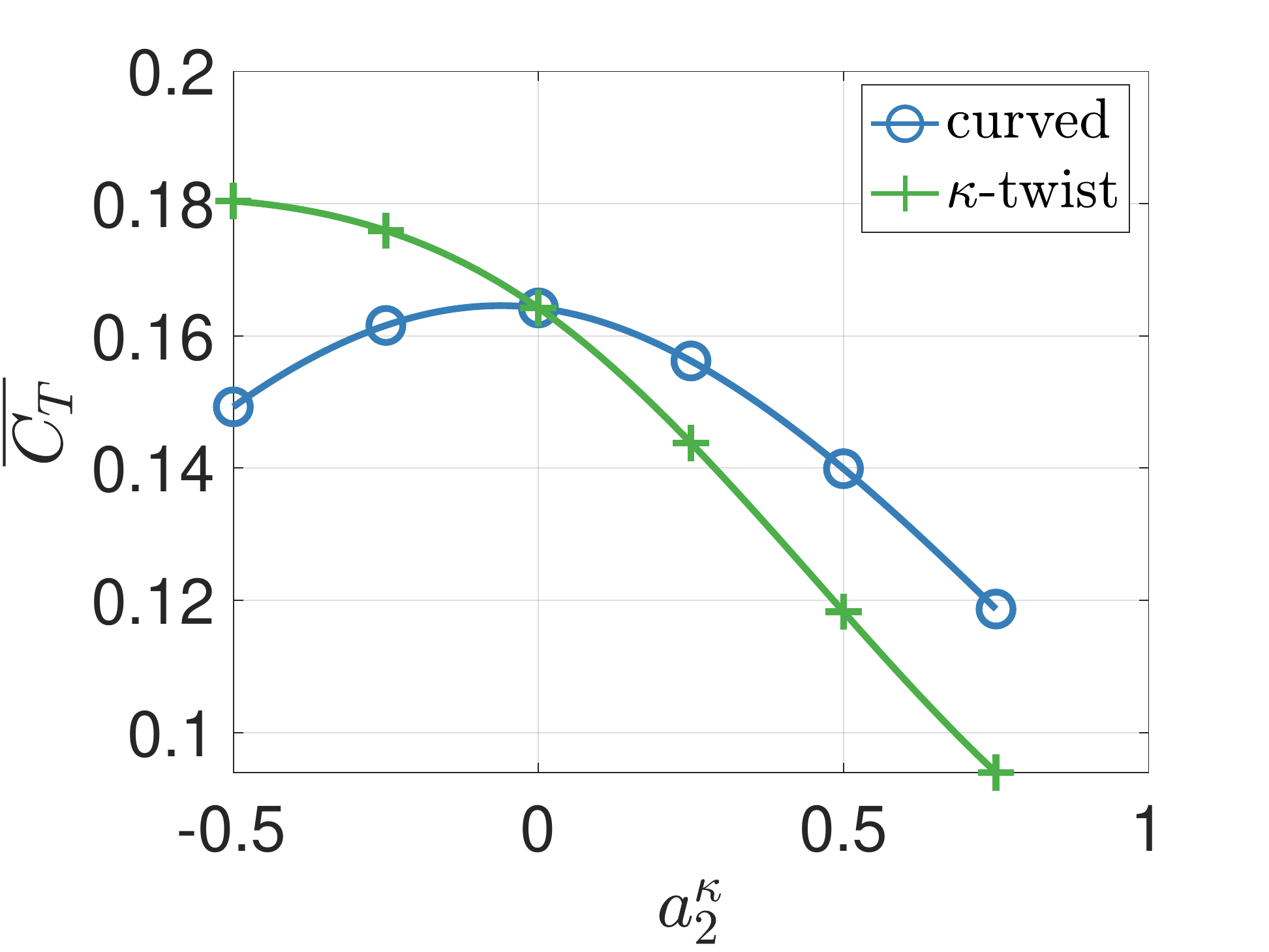}
\includegraphics[width=0.32\linewidth]{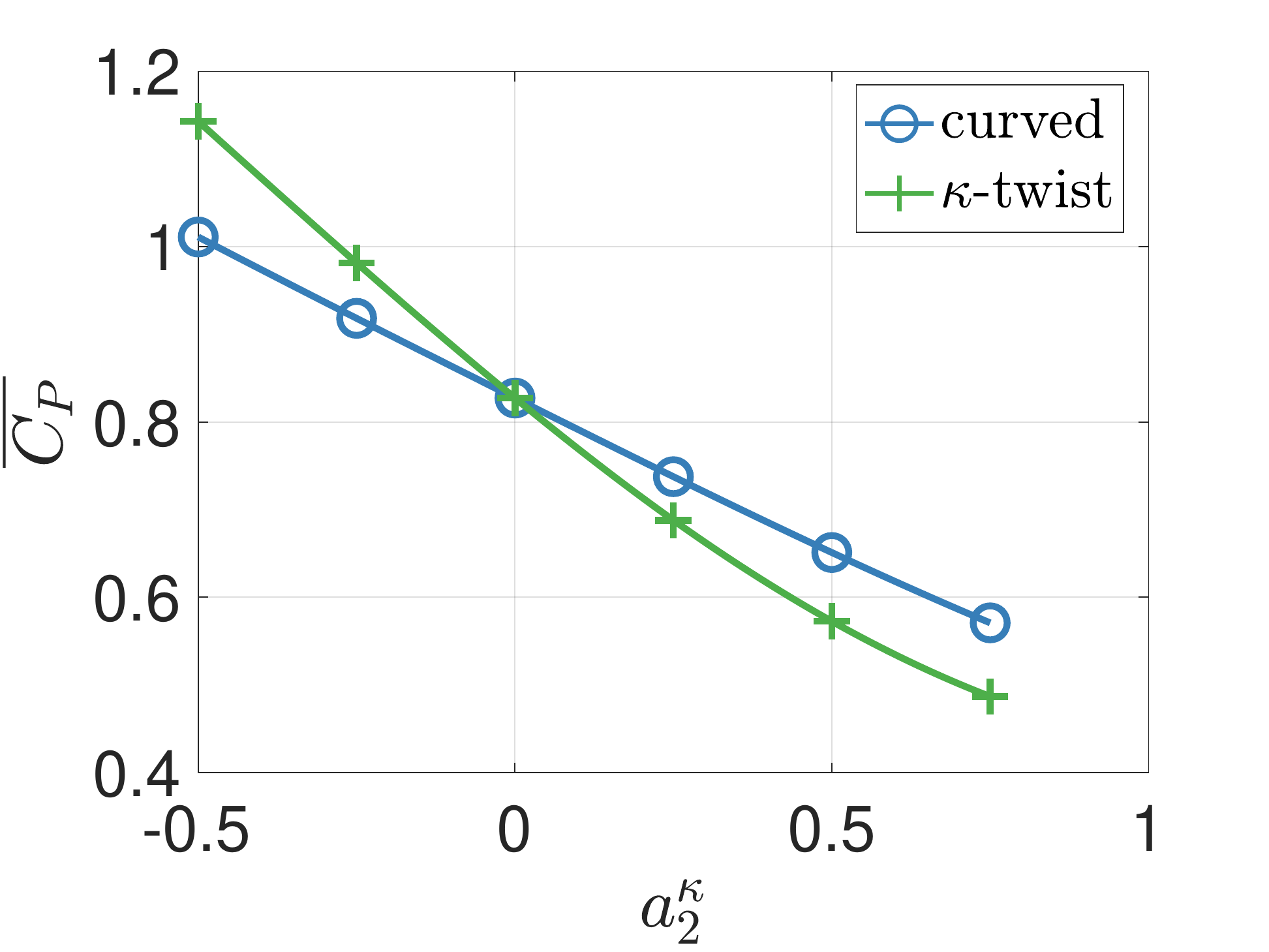}
\includegraphics[width=0.32\linewidth]{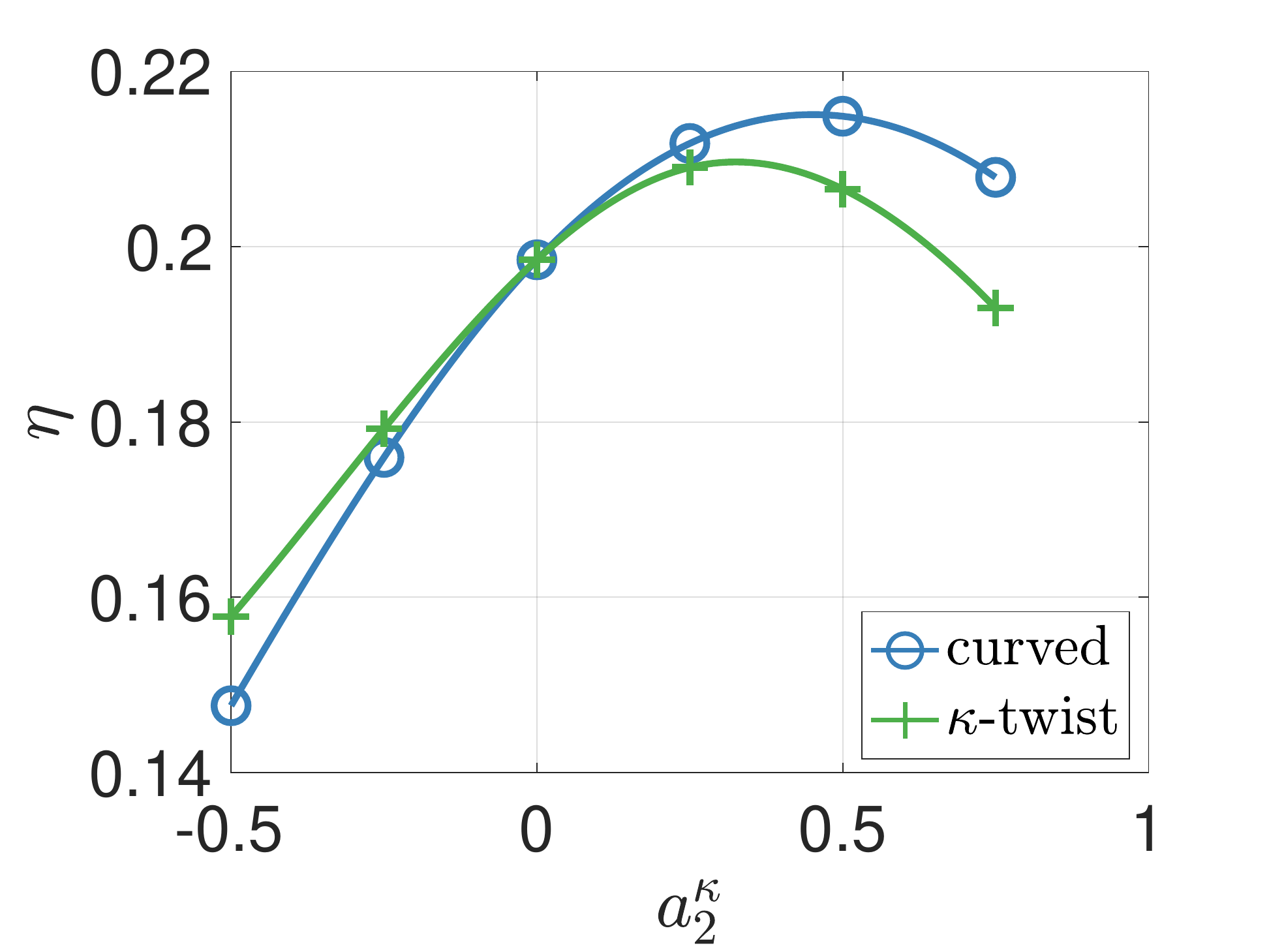}
\caption{Cycle-averaged mean thrust coefficient (\textit{left}), mean power coefficient (\textit{middle}), and efficiency (\textit{right}) as a function of the spanwise curvature parameter $a_2^\kappa$, for the fin with curvature variations (\textit{blue}) and the $\kappa$-twist configuration (\textit{green}).}
\label{fig:caudalfin_ktwist}
\end{figure}

Figure~\ref{fig:caudalfin_ktwist} compares the behavior of the deformed fin with that of the $\kappa$-twisted fin, across the range of $a_2^\kappa$ values considered. The qualitative trends are similar, with increasing $a_2^\kappa$ values increasing thrust, decreasing power, and increasing efficiency for both the curving and the $\kappa$-twist fins. This demonstrates that the spanwise twist is the predominant factor underlying these hydrodynamic characteristics, rather than the actual curvature of the rays. We observe a slight increase in peak efficiency of the curving fin compared with the $\kappa$-twist configuration indicating that here, again, the camber can improve efficiency. 

\begin{figure}
\centering
\includegraphics[width=0.49\linewidth]{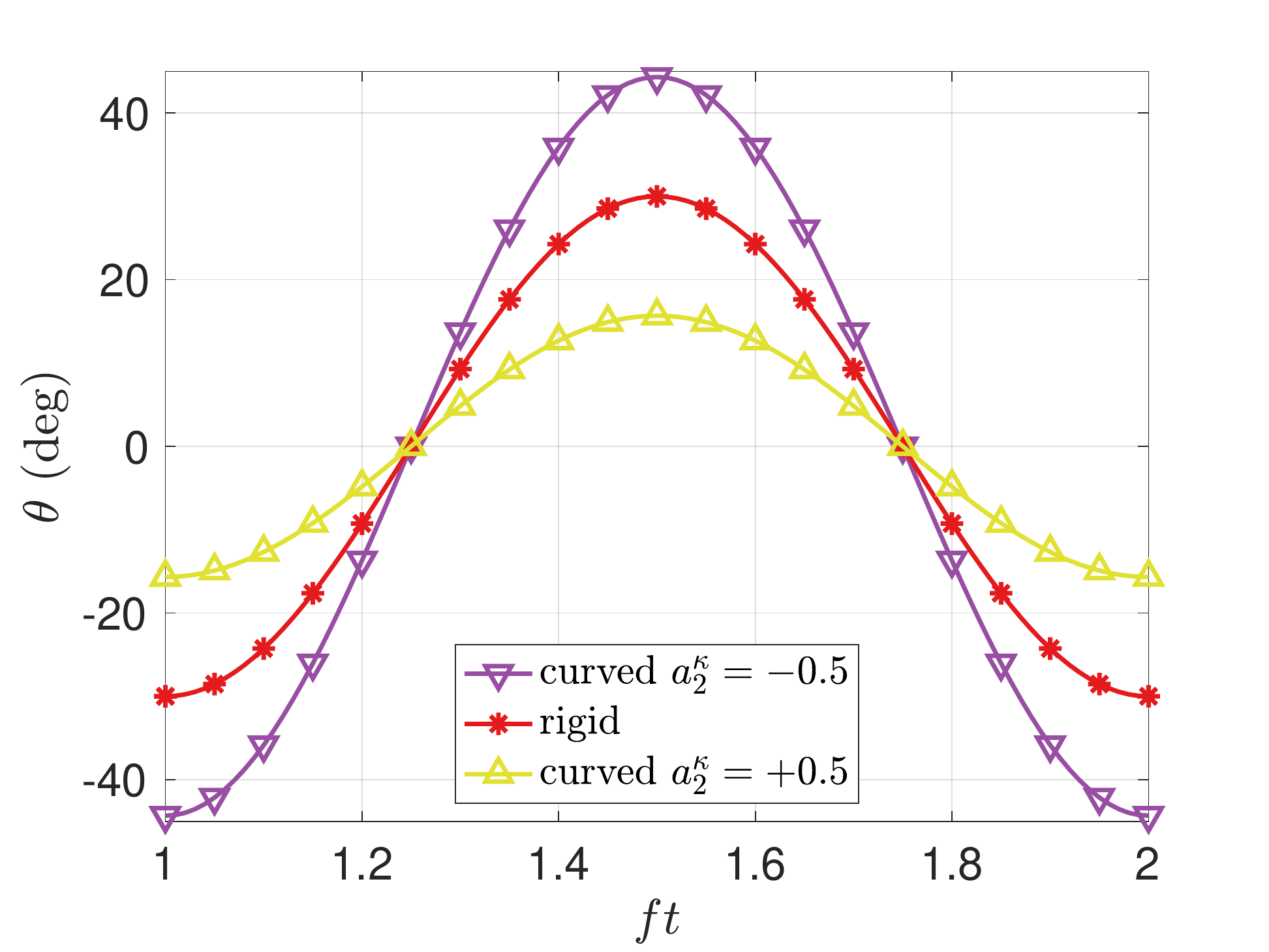}
\includegraphics[width=0.49\linewidth]{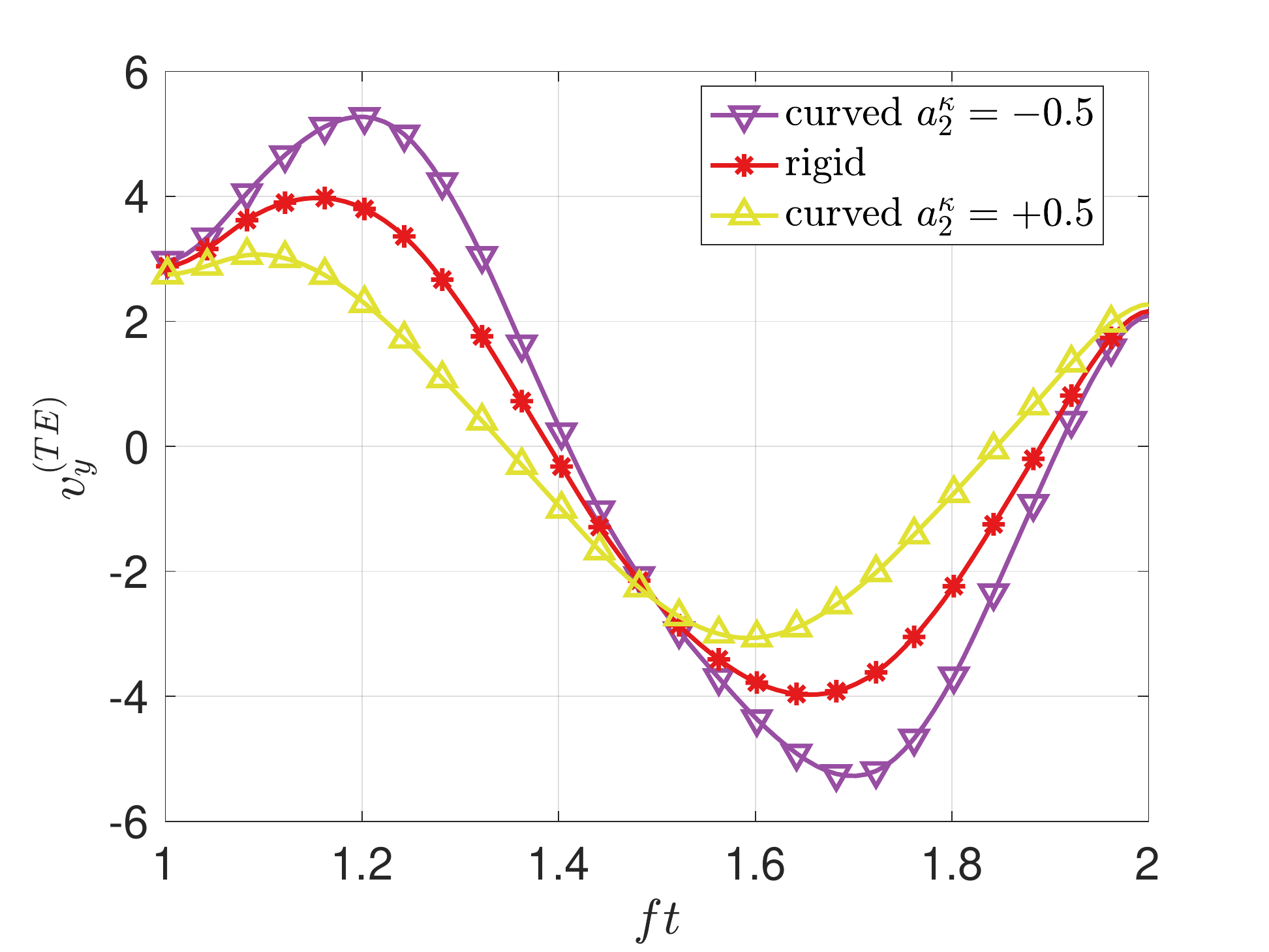}
\caption{Pitch angle (\textit{left}) and trailing edge lateral velocity (\textit{right}) of the top ray during a flapping cycle for the rigid fin (\textit{red}), and the fin with spanwise curvature variations $a_2^\kappa = -0.5$~(\textit{purple}) and $a_2^\kappa = +0.5$~(\textit{yellow}).}
\label{fig:caudalfin_ktwist_vyTE}
\end{figure}

\begin{figure}
\centering
\includegraphics[width=\linewidth]{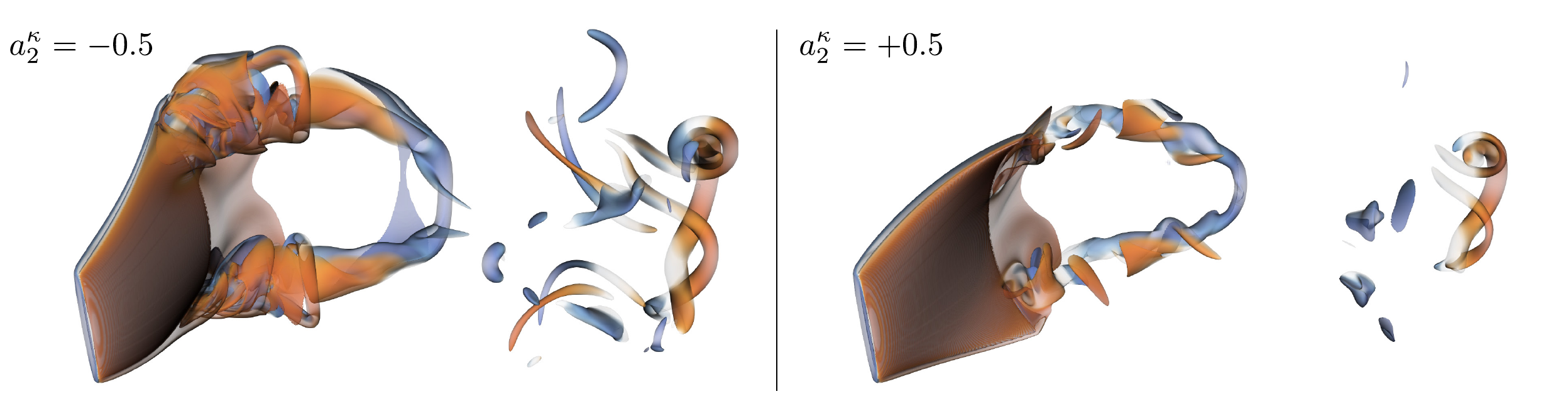}
\caption{Vorticity field of the fin for the sole spanwise curvature configuration with $a_2^\kappa=-0.5$~(\textit{left}) and $a_2^\kappa=0.5$~(\textit{right}), both at $f t =1.5$ and with $a_0^\kappa = 0$. The vorticity structures visualize, and are colored by, $\omega_z$. }
\label{fig:caudalfin_flow_spanwise}
\end{figure}

To understand the effect of $\kappa$-twist kinematics on the performance, we can examine equation~\eqref{eq:kappa_twist} further. For our spanwise curvature parametrization, the spanwise curvature variations are in phase with pitch but of the opposite sign. Positive values of $a_2^\kappa$ then decrease the effective pitch angle, and vice versa, with the maximum effect noticeable at the top and bottom of the fin, away from the centerplane.  This is observed in figure~\ref{fig:caudalfin_ktwist_vyTE}, showing that the pitch angle and TE velocity amplitude of the top ray during a flapping cycle significantly reduces when $a_2^\kappa$ is increased. 
Consequently, since the outer parts of the fin undergo smaller pitching amplitudes, the associated power reduction is observed predominantly in the pitching component $C_P^M$. Further, the reduced power and increased efficiency with increasing spanwise curvature parameter are consistent with the smaller vortical signature of the wake, as shown in figure~\ref{fig:caudalfin_flow_spanwise}. The twisted configuration with $a_2^\kappa = 0.5$ leads to significantly smaller tip vortices shed from the outer edges of the fin, compared with both rigid and $a_2^\kappa = -0.5$. Lastly, we note that the qualitative deformation of the fin when $a_2^\kappa > 0$ is intuitively consistent with the elastic deformation of a finite-span flapping fin: the outer edges will curl inwards during the heave reversal, lagging behind the central rays of the fin. Together with the previous results, this provides further indication that the curvature variations of passively deforming 3D fins can lead to higher propulsive efficiency that those of rigid fins, as measured solely through hydrodynamic performance.

\section{Concluding remarks} \label{s:conclusions}

Our results describe and analyse the hydrodynamic effects of leading-edge actuated curvature variations on flapping fin performance. We have demonstrated that such actuation can lead to an increase in efficiency by about $18$\% and mean thrust coefficient by about $15\%$ compared to a rigid fin. Within our parametrization, thrust is maximized when considering some degree of positive curvature, both chordwise and spanwise, while efficiency benefits from negative chordwise and positive spanwise curvature deformations. Throughout, the chordwise parameter dominates the hydrodynamic performance, with spanwise curvature variations only providing small additional changes in performance. 

Leading-edge curvature actuation not only introduces a camber in the fin cross-section, but also affects the trailing-edge kinematics. To investigate these two effects separately, we investigated rigid fins without camber whose pitch kinematics are tuned to the trailing edge kinematics of the curving fin. The analysis shows that the increase in thrust due to chordwise curvature against the flow can also obtained by a rigid fin with modified pitching kinematics; in fact, the rigid fin outperforms the curving fin, because the camber associated with the chordwise curvature variations is hydrodynamically disadvantageous. On the other hand, the increase in efficiency due to curvature with the flow is predominantly caused by the hydrodynamically advantageous camber in this regime, together with a small effect of spanwise twist that reduces the intensity of the shed tip vortices. 

Overall, throughout this work we have found that the performance benefits of fins with leading-edge curvature actuation can in large part be reproduced by rigid fins with suitably adapted pitch kinematics, ignoring the small benefits of camber changes on efficiency. This poses an interesting question, namely to compare the benefits of these different types of actuation: the trade-off between modulating the phase and amplitude of leading-edge curvature variations, versus those of the pitch kinematics. In nature, the kinematics of the flapping fin are rarely as simple as the idealized case considered here, since lateral motions and body undulations combine to give rise to what we model here as heaving and pitching motions in a uniform inflow. A possible benefit of leading-edge curvature actuation is then that it provides a localized approach that can be controlled independent of the body and swimming motions. This could improve the swimmer's versatility and responsiveness, enabling it to use local muscle actuation to deliver more thrust or reduced power without adapting the body undulations that give rise to the pitching kinematics. Further, specifically for the high-efficiency  curvature regimes considered here, part of our imposed deformations could be realized passively through elastic deformations to the hydrodynamic loading, making such swimming modes simpler to control. Taken together, a combination of the right structural design of the fin together with leading-edge curvature actuation could provide a simple, versatile, and responsive way to achieve the hydrodynamic benefits described in this work. 

Lastly, we note two distinct other contributions of this work. First, we have presented a geometric and numerical framework to construct ray-based fin-shapes with imposed ray curvature profiles. Second, we have demonstrated that rigid fins with modified kinematics can be used as a qualitative and, to some extent, quantitative proxy for fins with leading-edge curvature actuation. Both contributions will be used in future studies. This future work will be aimed at further separate the effect of trailing edge kinematics from the changes in camber of the fin, and to design curvature actuation patterns that can fully exploit the benefits of both.

\vspace{0.5cm}
\noindent
\textbf{Acknowledgments}
This work was supported by the Chang Fund from the MIT School of Engineering Research Support Committee. WvR thanks the American Bureau of Shipping for support through a Career Development Chair, and the MIT Sea Grant College Program for support through the Doherty Professorship in Ocean Utilization.



\bibliographystyle{wim}
\bibliography{bib}

\end{document}


\maketitle
\thispagestyle{empty}

\appendix
\vspace{-15pt} 
\section{Verification} 
\label{app:verification}

To evaluate the accuracy of the numerical scheme, we investigate the flow around an ellipsoid heaving and pitching around its centroid, and compare our force coefficient results to those presented in \citet{dong2006}. Rather than representing the ellipsoid directly through its continuum geometric description, we use our algorithm as described above to the describe the shape. Specifically, twenty-one rays are distributed along the ellipsoid height such that
\begin{align}
    v_i = 0.5 \cos \left( \pi \frac{N_r-i}{N_r -1}  \right) \,, \quad i=1..N_r \,.
\end{align}
The aspect ratio of the ellipse is $a_z / a_x = 4$, where $a_x$ and $a_z$ are the ellipsoid length and height, respectively. We adopt the same configuration as \citep{dong2006} by setting $\Rey = U_{\infty} a_x/\nu = 200$, $\Str=0.3$, $\tilde{A}_y=0.5$, $A_{\theta}=30^{\circ}$, and $\varphi_{\theta}=-90^{\circ}$.  

Figure~\ref{fig:ct_cl_dong} compares the reported time-history of thrust coefficients with the values from our computational model. We include the time-series obtained with various grid resolutions, compared with the data taken from the reference paper. 
\begin{inplacefigure}
\centering
\includegraphics[width=\linewidth]{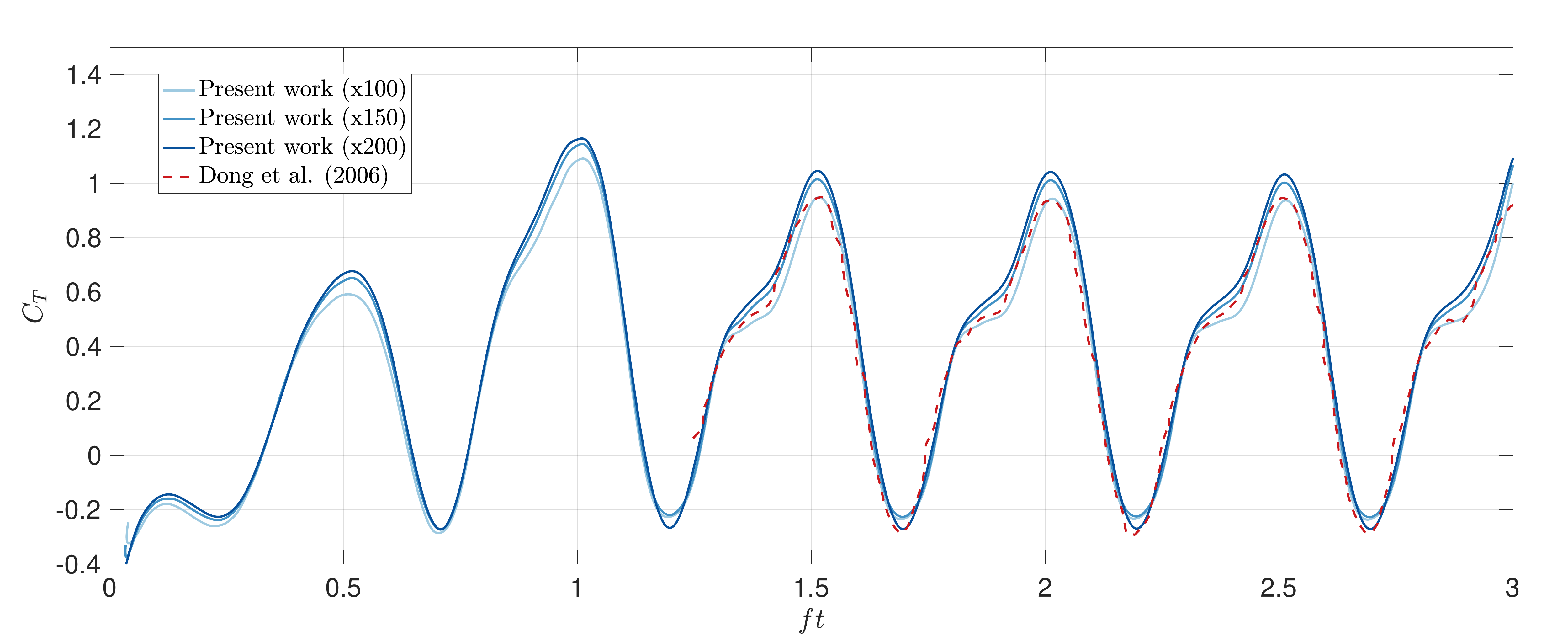} 
\caption{Comparison between our results (\textit{solid}) and the results adapted from \citet{dong2006} (\textit{red dashed}) of the thrust coefficient time-series for a heaving and pitching ellipse with aspect ratio $a_z/a_x  = 4$. 
}
\label{fig:ct_cl_dong}
\end{inplacefigure}

We obtain a very similar match, although some differences appear in peak amplitude for both thrust and lift coefficients. Our computed forces are, however, converged, as shown by the grid resolution study. Together with the previous validations of the methodology \citep{gazzola2011, bernier2019}, this validation case provides further confidence in our results.

\section{Chordwise curvature parametrization} 
\label{app:chordwise_curv}

We can interpret a configuration with sole chordwise curvature 
as a deformed shape that is purely characterized by the cross section of the mid-surface by the $x-y$ plane. 
Under the assumption of uniform $\kappa^n$ along each ray adopted in this work, this cross-section follows a circular trajectory \footnote{Note that the deformed cross-section remains at the $x-y$ plane since there is no spanwise curvature, so $\kappa^t(u,v=0,t)=\kappa^g(u,v=0,t)=0$}. This implies that the deformed mid-surface must be embedded in a cylinder of radius~$C/a_0^{\kappa}$ as shown in figure~\ref{fig:geom_chordwisecurv}, where $a_0^{\kappa}$ is the coefficient modulating the non-dimensional chordwise curvature (section~3.2). 
However, we show here that this shape cannot be generated with constant $\kappa^n$ across rays when the ray angle $\beta \neq 0$, which means  a correction factor is needed in the definition of $\kappa^n$ to account for the direction of each ray.
\begin{inplacefigure}
\centering
\includegraphics[height=2.0in]{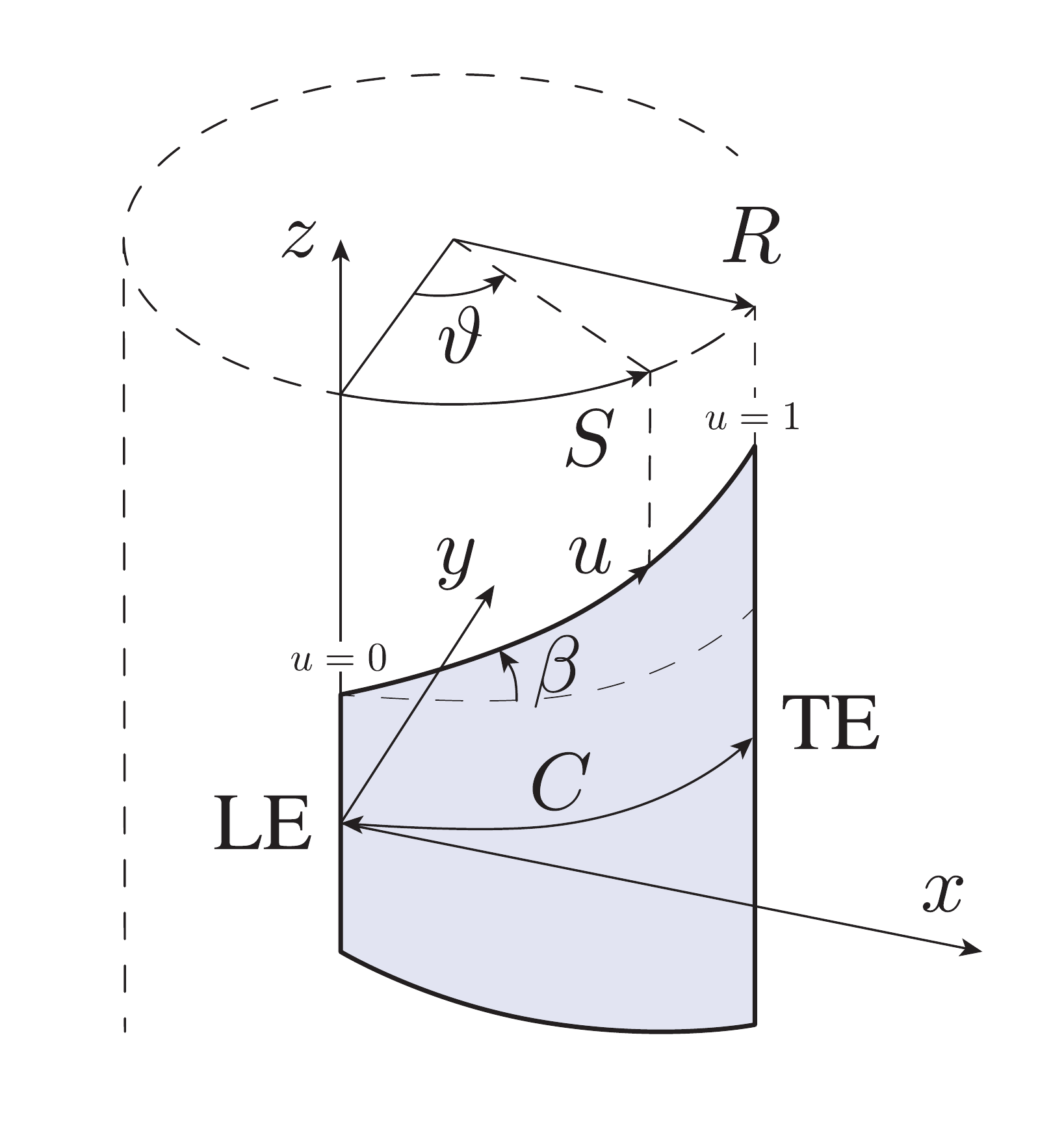}
\caption{The fin shape (\textit{in blue}) follows the surface of a cylinder with radius $C/a_0^\kappa$ in a sole chordwise curvature configuration as determined by $a_0^\kappa$.}
\label{fig:geom_chordwisecurv}
\end{inplacefigure}

Starting from equations~(2.3)-(2.4), 
we recall that the ray shape is reconstructed starting from the LE through integration along its chord. The LE is aligned with the $z$-axis, and, consistent with the rest of this work, we consider the case where the TE is also described by a straight line parallel to the $z$-axis.

Denoting $\vartheta$ as the angular coordinate over the cylinder containing the mid-surface, as shown in figure~\ref{fig:geom_chordwisecurv}, we can establish the following geometrical relation:
\begin{align}
    \vartheta &= \frac{S}{R} = \frac{u\, c(v) \cos \beta(v)}{C/a_0^\kappa} = u \, a^{\kappa}_0 \,,
\end{align}
since, for a cylindrical deformation of a fin with straight TE, we can directly write $C = c(v) \cos \beta(v)$. Using the definition of a cylindrical surface, we can determine  the desired tangent and normal vectors for any ray at inclination angle $\beta(v)$ as
\begin{align}
    \hat{\bt} &= \cos(\beta) \cos(\vartheta) \hat{\bx} + \cos(\beta) \sin( \vartheta) \hat{\by} + \sin(\beta)\hat{\bz} \,,\\
    \hat{\bn} &= -\sin\left( \vartheta \right) \hat{\bx} + \cos\left( \vartheta \right) \hat{\by} \,.
\end{align}
From this, we can determine analytically the non-dimensional normal curvature $\kappa^n(v)$ using its geometrical definition from equation~(2.2), 
\begin{align}
    \kappa^n &= \frac{\mathrm{d}\hat{\bt}}{\mathrm{d}u} \cdot \hat{\bn} 
    = a^{\kappa}_0 \cos(\beta) \,,
\end{align}
In dimensional form this expression becomes,
\begin{align}
    \frac{\kappa^n}{c(v)} &= \frac{a^{\kappa}_0}{c(v)} \cos(\beta(v)) = \frac{a^{\kappa}_0}{C} \cos^2(\beta(v))  \,,
\end{align}
The above derivation demonstrates that for a cylindrical fin deformation, the  correction factor $\cos(\beta(v))$ in equation~(3.5) 
is required to account for the particular direction of each ray.

\section{Grid convergence} 
\label{app:cell_refinement}

The trapezoidal fin described in section~3 
is used here to evaluate the grid convergence of the numerical scheme. 
In particular, we focus on three characteristic configurations from the range of simulations analyzed in section~4: 
the rigid fin, the curved fin that yields the largest efficiency ($a_0^{\kappa}=-0.2$ and $a_2^{\kappa}=0.25$), and the curved fin that produces the larger thrust ($a_0^{\kappa}=0.3$ and $a_2^{\kappa}=0.1$).

Table~\ref{tab:grid_convergence} summarizes the cycle-averaged thrust and power coefficients computed with four different grid resolutions of 100, 150, 200, and 250 points along $C$. 
We include also in table~\ref{tab:grid_convergence} the difference each in metric with respect to the finest resolution, defined for a generic variable $\phi$ as $\Delta_{\phi} = |\phi/\phi_{n_x/C = 250} -1|$. 
In addition, figure~\ref{fig:ct_cp_convergence} shows the convergence of time-series of thrust and power coefficients computed with each grid resolution for the configuration generating the maximum thrust, which yields the largest flow perturbations.  
\begin{inplacetable}
\small
  \begin{center}
  \begin{tabular}{@{}c@{}|@{}c@{}|@{}c@{}}
  rigid:  & max. thrust:  &  max. efficiency:  \\
  $a_0^{\kappa}=0$, $a_2^{\kappa}=0$ &
  $a_0^{\kappa}=0.3$, $a_2^{\kappa}=0.1$ &
  $a_0^{\kappa}=-0.2$, $a_2^{\kappa}=0.25$ \\
  \hline 
  \begin{tabular}{@{}c|cc|cc}
      $n_x/C$  & $\overline{C_T}$ &  $\Delta_{C_T}$  &   $\overline{C_P}$ &  $\Delta_{C_P}$ \\[3pt]
       100  & 0.122  &  28.7\%  & 0.832 & 0.7\% \\ 
       150  & 0.151  &  12.0\%  & 0.829 & 0.3\% \\ 
       200  & 0.164  &   4.3\%  & 0.827 & 0.1\% \\ 
       250  & 0.172  &    --    & 0.827 & -- \\ 
  \end{tabular} &
  \begin{tabular}{cc|cc}
      $\overline{C_T}$ &  $\Delta_{C_T}$  &   $\overline{C_P}$ &  $\Delta_{C_P}$ \\[3pt]
       0.140  &  29.6\%  & 1.344 & 0.1\% \\
       0.174  &  12.7\%  & 1.341 & 0.3\% \\
       0.189  &   4.8\%  & 1.343 & 0.2\% \\
       0.199  &    --    & 1.346 &  -- \\
  \end{tabular} & 
  \begin{tabular}{cc|cc@{}}
      $\overline{C_T}$ &  $\Delta_{C_T}$  &   $\overline{C_P}$ &  $\Delta_{C_P}$ \\[3pt]
       0.076  &  35.7\%  & 0.479 & 0.2\% \\ 
       0.100  &  15.0\%  & 0.478 & 0.0\% \\ 
       0.112  &   5.4\%  & 0.477 & 0.0\% \\ 
       0.118  &    --    & 0.477 &  -- \\
  \end{tabular}
  \end{tabular}
  \caption{Thrust and power coefficient variation with grid resolution. We report both the values of thrust and power coefficient for each case, as well as the difference with respect to the finest resolution simulated. }
  \label{tab:grid_convergence}
  \end{center}
\end{inplacetable}
\begin{inplacefigure}
\centering
\includegraphics[width=0.49\linewidth]{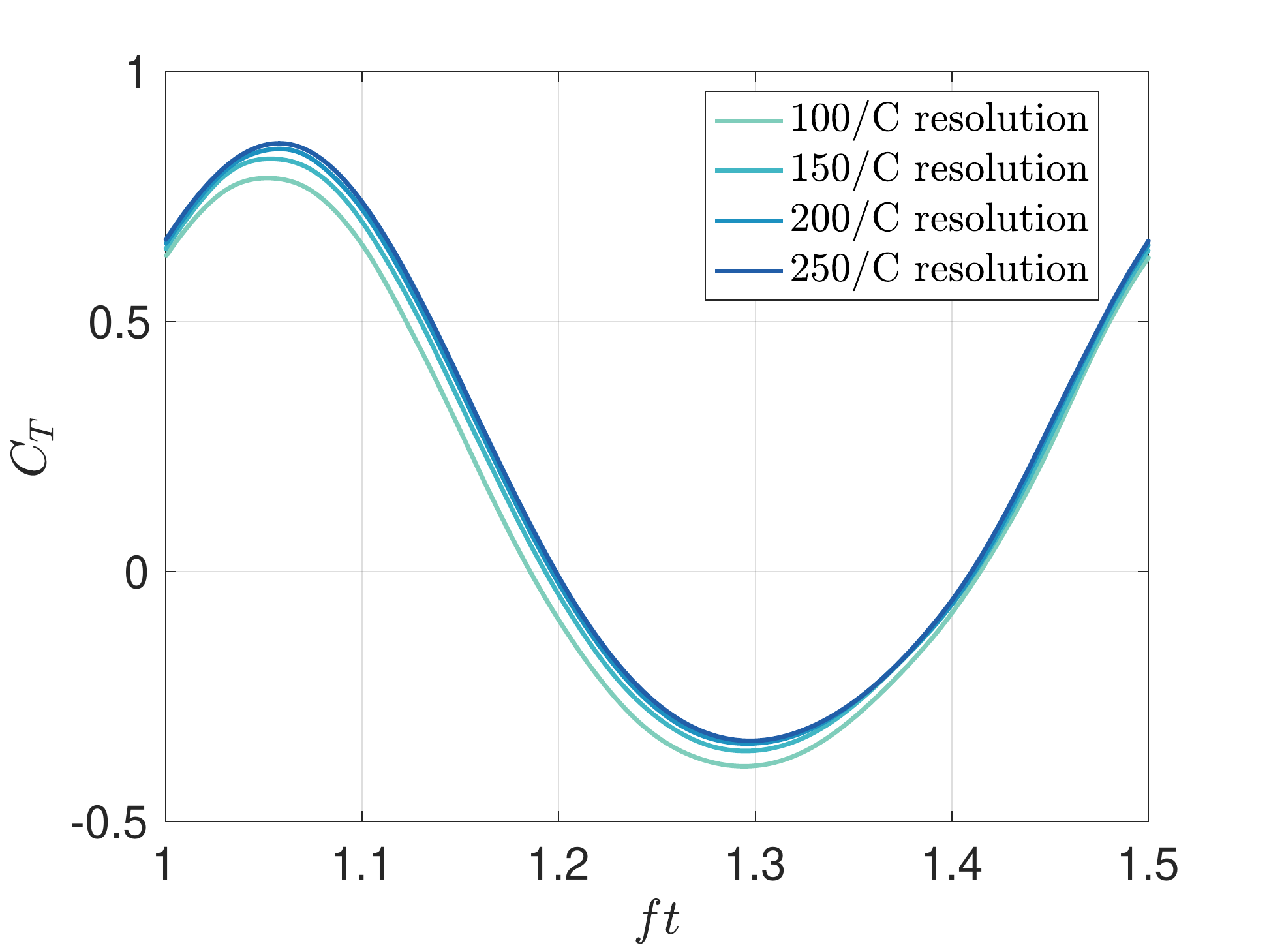} 
\includegraphics[width=0.49\linewidth]{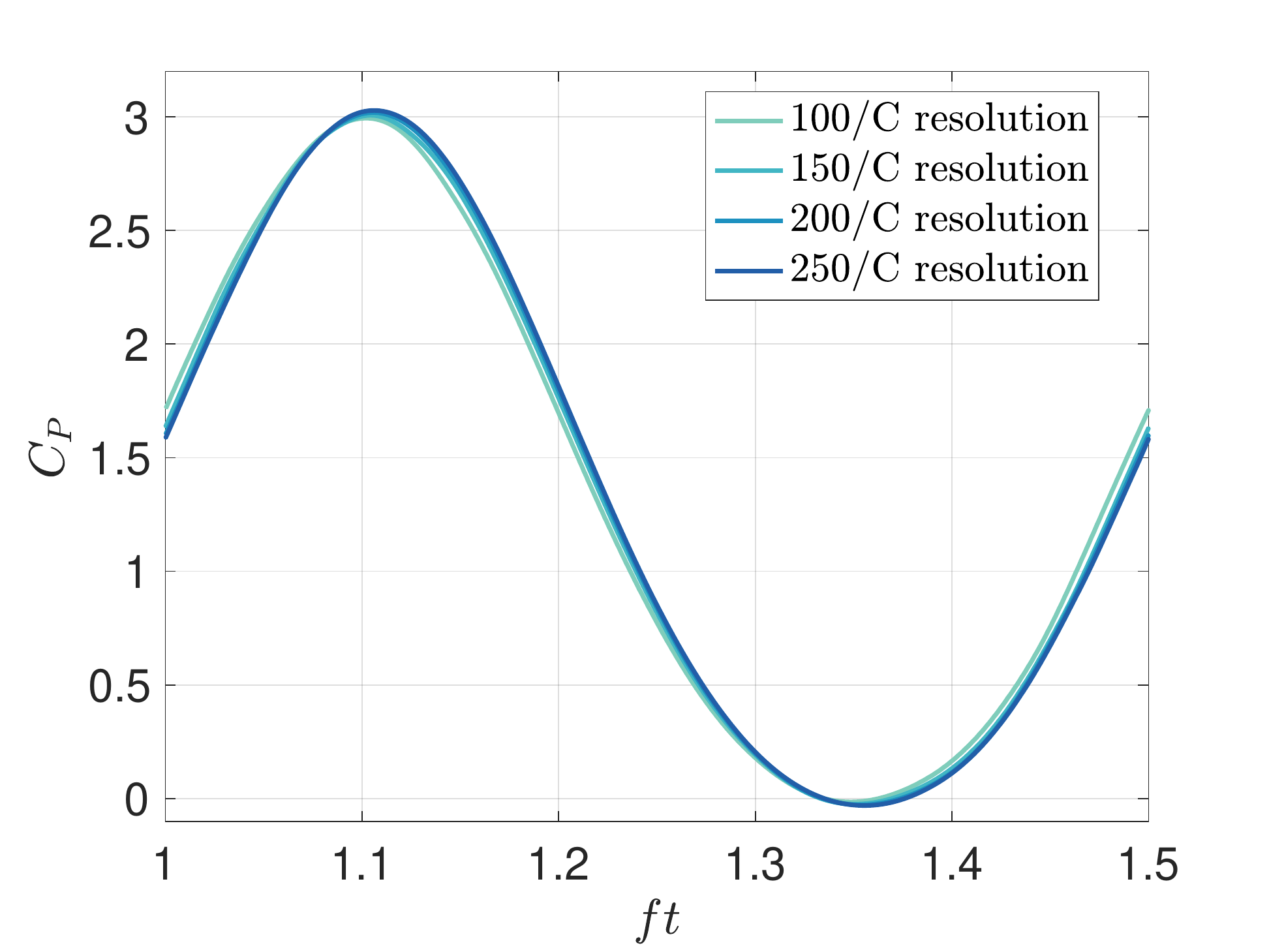} 
\caption{Thrust (\textit{left}) and power (\textit{right}) coefficients of the fin with various grid resolutions for the maximum thrust conditions ($a_0^{\kappa}=0.3$ and $a_2^{\kappa}=0.1$). }
\label{fig:ct_cp_convergence}
\end{inplacefigure}

The results show that the power coefficient is very robust to the resolution, both in terms of its time evolution as well as the time average. The time evolution of the thrust coefficient is also converging well within this resolution range, but the average value of a cycle is still sensitive to resolution changes. Based on these data, we have chosen to use 200 elements per chord as a balance between computational speed and accuracy. We emphasize that we are specifically interested in the effect of the different parameters, and the data in this section shows that this resolution is sufficient to predict trends in parameter variations, as well as a sufficiently close quantitative estimate of the metrics. The fact that the power coefficient is especially insensitive to the resolution further gives us confidence that our computed efficiency, defined as $\eta = \overline{C_T}/\overline{C_P}$, also provides sufficiently accuracy to support the results of this work.

\section{Constant thrust contour plot}
\label{app:design_contours}

Rather than purely maximizing thrust or efficiency, a more relevant design metric is to provide a given thrust to maintain steady incident stream velocity $U_{\infty}$ with the least power consumption, \textit{i.e.} with the largest efficiency possible. 
Graphically, we can visualize the potential gains from curvature under this design approach by observing the power or efficiency variations along a given thrust contour line, as shown in figure~\ref{fig:caudalfin_ctm_eff_cpwm}. Compared to the rigid case, we can appreciate how curving the fin provides a marginal increase in efficiency. We can attain more substantial gains if the fin shape can be controlled, in which case we can tailor the design to yield the required thrust under the conditions of optimum efficiency. The red lines in figure~\ref{fig:caudalfin_ctm_eff_cpwm} indicate the conditions that maximize the efficiency, or equivalently  minimize the required power to actuate the fin, for a given thrust. The shaded regions bound the region where, for each value of $C_T$, the efficiency is within $95$\% of its maximum value for that specific $C_T$. 

\begin{inplacefigure}
\centering
\includegraphics[width=0.49\linewidth]{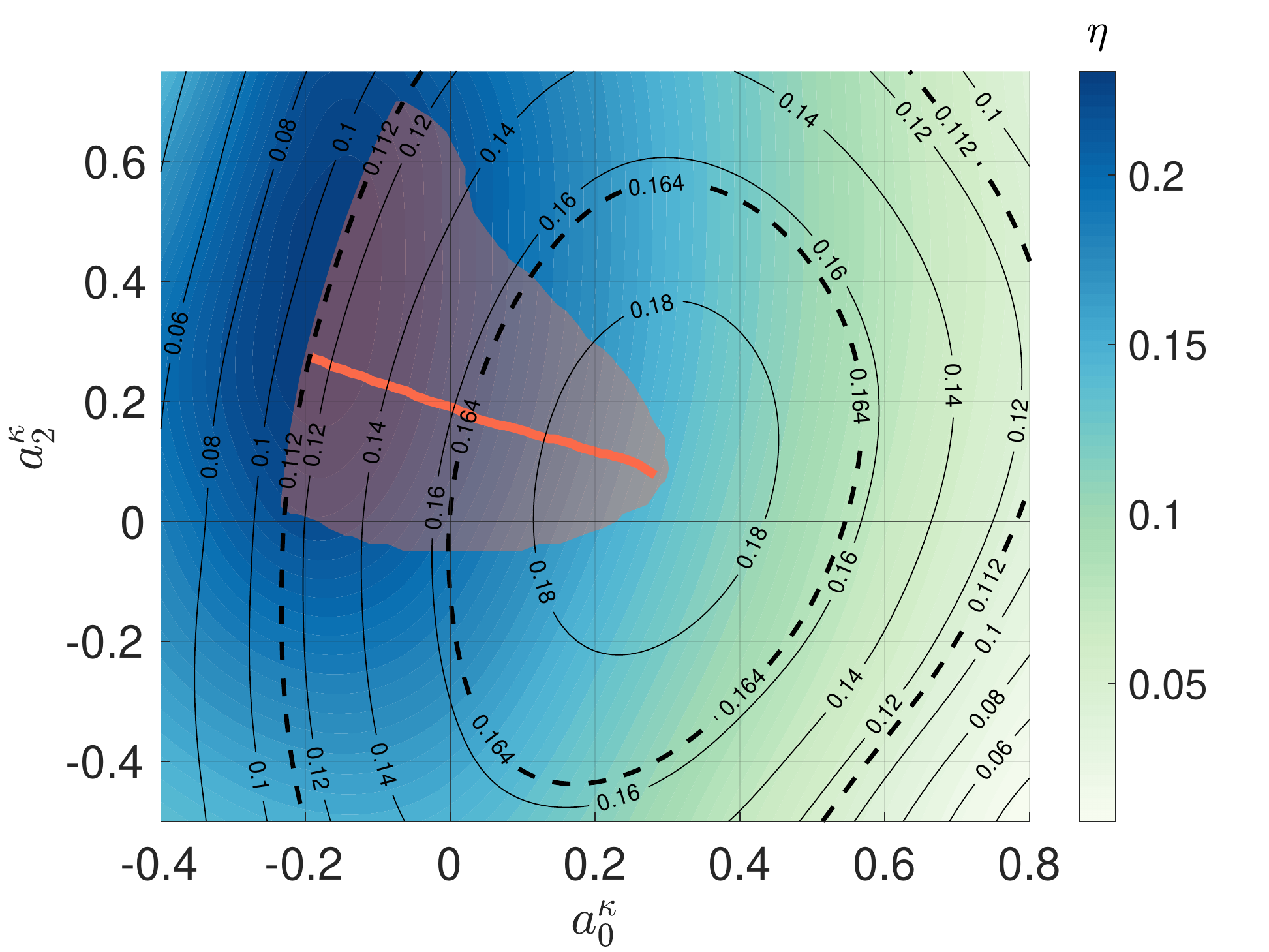}
\includegraphics[width=0.49\linewidth]{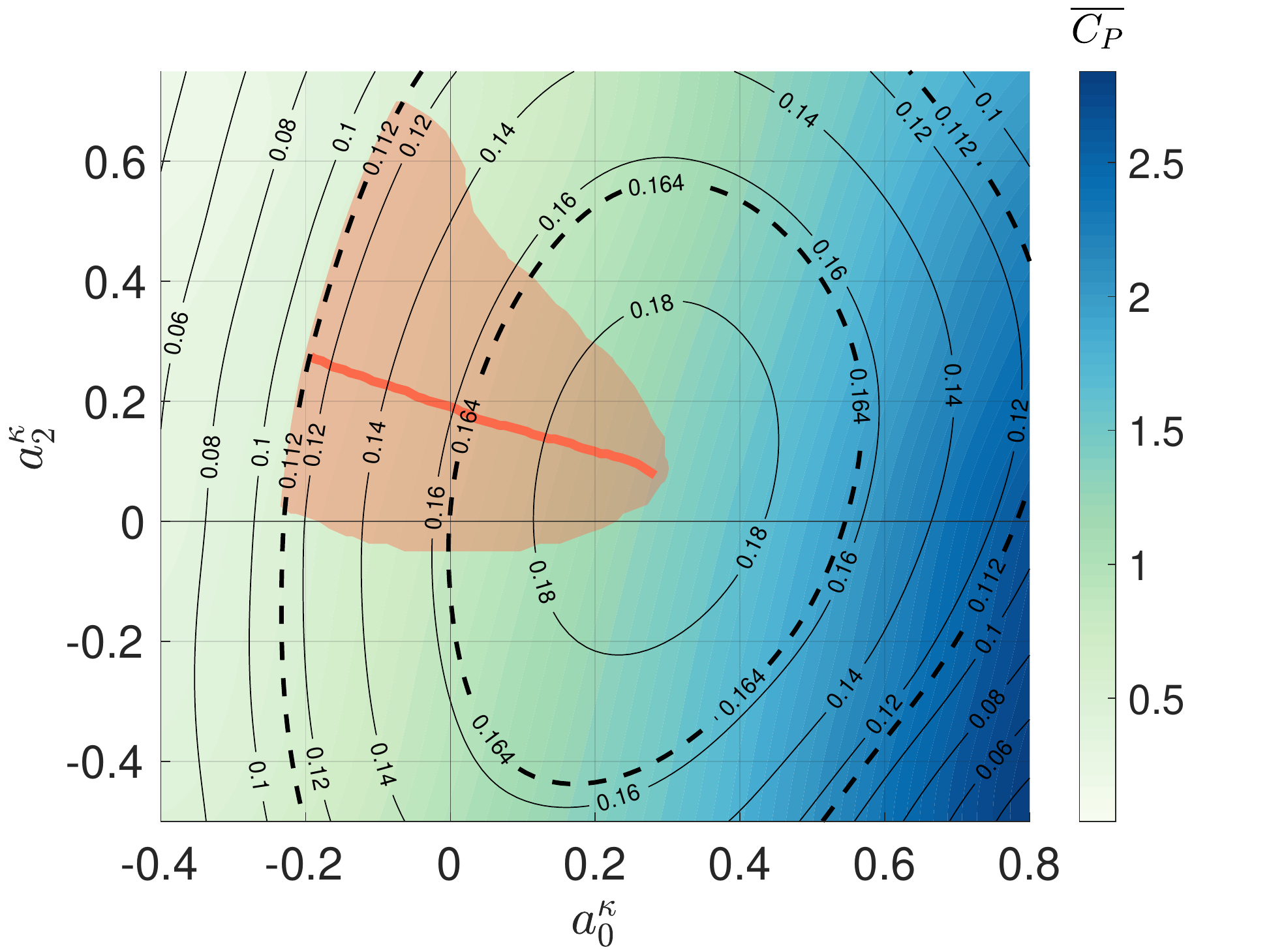}
\caption{Efficiency (\textit{left}) and cycle-averaged power coefficient (\textit{right}) contours, overlapped with thrust contour lines, as a function of the two curvature parameters $a_0^\kappa$ and $a_2^\kappa$. Dashed lines highlight the thrust contour lines crossing the rigid and maximum efficiency configurations. Red lines connect that points that, for each of the $C_T$ contours, achieve maximum efficiency (or equivalently the minimum power), varying between the $C_T$ of maximum thrust and the $C_T$ of the maximum efficiency solutions. Shaded red areas indicate the range where efficiency is higher than 95\% of the maximum efficiency for each thrust value along the red line.} 
\label{fig:caudalfin_ctm_eff_cpwm}
\end{inplacefigure}

\clearpage
\section{Square fin parametric analysis}
\label{app:square_fin}

The impact that the fin taper from its trapezoidal configuration has on thrust and efficiency can be assessed by computing the flow around an square fin, characterized by $\beta(v)=0$ and $H=C$, and whose results are displayed in figures~\ref{fig:square_ctm_eff_cpm}-\ref{fig:square_flow}. 

Overall, we observe that the same trends of the trapezoidal fin are kept, with a small decrease in the cycle-averaged thrust magnitudes of the square fin. 
\begin{inplacefigure}
\centering
\includegraphics[width=0.49\linewidth]{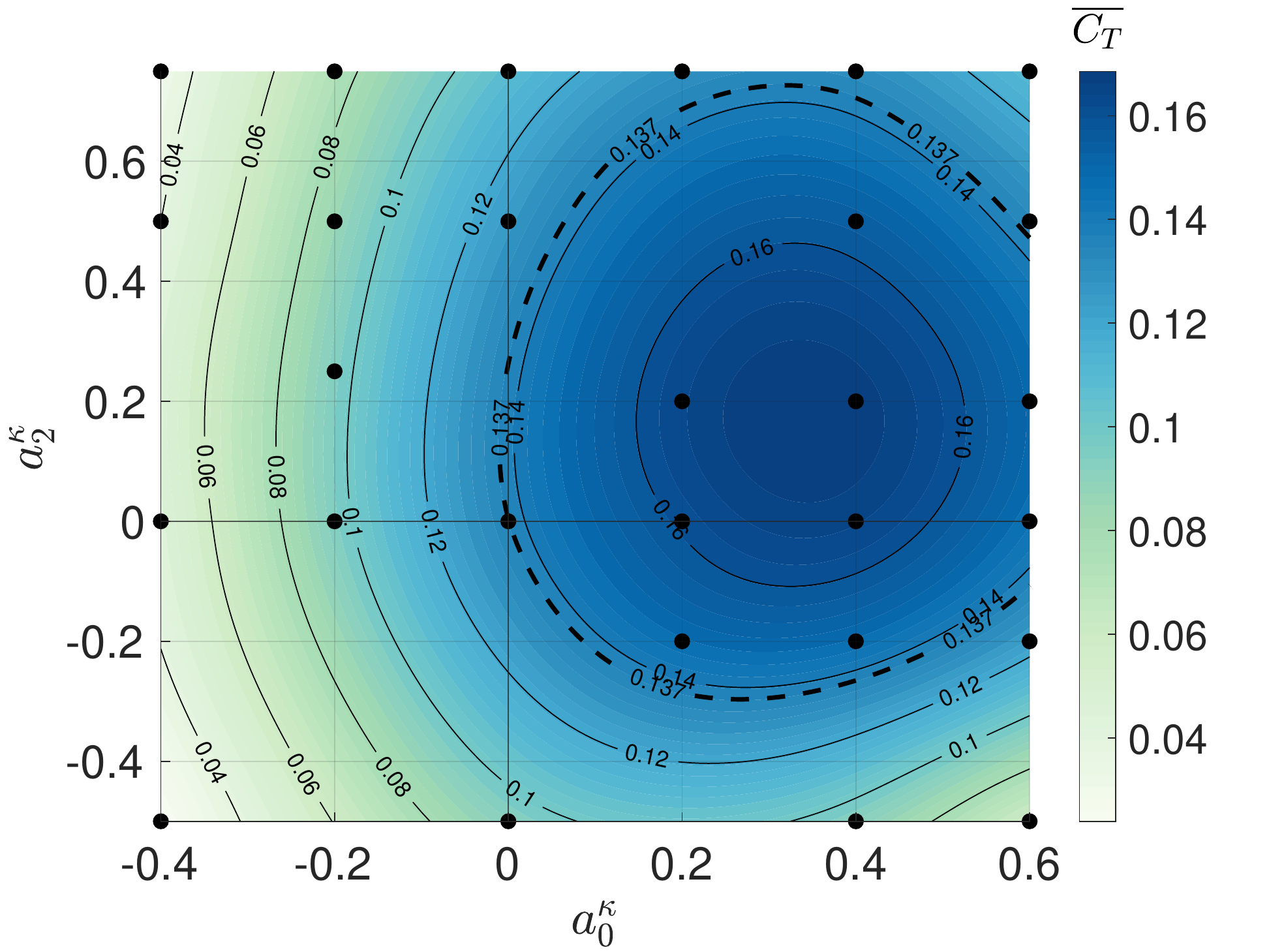}
\includegraphics[width=0.49\linewidth]{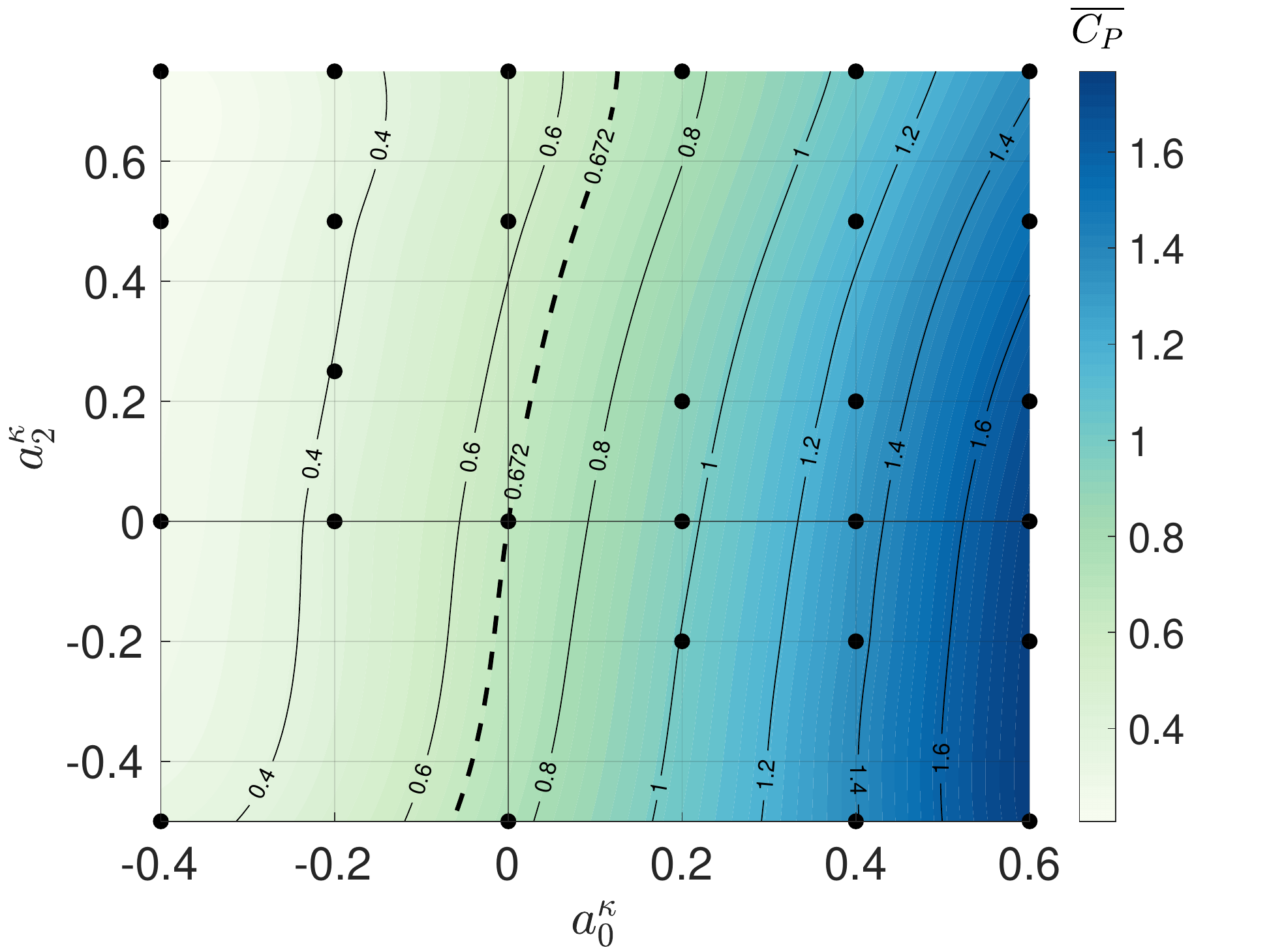}
\includegraphics[width=0.49\linewidth]{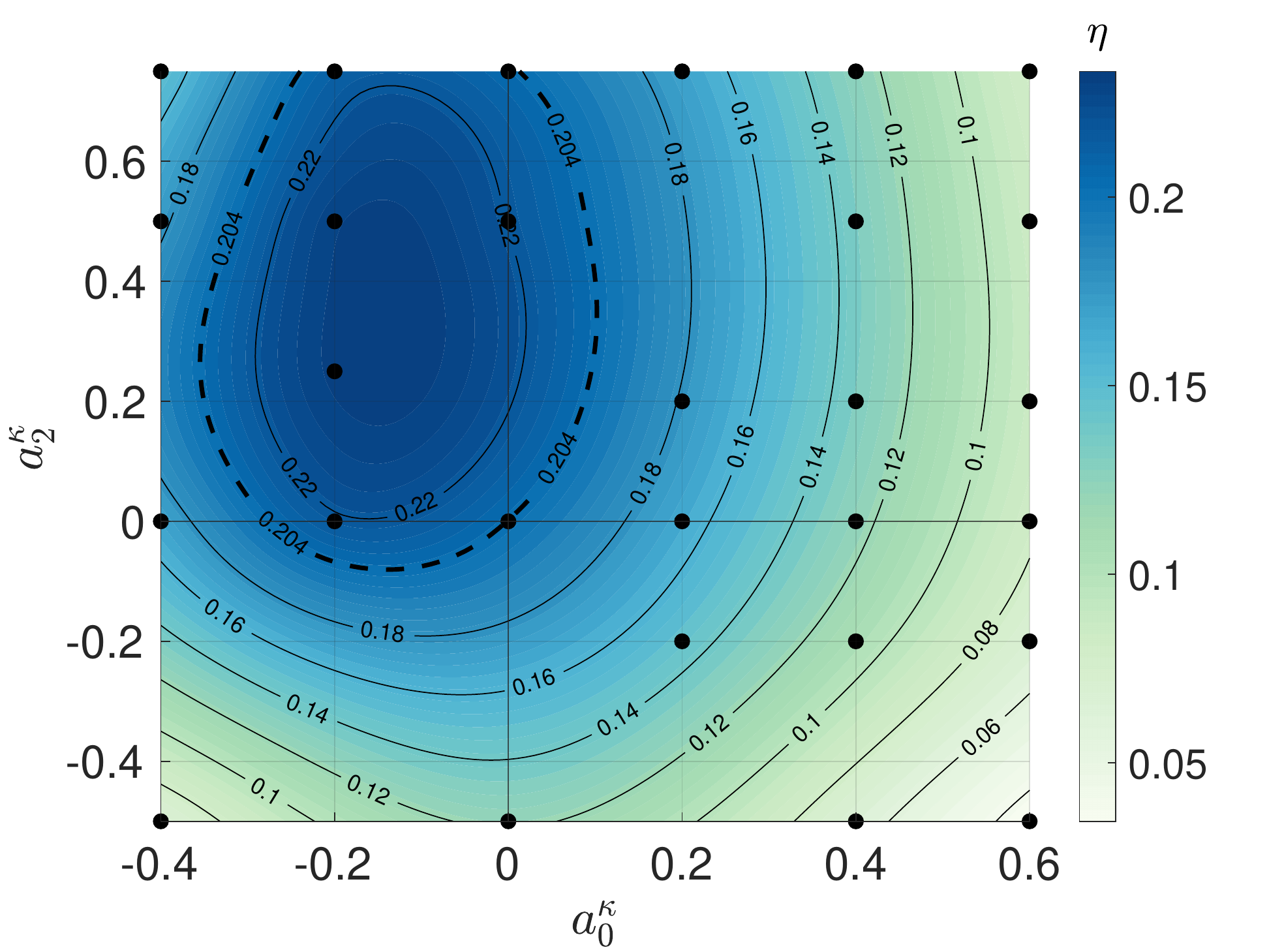}
\caption{Cycle-averaged thrust (\textit{top-left}) and power (\textit{top-right}) coefficients and efficiency (\textit{bottom}) results from Navier-Stokes simulations of a square fin (\textit{black dots}), and an interpolated contour plot based on these results, as a function of the two curvature parameters $a_0^\kappa$ and $a_2^\kappa$.}
\label{fig:square_ctm_eff_cpm}
\end{inplacefigure}
\begin{inplacefigure}
    \centering
    \includegraphics[width=\linewidth]{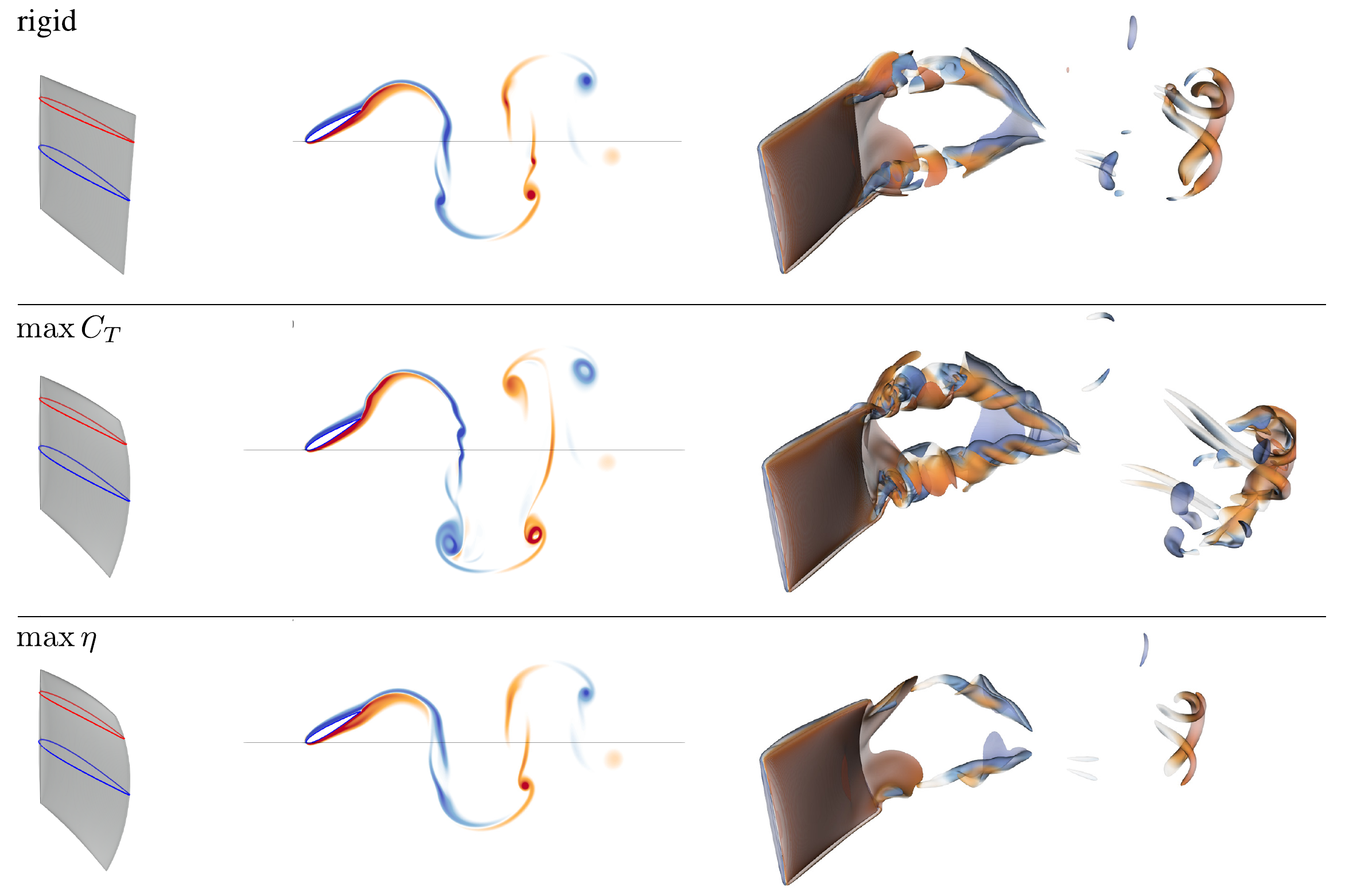}
    \caption{Square fin shape (\textit{left}), vorticity field at $v=0$ (\textit{middle}), and 3D vorticity field (\textit{right}) for the rigid configuration (\textit{top}, $a_0^\kappa=0.0$, $a_2^\kappa=0.0$), the maximum computed thrust configuration (\textit{middle}, $a_0^\kappa=0.4$, $a_2^\kappa=0.2$), and the maximum computed efficiency configuration (\textit{bottom}, $a_0^\kappa=-0.2$, $a_2^\kappa=0.25$), all at $ft=1.5$. Both the 2D and 3D vorticity fields visualize, and are colored by, $\omega_z$. }
    \label{fig:square_flow}
\end{inplacefigure}

\clearpage
\section{Effect of pitch kinematics variations on rigid fin performance}
\label{app:rigid-body}

Figure~\ref{fig:caudalfin_kpitch_amplitude_phase} shows the effect of changing the pitch amplitude (top) or the pitch phase (bottom) on the thrust and power coefficients, and the efficiency (left-to-right). In all cases the fin undergoes harmonic heave and pitch motions without curvature deformations. 

\begin{inplacefigure}
\centering
\includegraphics[width=0.32\linewidth]{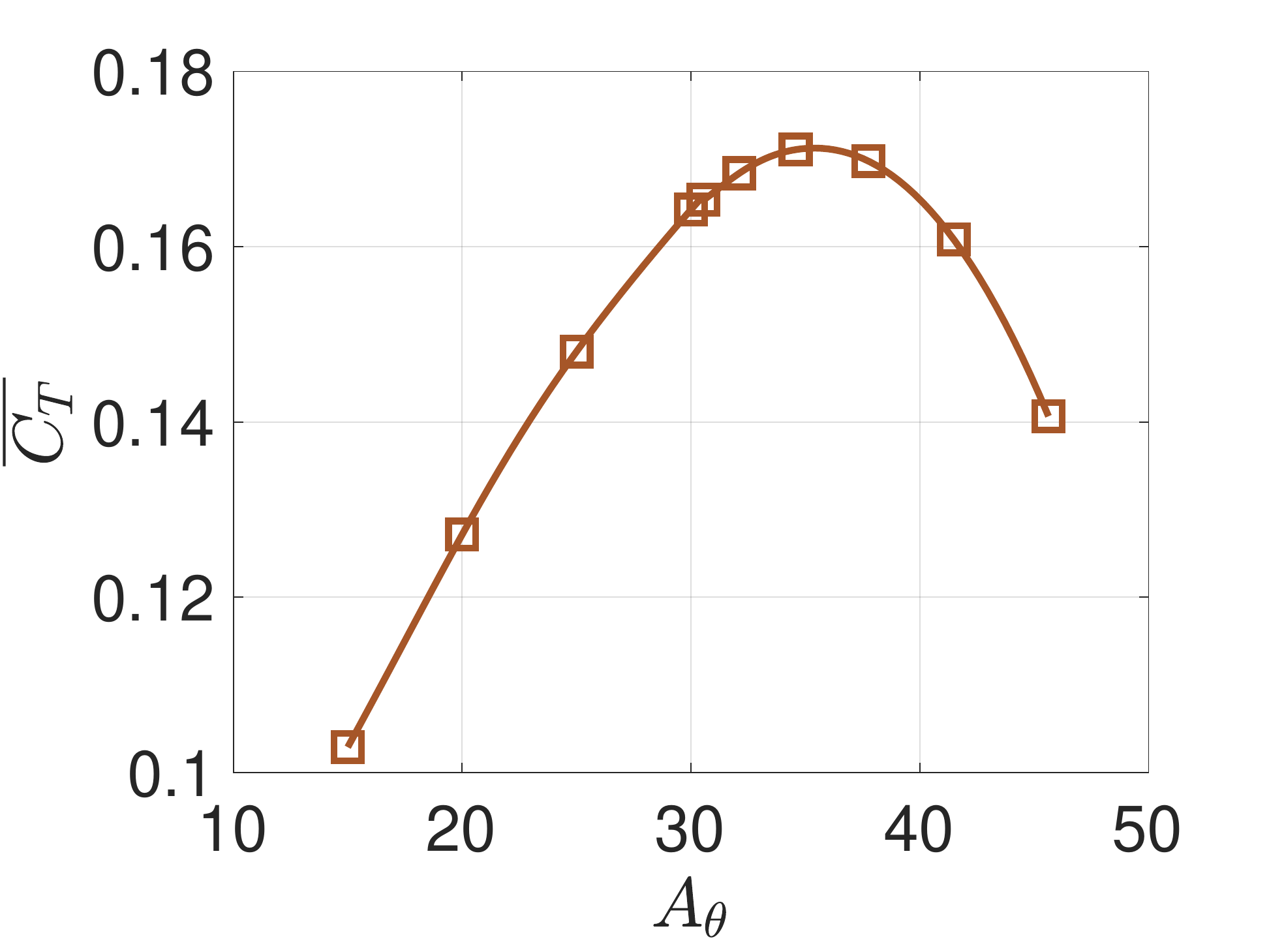}
\includegraphics[width=0.32\linewidth]{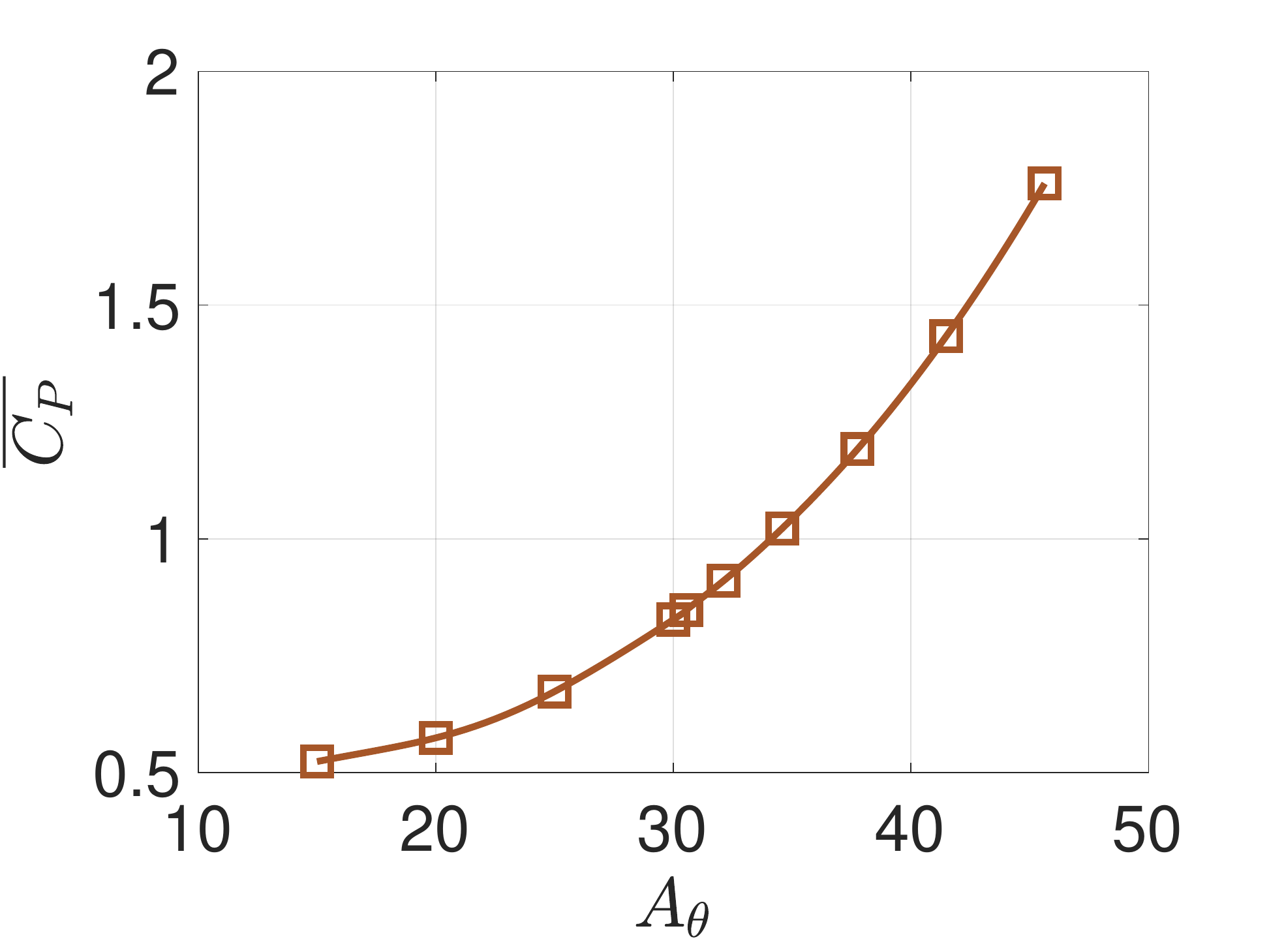}
\includegraphics[width=0.32\linewidth]{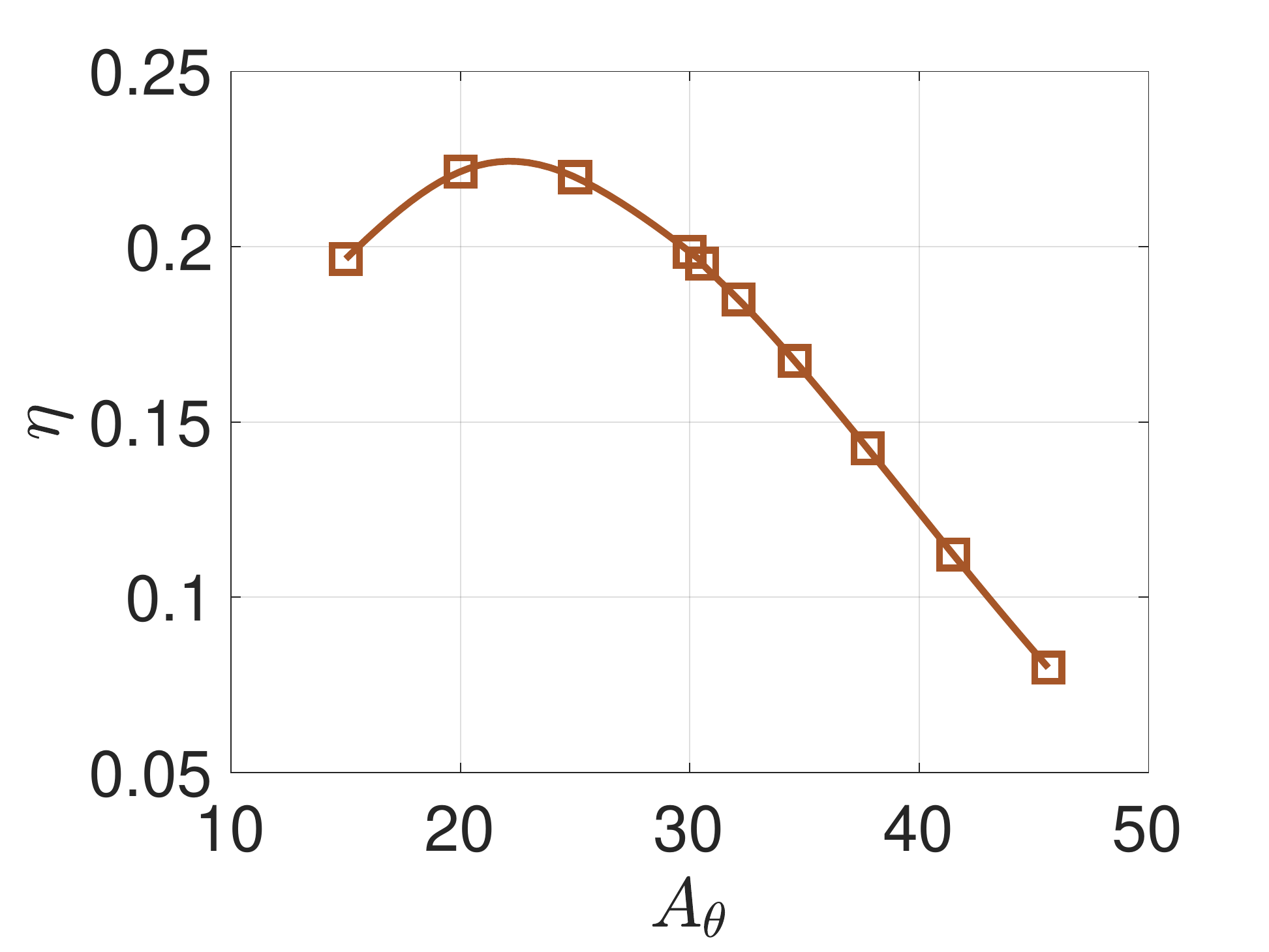}
\includegraphics[width=0.32\linewidth]{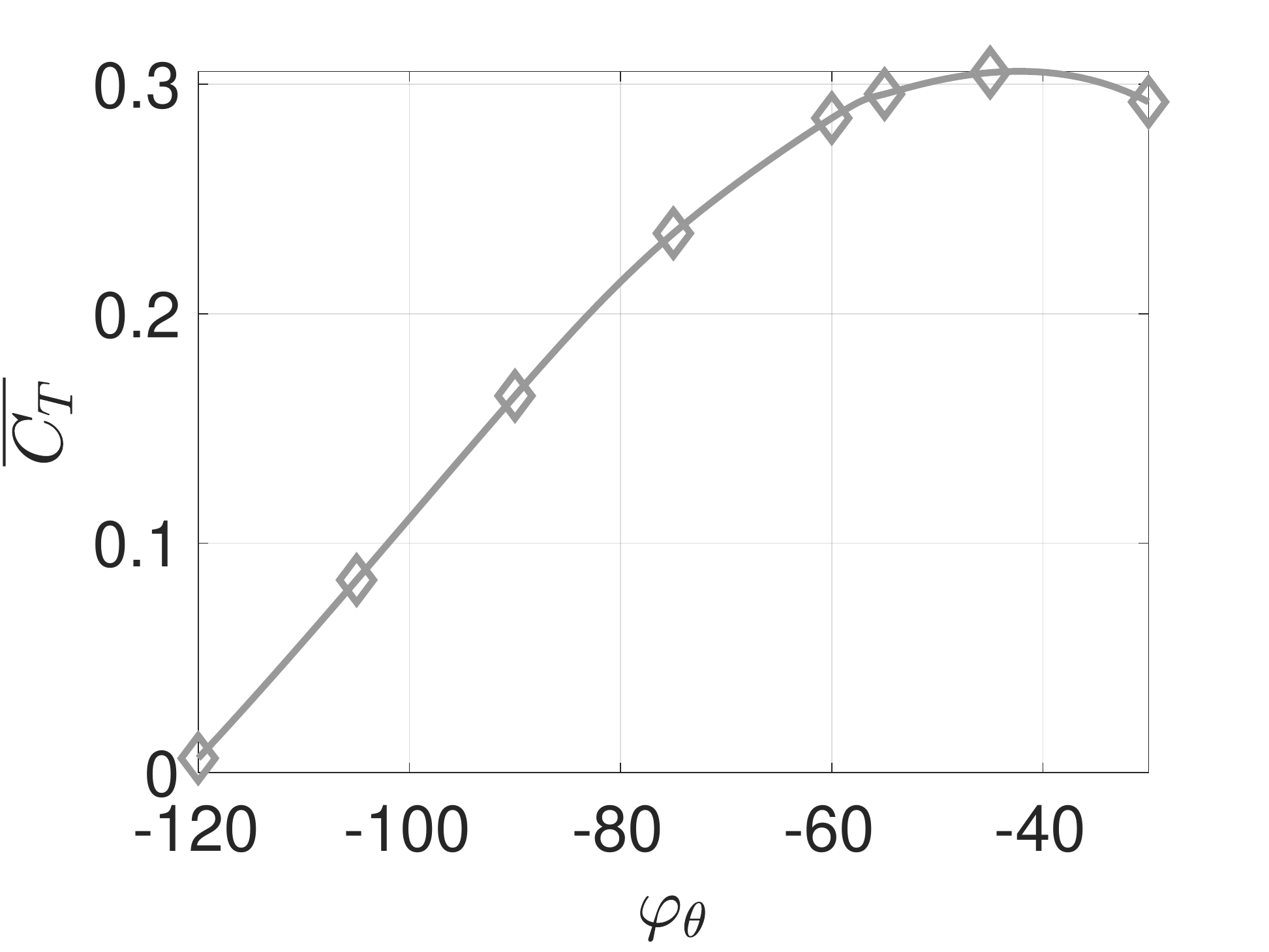}
\includegraphics[width=0.32\linewidth]{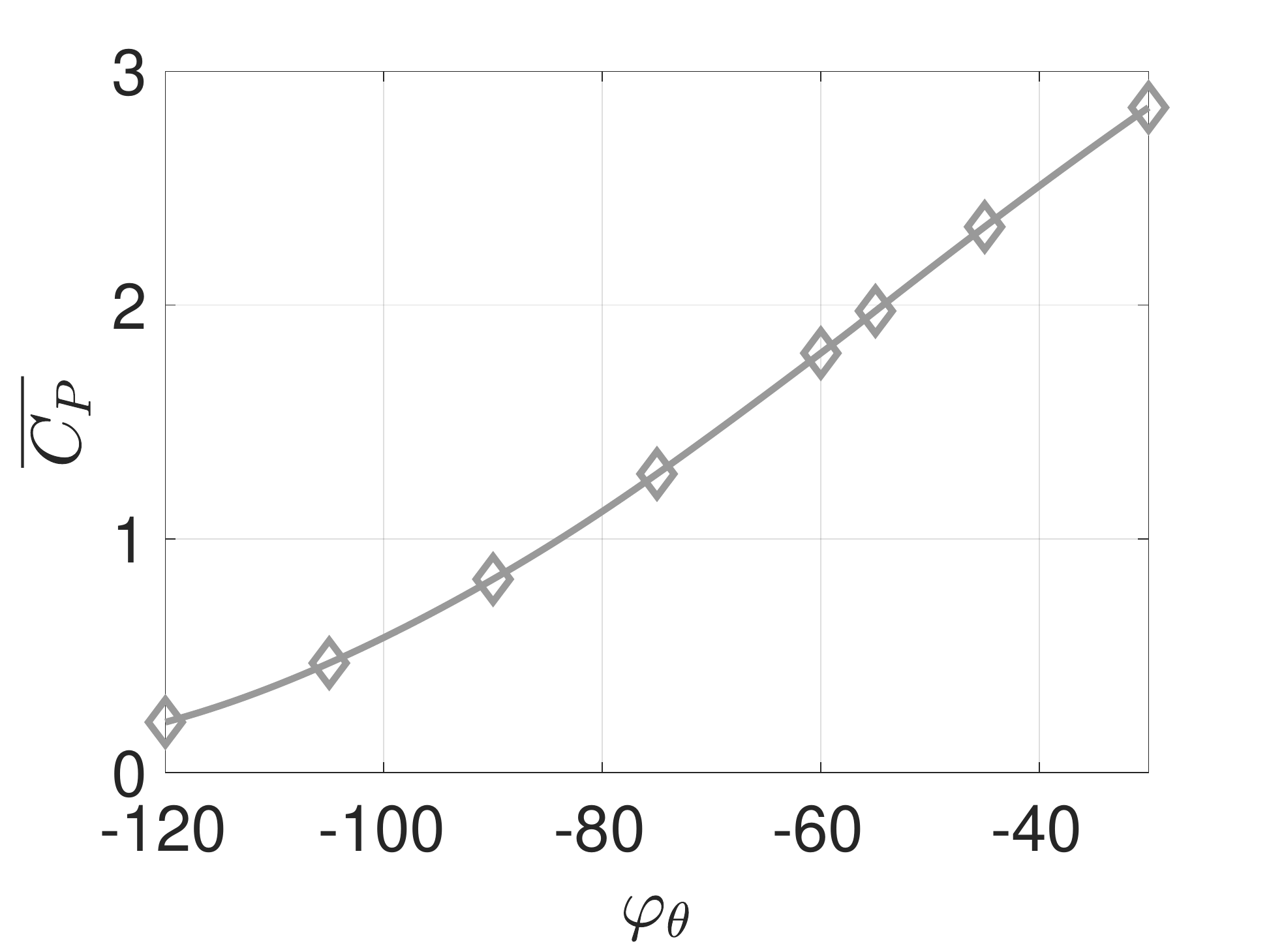}
\includegraphics[width=0.32\linewidth]{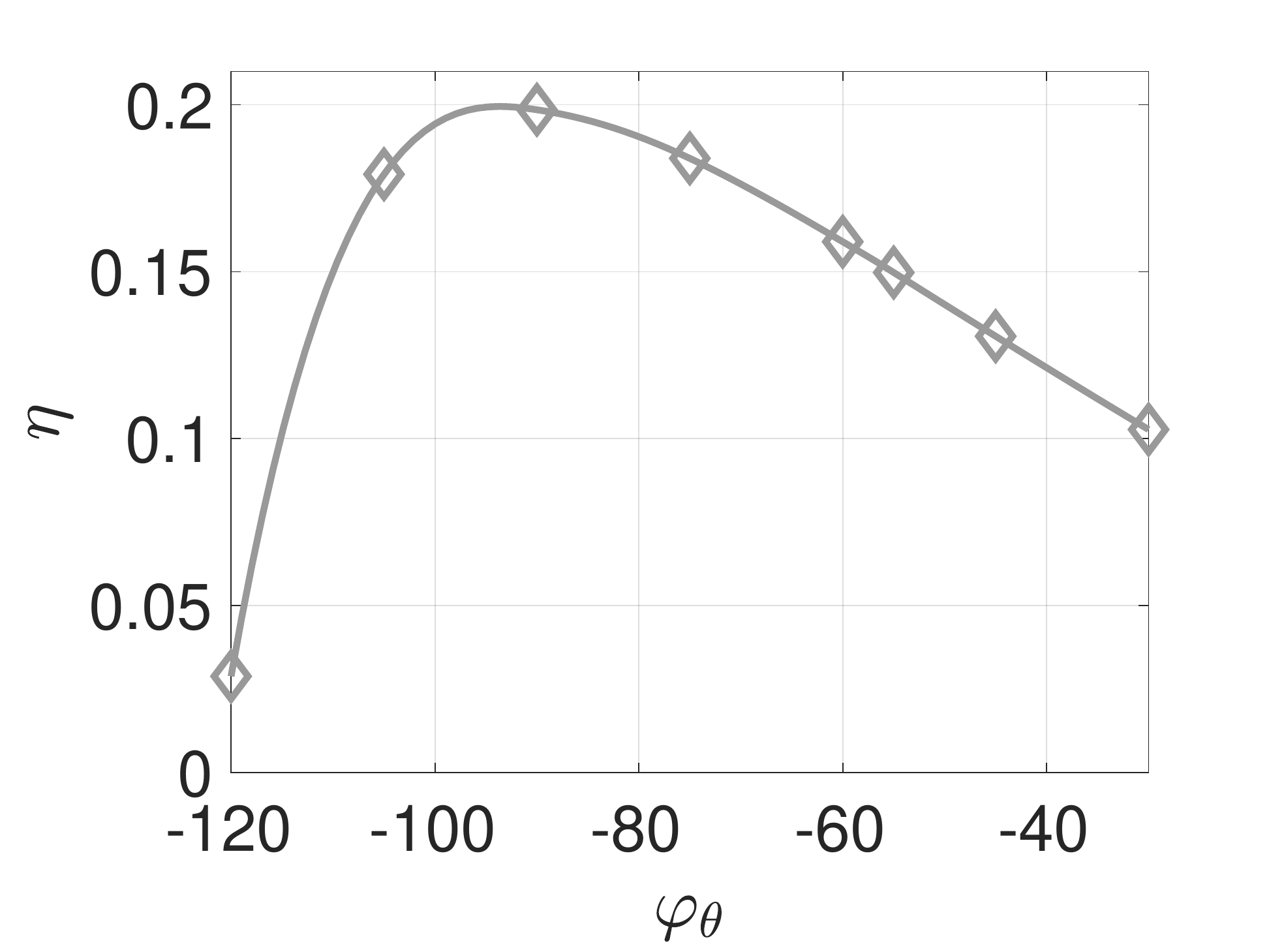}
\caption{Cycle-averaged thrust coefficient (\textit{left}), power coefficient (\textit{center}), and efficiency (\textit{right}) of the rigid fin as a function of the pitch angle amplitude (\textit{top}) and phase angle with respect to heave (\textit{bottom}).}
\label{fig:caudalfin_kpitch_amplitude_phase}
\end{inplacefigure}

\section{Thrust and lift coefficient time series}
\label{app:thrust-lift-timeseries}

Figure~\ref{fig:caudalfin_kpitch_ct_cp} shows the time evolution of thrust (left) and lift (right) coefficients for the rigid fin (in red), the fin with curvature variations $a_0^\kappa = 0.8$ and $a_2^\kappa = 0$ (in blue), and the $\kappa$-pitch configuration with $a_0^\kappa = 0.8$ (in yellow).  
\begin{inplacefigure}
\centering
\includegraphics[width=0.49\linewidth]{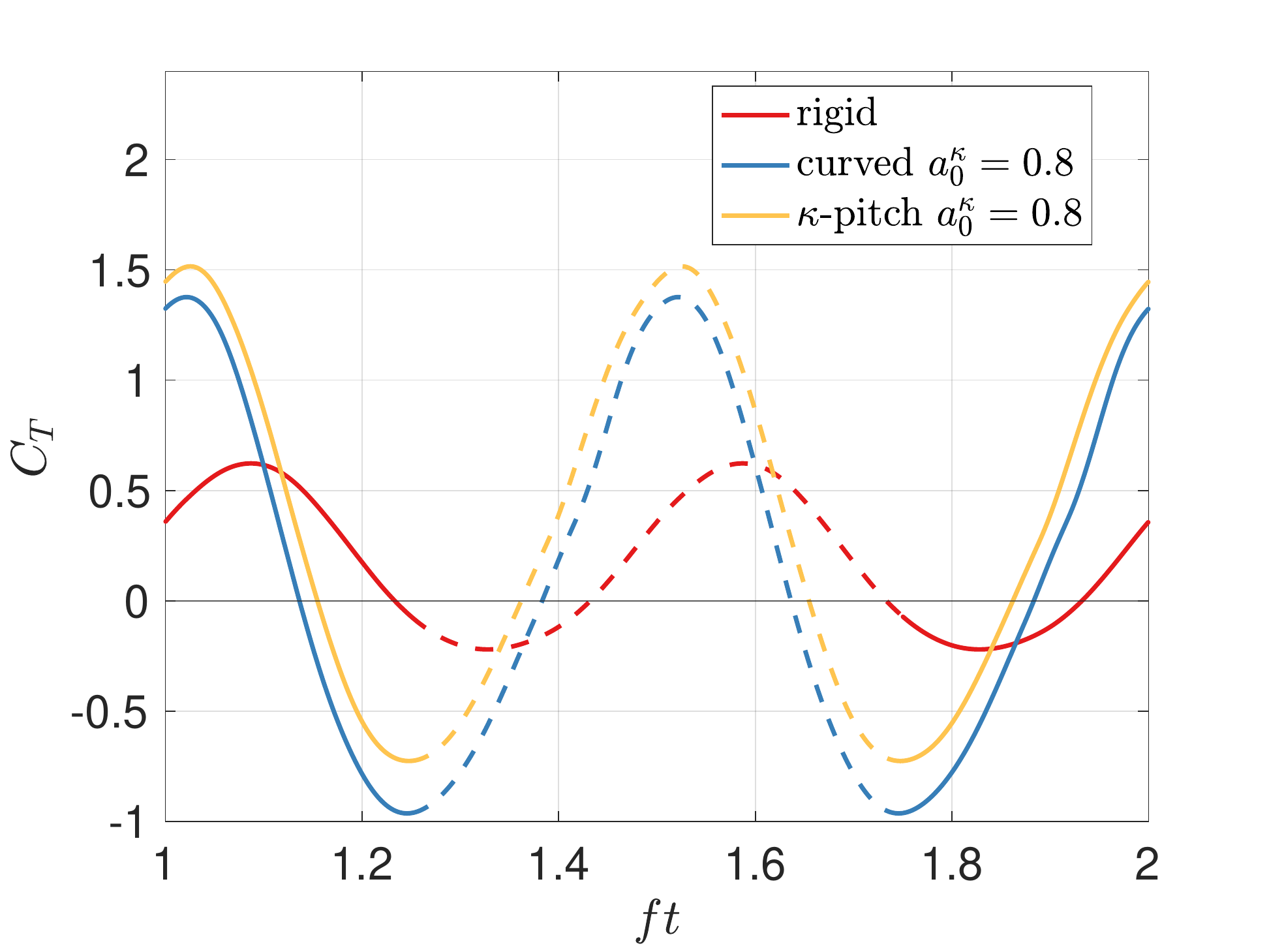}
\includegraphics[width=0.49\linewidth]{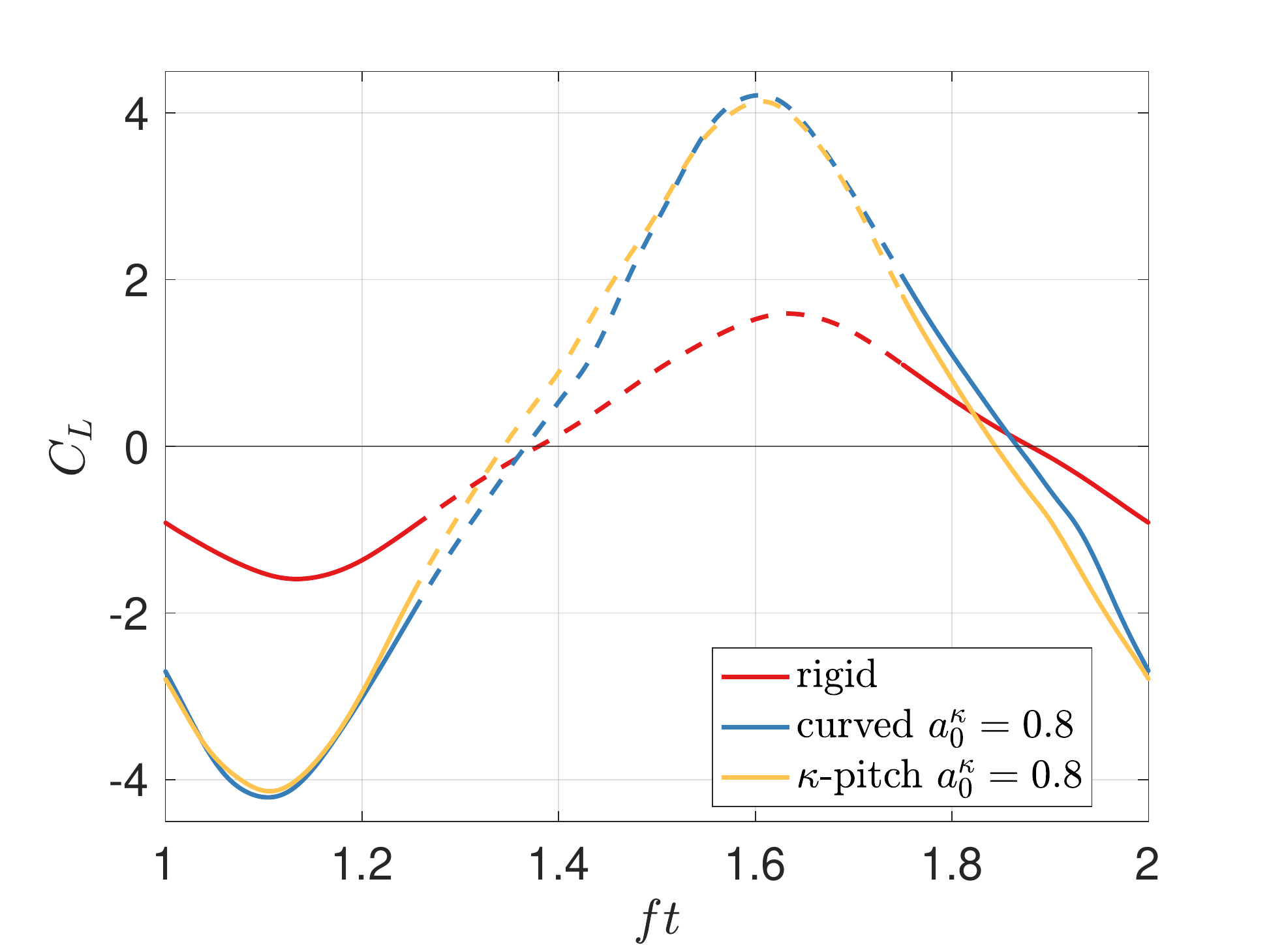}
\caption{Thrust (\textit{left}) and lift (\textit{right}) coefficients variation during a flapping cycle of the rigid, curved, and $\kappa$-pitch configurations with $a_0^{\kappa}=0.8$. Solid and dashed lines identify the upstroke and downstroke half-cycles, respectively. }
\label{fig:caudalfin_kpitch_ct_cp}
\end{inplacefigure}

\clearpage
\bibliographystyle{wim}
\bibliography{bib}